\documentclass[letterpaper,12pt]{article}
\usepackage[margin=1.1in, letterpaper]{geometry}
\usepackage{graphicx} 
\usepackage{amsmath}
\usepackage{amsfonts}
\usepackage{amssymb}
\usepackage{mathrsfs}
\numberwithin{equation}{section}
\usepackage{circuitikz}
\usepackage{tikz-3dplot}
\usepackage{rotating}
\usetikzlibrary{shapes.geometric}

\usepackage[sectionbib, square, compress, comma, sort]{natbib}
\setcitestyle{numbers}

\usepackage[linktocpage=true]{hyperref}
\hypersetup{
    colorlinks=true,
    linkcolor=blue,
    citecolor=blue,
    filecolor=magenta,      
    urlcolor=blue,
    pdftitle={Topological Manipulations on R Symmetries of Abelian Gauge Theory},
    pdfpagemode=FullScreen,
    }

\usepackage[skip=4pt plus5pt, indent=20pt]{parskip}

\newcommand\R{\mathbb{R} }
\newcommand\Z{\mathbb{Z} }
\newcommand\X{\mathcal{X} }
\newcommand\V{\mathsf{V} }
\newcommand\W{\mathsf{W} }
\newcommand\Y{\mathsf{Y} }
\newcommand\T{\mathbb{T} }

\newcommand\B{\mathsf{B} }
\newcommand\C{\mathsf{C} }

\title{\vbox{ \centerline{Topological Manipulations On $\R$ Symmetries
    }
    \bigskip 
    \centerline{Of Abelian Gauge Theory}
    \bigskip
    }}

\author{  
    \phantom{$^{| 1 \rangle, |2 \rangle }$}Burak Oğuz$^{| 1 \rangle, |2 \rangle }$ \vspace{4mm} 
    \\ {\small{\it {$^{| 1 \rangle }$Middle East Technical University, 06800 Ankara, Turkey }} } \vspace{2mm}
    \\ {\small{\it {$^{| 2 \rangle }$Abdus Salam International Centre for Theoretical Physics}} }
    \\ {\small{\it {Strada Costiera 11, Trieste 34151, Italy}} } \vspace{2mm}
    \\ { \small { \href{https://mail.google.com/mail/u/0/?fs=1&tf=cm&source=mailto&to=boguz@ictp.it}{boguz@ictp.it} } }
    }
\date{}

\begin{document}

\maketitle

\begin{abstract}
    Performing topological manipulations is a fruitful way to understand global aspects of Quantum Field Theory (QFT). Such modifications are typically controlled by the notion of Topological QFT (TQFT) coupling across different codimensions. Motivated by the recent developments involving non-compact TQFTs as the Symmetry Topological Field Theory (SymTFT) for continuous symmetries, we realize topological manipulations on global $\R$ symmetries via TQFT coupling in the simple context of non-compact abelian gauge theory. Namely, by inserting the background fields for $\R$ symmetries into non-compact TQFTs on spacetime, we study the topological gaugings. Furthermore, we explore topological defects in non-compact theories by employing the said manipulations on the half-space, which are analogous to duality defects in compact gauge theories. We examine the action of these defects on the local and extended operators, and discuss their algebra. As opposed to the duality defects, our defects act invertibly on the spectrum. To further understand their role, we discuss the way they mix with other topological defects and the resulting global symmetries. We also comment on possible applications of these ideas to the bosonic String Theory. After studying defects, we provide a detailed inspection of manipulations on $\R^{(-1)}$ and $\R^{(d-1)}$ symmetries in abelian gauge theory. Notably, we develop a novel deformation tailored for flat gauging subgroups of $\R^{(d-1)}$ symmetries, and provide an extension of it for $\R^{(p)}$ symmetries. We show that this new simple deformation captures all the topological boundary conditions of the corresponding non-compact SymTFTs as the deformation parameter is varied, which  the ordinary non-compact BF coupling cannot do.

\end{abstract}

\vspace{5mm}

April, 2025
\pagebreak

\tableofcontents

\section{Introduction}

\phantom{met} The study of global structures in Quantum Field Theory (QFT), pertaining to their generalized global symmetries \cite{Gaiotto:2014kfa} and 't Hooft anomalies, has become a prominent line of research over the years. Many novel insights were gained from the exploration of these structures across different contexts \cite{Gaiotto:2014kfa, Seiberg:2010qd, Banks:2010zn, Kapustin:2014gua, Cordova:2022ruw, Chen:2011pg, Gaiotto:2017yup, Kapustin:2013uxa, McNamara:2020uza, Harlow:2018tng, Tanizaki:2019rbk, Komargodski:2020mxz, Gaiotto:2020iye, Bhardwaj:2017xup, Apruzzi:2021nmk, Freed:2022qnc, Bhardwaj:2023ayw, Nguyen:2021naa, Nguyen:2021yld, Unsal:2020yeh, Aloni:2024jpb, Antinucci:2024zjp, Brennan:2024fgj, Kaidi:2022cpf, Freed:2017rlk, Choi:2021kmx, Choi:2022zal, Kaidi:2021xfk, Cordova:2019jnf, Cordova:2019uob, McGreevy:2022oyu, Gomes:2023ahz, Bhardwaj:2023kri, Shao:2023gho, Schafer-Nameki:2023jdn}. In these advancements, Topological QFTs (TQFT) play a prominent role, as the global data of QFTs can be probed and manipulated by stacking TQFTs along submanifolds of certain codimensionality in spacetime \cite{Kapustin:2014gua, Shao:2023gho, Schafer-Nameki:2023jdn}. 

Suppose $\X$ is a QFT endowed with the topological data of a generalized global symmetry $G$. On the symmetry of $\X$, one can perform some topological manipulations, thereby producing new QFTs with different global structures $\X' = \phi(\X)$, where $\phi$ denotes the said manipulations as a map $\phi : \X \mapsto \X'$ (we are borrowing this notation from \cite{Kaidi:2022cpf}). Such a global modification can, for example, be the gauging of a discrete subgroup $\Gamma$ of the symmetry $G$ in $\X$. Other manipulations include generalizations of the gauging concept, such as higher-gauging and half-gauging \cite{Shao:2023gho}, that utilize TQFT coupling not on the entire space-time, but on some submanifold of appropriate codimensionality. In this paper, the primary tools we operate upon are the non-compact TQFTs \cite{Antinucci:2024zjp, Brennan:2024fgj}, and the type of topological manipulations they bring about upon coupling to a QFT.

Recent years witnessed significant developments from a codimension-$(-1)$ (or, equivalently, $d+1$ dimensional) TQFT coupled to a QFT in $d$ dimensions, which goes by the name of Symmetry Topological Field Theory (SymTFT) \cite{Gaiotto:2020iye, Apruzzi:2021nmk, Burbano:2021loy, Freed:2022qnc, Bhardwaj:2023ayw, Shao:2023gho, Schafer-Nameki:2023jdn, Bhardwaj:2023kri, Apruzzi:2022rei, Antinucci:2022vyk, Kaidi:2023maf, Zhang:2023wlu, Cordova:2023bja, Antinucci:2023ezl, Sun:2023xxv, Freed:2022iao, Putrov:2025xmw, Putrov:2024uor, Najjar:2024vmm, Antinucci:2024zjp, Brennan:2024fgj, Gagliano:2024off, Kaidi:2022cpf, DelZotto:2024tae, Kapustin:2010if, Baume:2023kkf, Heckman:2024oot, Cvetic:2024dzu, Apruzzi:2023uma}, which has a close relationship to holography \cite{Witten:1998wy}. Via the SymTFT picture, the global structures of the QFT on $M_d$, and the manipulations thereof, are understood from a broad perspective via the choice of boundary conditions of the TQFT on the slab $M_d \times I$, with $I$ being an interval. The initial wave of ideas along this direction pertains to finite symmetries, such as $\Z_N$. 

A standard example for a SymTFT of $\Z_N$ symmetry, discussed in \cite{Gaiotto:2014kfa}, concerns Yang-Mills gauge theory with gauge groups $SU(N)$ and $SU(N)/\Z_N$. The latter is obtained by gauging the 1-form center $\Z_N$ symmetry of the former, and in the codimension-$(-1)$ $\Z_N$ SymTFT picture, this gauging corresponds to changing the Dirichlet boundary conditions to Neumann boundary conditions \cite{Gaiotto:2014kfa}. Mixed boundary conditions of the SymTFT for $SU(N)$ gauge theory capture other discrete gaugings of the form $SU(N)/ \Z_M$.

In the past year, these ideas have evolved into intriguing advancements toward understanding the SymTFT of more complicated symmetries, governing the set of topological manipulations on them. In particular, the SymTFT for continuous symmetries has been explored in \cite{Antinucci:2024zjp, Brennan:2024fgj, Gagliano:2024off}, which involves the coupling of non-compact TQFTs. There are also exciting discussions on SymTFTs for non-abelian symmetries \cite{Bonetti:2024cjk, Jia:2025jmn}, and codimension-$(-1)$ couplings of not TQFTs but of free theories \cite{Apruzzi:2024htg, Cvetic:2024dzu}, referred to as Symmetry Theories (SymTh).\footnote{Additionally, it was argued in \cite{Antinucci:2024zjp} that for dynamical manipulations -say, gauging a 4d complex scalar to obtain scalar QED- that introduce new local degrees of freedom, there is a controllable map between the corresponding SymTFTs of the two theories.} In addition to group-like symmetries, the notion of SymTFT for non-invertible and categorical symmetries has gone through substantial progress as well \cite{Kaidi:2022cpf, Kaidi:2023maf, Putrov:2024uor, Putrov:2025xmw, DelZotto:2025yoy, Lin:2025oml, Bhardwaj:2025jtf, Apruzzi:2024cty, Argurio:2024ewp, Bhardwaj:2024ydc, Bhardwaj:2024kvy, Franco:2024mxa, Bhardwaj:2024qrf}, with many more to come.

\textbf{Topological Manipulations on $\R$ Symmetries: } Our concern in this paper will be the study of non-compact TQFTs encoding the topological manipulations on $U(1)$ and $\R$ global symmetries, developed in \cite{Antinucci:2024zjp, Brennan:2024fgj}. We will study these global modifications in the specific examples of 2d real-valued scalar and the $4d$ Maxwell theory with $\R$ gauge group. In the compact versions of these theories, the topological manipulations have been well understood \cite{Fuchs:2007tx, Witten:1995gf}, where extensive use of T-duality and electric-magnetic duality is made. Such ideas are being revisited under the new paradigm of generalized global symmetries and dualities \cite{Choi:2021kmx, Shao:2023gho, Thorngren:2021yso, Argurio:2024ewp, Gaiotto:2014kfa, Meynet:2025zem}. Employing the analogues of these manipulations in the non-compact versions, we will discuss local (codimension-0) realizations of the topological boundary conditions of the non-compact SymTFT of \cite{Antinucci:2024zjp, Brennan:2024fgj}. By a codimension-0 realization, we mean that we realize these topological manipulations by a TQFT coupling \emph{on the spacetime} $M_d$, \emph{not on the slab} $M_d \times I$. The said manipulations are very simple to understand by virtue of the fact that path integrals are easily performed\footnote{Considering a non-compact gauge group brings about numerous issues. For example, when normalizing the path integral of the non-compact theories we typically need to divide by the volume of the group $\R$, which is infinity. Also, the partition function of the non-compact TQFT diverges \cite{Antinucci:2024bcm}. We will overlook these issues in our study of non-compact gauge theories. I would like to thank Anatoly Dymarsky for pointing out non-compactness issues.}. There are tangential works on the manipulations of $\R$ symmetries in the literature \cite{Argurio:2024ewp, Paznokas:2025epc, Arbalestrier:2025poq}. See also the more recent work \cite{Perez-Lona:2025add} for related discussions on manipulations of non-compact gauge theories.

The relevant non-compact SymTFTs living on the slab $M_d \times I$ have the $\R / \R$ and $U(1)/\R$ form of BF coupling. The former concerns global $\R$ symmetries, where one of the $\R$ fields of the TQFT is viewed as the background coupling to a $p-$form symmetry on the left boundary $M_d$, and the second $\R$ is the conjugate field. With the choice of ordinary (Dirichlet and Neumann) topological boundary conditions, the resulting symmetry, upon interval contraction, is either an $\R^{(p)}$ symmetry or its quantum dual $\R^{(d-p-2)}$. \cite{Antinucci:2024zjp} discusses another boundary condition which corresponds to a $2\pi \Z \subset \R$ subgroup gauging, and the resulting theory on $M_d$ has a $U(1)^{(p)} \times U(1)^{(d-p-2)}$ global symmetry with mixed anomaly.

The latter, $U(1)/\R$ TQFT, concerns global $U(1)$ symmetries, where the $U(1)$ field is regarded as the background field coupled to the $p-$form symmetry on $M_d$, which is inserted into a SymTFT on $M_d \times I$ with an $\R$ valued conjugate. This time, Dirichlet/Neumann boundary conditions yield $U(1)^{(p)}$ and $\Z^{(d-p-2)}$ symmetries upon interval contraction, respectively. There is also a topological boundary condition that gauges a $\Z_n \subset U(1)$ subgroup, and the resulting symmetry on $M_d$ reads $U(1)^{(p)} \times \Z_n^{(d-p-2)}$ with a mixed anomaly.

The question we investigate, while being guided by the broad perspective of the SymTFT discussed above, is what kind of topological manipulations are obtained by local couplings of the $\R/\R$ and $U(1)/\R$ TQFTs to a QFT on spacetime. To gain perspective on this TQFT coupling \cite{Kapustin:2014gua}, we recall the discrete $\Z_p$ TQFT and its coupling to a QFT. The codimension-$(-1)$ coupling of the $\Z_p$ TQFT is a SymTFT for global $\Z_p$ symmetries, capturing all the topological manipulations on the symmetry from the choice of boundary conditions on the slab. The codimension-0 coupling, on the other hand, gauges the $\Z_p$ global symmetry. Typically, this $\Z_p$ TQFT coupling arises in the notion of subgroup gauging $\Z_p \subset U(1)$ for $U(1)$ symmetries (see, for example, \cite{Seiberg:2010qd, Tanizaki:2019rbk, Shao:2023gho}. Related results on modifying topological sectors have been noted in earlier works in the literature \cite{Pantev:2005rh,Pantev:2005wj,Pantev:2005zs} employing the language of stacks, which were applied to the quantization of Fayet-Iliopoulos terms in the same gerby languge in \cite{Hellerman:2010fv,Distler:2010zg}). Thus, the $\Z_p$ TQFT coupling plays the roles of SymTFT and discrete gauging: 
\begin{equation*}
\begin{aligned}
    \text{Codimension-}(-1) \text{ coupling of } \Z_p \text{ TQFT} \quad &\mapsto \quad \text{SymTFT for } \Z_p, \\
    \text{Codimension-} \phantom{-} 0 \phantom{-} \text{ coupling of } \Z_p \text{ TQFT} \quad &\mapsto \quad \text{Gauging } \Z_p \subset U(1) .
\end{aligned}
\end{equation*}
Given that noncompact TQFTs at codimension-$(-1)$ coupling are the SymTFT for continuous symmetries, we ask the corresponding manipulations at codimension-0 coupling
\begin{equation*}
\begin{aligned}
    \text{Codimension-}(-1) \text{ coupling of } \R/\R \text{ TQFT} \quad &\mapsto \quad \text{SymTFT for } \R, \\
    \text{Codimension-} \phantom{-} 0 \phantom{-} \text{ coupling of } \R/\R \text{ TQFT} \quad &\mapsto \quad \text{Gauging } ?
\end{aligned}
\end{equation*}
\begin{equation*}
\begin{aligned}
    \text{Codimension-}(-1) \text{ coupling of } U(1)/\R \text{ TQFT} \quad &\mapsto \quad \text{SymTFT for } U(1), \\
    \text{Codimension-} \phantom{-} 0 \phantom{-} \text{ coupling of } U(1)/\R \text{ TQFT} \quad &\mapsto \quad \text{Gauging } ?
\end{aligned}
\end{equation*}
We will fill in the question marks in the simple context of abelian gauge theory by leveraging the simplicity of the path integrals of 2d free boson and the 4d pure Maxwell theories. In particular, we will demonstrate that the insertion of $\R$ background fields into an $\R/\R$ TQFT amounts to a flat gauging of the $\R$ symmetry, upon which another $\R$ symmetry of the same rank appears. That the quantum dual symmetry has the same rank is essentially the statement of T- and $S-$duality under gauging the corresponding symmetries in the compact theories\footnote{In the non-compact boson and Maxwell, the radius and electric coupling are not physical so this does not correspond to a path integral duality in its strict sense. However, the invariance under gauging still has many essential aspects of duality in the compact versions, which we will use to construct topological defects.}. From the TQFT coupling perspective, the manipulations in the non-compact versions are viewed as the theories being invariant under flat gaugings of their $\R$ symmetries, since the quantum dual is $\textsf{Rep}(\R) = \equiv \widehat \R \simeq \R$.

On the other hand, we will demonstrate that the insertion of $\R$ background fields into the $U(1)/\R$ TQFT results in a $\Z \subset \R$ subgroup gauging as has been noted in \cite{Argurio:2024ewp} from a local TQFT coupling, and also has been observed from the SymTFT side \cite{Antinucci:2024zjp}. We read off this gauging by observing that the TQFT coupling produces a $U(1)$ symmetry, which from the Pontryagin relation $\textsf{Rep}(\Z) = \widehat \Z = U(1)$ implies that the gauged group is $\Z$. It is straightforward to deform 2d real scalar and 4d $\R$ Maxwell with the $U(1)/\R$ TQFT and then to integrate over the $\R$ field, which respectively yields the compact scalar and $U(1)$ Maxwell theory. The coupling constants of the latter pair can be tuned by rescaling the $U(1)/\R$ TQFT Lagrangian using the fact that $\R$ gauge fields have no period quantization, and therefore the parameter multiplying the action need not be an integer to ensure gauge invariance. At specific values of this parameter, one also observes further subgroup gaugings of the form $\Z_p \subset U(1)$. We also develop a different deformation of non-compact abelian gauge theories, which extend the possible gaugings of $\R$ symmetries to include $\Z_p \times \Z_q$ (where gcd$(p,q)=1$) mixed gaugings in the resulting compact theories. We first develop this deformation when discussing manipulations on $\R^{(d-1)}$ symmetries because for a $(d-1)$-form symmetry background field a BF coupling is not possible in $d-$dimensions. This new deformation turns out to be richer than a BF coupling for $\R$ symmetries. Motivated by this comprehensive new deformation for $(d-1)-$form symmetries, we discuss generalizations to arbitrary rank higher symmetries. Our deformation includes a free real parameter, which we identify with the SymTFT parameter that labels the infinite family of Lagrangian algebras \cite{Antinucci:2024zjp}.  

\textbf{Topological Defects in $\R$ Gauge Theories:} An interesting consequence of the invariance under gauging the $\R$ shift symmetry is that by performing this gauging on the half-space with Dirichlet boundary conditions, a topological defect may be defined. These defects are analogues of the duality defects in the compact theories \cite{Choi:2021kmx, Kaidi:2021xfk}, constructed by gauging the $\Z_p \subset U(1)$ subgroup at specific values of coupling constants. These duality defects obey the fusion algebra of a Tambara-Yamagami category over $\Z_p$, denoted $\mathsf{TY}(\Z_p)$ \cite{Choi:2021kmx, Kaidi:2021xfk}, therefore they generate non-invertible symmetries. 

In our construction of the topological defect, we are gauging $\R$, so one may naturally expect the relevant fusion category to be a Tambara-Yamagami category over the group $\R$, namely $\mathsf{TY}(\R)$. This expectation turns out to be wrong, and the non-compact gauge theory defects act invertibly on the spectrum. We determine the action of the defect on the spectrum, and verify that our derivation also gives the correct result for the duality defects in the literature \cite{Choi:2021kmx}. The invertibility of the defect has mainly to do with the fact that in the 4d non-compact gauge theory, there are no topological surfaces (or lines in 2d boson) that can be attached to Wilson lines (or local operators in 2d boson). In \cite{Choi:2021kmx}, it was shown that the action of the duality defect on Wilson lines attaches surfaces, and converts the lines into a non-genuine line operators. The non-existence of such surfaces in the non-compact versions has to do with the mathematical statement that $\R$-bundles cannot be non-trivial as opposed to $U(1)$ bundles. Physically, this means that the $\R-$gauge theory cannot have Dirac monopoles whereas the $U(1)$ gauge theory can admit them at the cost of introducing non-trivial topology. 

Even though the half-space gauging does not give a non-invertible symmetry, it still seems to have an interesting topological nature. This we infer by inspecting the mixing between the shift generators and the defect obtained by half-space gauging. The two generators fail to commute due to a phase, which very much has a 't Hooft anomaly like flavor. We argue that the resulting global symmetry structure in the 2d non-compact boson is given by a $\textsf{Heis}(\R \times \R)$ group. We suspect that the corresponding statement for the 4d $\R$ Maxwell involves a higher-group, since the mixing is between codimension-1 defects and codimension-2 defects, both of which act on Wilson lines. 

Additionally, we discuss condensation defects that gauges the $\R$ symmetry on codimension-1. As we will show, this defect acts trivially on all operators, and the triviality is equivalent to the fact that the topological defect discussed above is invertible. We also comment on a condensation defect obtained by gauging $\Z \subset \R$ on codimension-1, which seems to act non-trivially, but we do not determine its properties in detail in the paper.

\textbf{$(-1)-$form $\R$ Symmetries and Their Topological Manipulations:} There is an almost trivial yet highly intriguing type of symmetry uncovered from the perspective of \cite{Gaiotto:2014kfa} which defines a symmetry as a topological defect. This is the notion of a $(-1)-$form symmetry, which is closely related to instantons and D$(-1)$-branes. These exotic symmetries have been studied from various perspectives in the literature \cite{Seiberg:2010qd, Tanizaki:2019rbk, McNamara:2020uza, Cordova:2019uob, Cordova:2019jnf, Heidenreich:2020pkc, Lin:2025oml}, see also the recent interesting work \cite{Perez-Lona:2025add} on the topological gauging of $\R^{(-1)}$ symmetries. The $(-1)-$form symmetries are quantum dual to $(d-1)-$form symmetries, which have several irregularities as well. $(d-1)-$form symmetries lead to the notion of decomposition, meaning the theory is a direct sum of local theories \cite{Vandermeulen:2022edk, Sharpe:2022ene}, which has been referred to as the universes in \cite{Hellerman:2006zs, Tanizaki:2019rbk, Komargodski:2020mxz}.

The abundance and the almost triviality of the $(-1)-$form symmetry come about because any top form on a spacetime manifold is closed; therefore, any local 0-form object $j_0$ can be regarded as a ``scalar Noether current'' such that $d * j_0 = 0$. This is not to say that $*j_0$ is necessarily locally exact, but one can still define a $(-1)-$form symmetry charge $Q \propto \int *j_0$ and since the integral is over the entire spacetime, there is no topological deformation to consider for $Q$ inside a correlation function. While $(-1)-$form symmetries are devoid of charged objects and many of the familiar concepts for an ordinary symmetry, the notion of coupling a background 0-form field to a $(-1)-$form symmetry exists, which allows for interesting deformations, such as a flat gauging. Since a background field for $(-1)-$form symmetry exists, it is possible to consider the SymTFT coupling, and obtain the set of topological manipulations as boundary conditions on the codimension-$(-1)$ TQFT coupling \cite{Antinucci:2024zjp, Aloni:2024jpb, Lin:2025oml, Najjar:2024vmm, Najjar:2025htp, Yu:2024jtk}. It was also demonstrated that $(-1)-$form symmetries can be spontaneously broken \cite{Aloni:2024jpb}.

Any Lagrangian field theory has $(-1)-$form symmetries, of which the currents are the terms of the Lagrangian itself. Considering a single term, one therefore has a $\R^{(-1)}$ symmetry, as the charge of the symmetry, the action, can attain any real value unless the term has a topological nature in which case it typically is a $U(1)^{(-1)}$ symmetry. The theta term in 4d gauge theory is an example of a $U(1)^{(-1)}$ symmetry. Turning on a 0-form background field $b_0$ for the $(-1)-$form symmetry, we are probing the path integral. We may further insert the field $b_0$ into TQFT terms to topologically manipulate the QFT, or deform the theory with the kinetic term of a scalar field. For instance, performing the latter for the instanton $(-1)-$form symmetry of 4d gauge theory produces the axion \cite{Reece:2023czb}. 

There are many possibilities once a background field is turned on. Staying within our context of abelian gauge theory, we will be concerned with the $(-1)-$ and $(d-1)-$form symmetries of $p-$form $\R$ gauge theories in $d= p+1$ dimensions. From the topological boundary conditions on the corresponding SymTFT \cite{Lin:2025oml, Aloni:2024jpb, Antinucci:2024zjp}, we obtain the global variants of these symmetries. While attempting to gauge $(d-1)-$form symmetries, we are led to a novel deformation than the ordinary BF coupling, since the background field in that case is a top-form and so there is no direction to take derivatives. This deformation achieves the flat gaugings of various subgroups inside $\R$, and covers all the Lagrangian algebras of the corresponding SymTFT \cite{Antinucci:2024zjp}. We also discuss this new deformation for arbitrary rank symmetries for $p-$form gauge theory in arbitrary dimensions.

\textbf{Organization of the Paper:} Section \ref{sec 2: non-invertible defects in U(1) gauge theory} reviews the non-invertible duality and condensation defects in 2d compact boson and 4d $U(1)$ Maxwell's theory \cite{Choi:2021kmx}. Section \ref{sec 3: R scalar in 2d} is devoted to the non-compact free boson in 2d, and the topological manipulations on the $\R^{(0)}$ shift symmetry. We construct topological defects and study their action. In \ref{subsec 3.7: small discussion on bosonic string theory}, we make some comments on bosonic string theory and sigma models. Section \ref{sec 4: R maxwell} discusses the non-compact Maxwell theory, and manipulations on the $\R^{(1)}$ symmetry and corresponding topological defects. In \ref{sec 5: subgroup gauging of (-1) form R symmetries}, we study manipulations of $(-1)-$ and $(d-1)-$form symmetries and develop a new gauging method for $(d-1)-$form symmetries, since the corresponding background field is a top form and cannot be entered into a BF-like theory in $d-$dimensions. In section \ref{sec 6: p-form non-compact gauge theory in arbitrary dimension}, we generalize the gauging method to the $p-$form gauge theory in arbitrary dimensions, and also comment on condensation defects. 

\textbf{Notation and Conventions:} In compact gauge theory, most of the fields and operators are denoted with sans-serif letters, whereas in the non-compact one, they are denoted by ordinary letters. When necessary, we will specify the rank of the gauge field explicity by subscripts. Given a differential form $\alpha$ on $M_d$, we denote by $|\alpha|^2$ the Hodge inner product $( \alpha, \alpha ) = \int_{M_d} \alpha \wedge * \alpha$. When convenient, we will drop the absolute signs and write $\alpha^2$. Throughout the paper, the spacetime is in Euclidean signature, unless explicitly stated otherwise.

\section*{Acknowledgments}

I am grateful to Justin Kulp for detailed feedback on the draft and for discussions, and also to the anonymous JHEP referee for a careful reading of the manuscript and helpful criticisms on the paper. Additionally, I thank Francesco Benini and Nicola Dondi for clarifying discussions on condensation defects, Lorenzo di Pietro for an illuminating discussion on the defect commutation relations in non-compact boson, and Pavel Putrov for comments on the normalization of non-finite defects. I also thank Mithat Ünsal for making me aware of the works \cite{Seiberg:2010qd, Tanizaki:2019rbk}, which indirectly led me to this paper. 

\section{Non-Invertible Defects in \texorpdfstring{$U(1)$}{U (1)} Gauge Theory}\label{sec 2: non-invertible defects in U(1) gauge theory}

\subsection{Compact Boson in \texorpdfstring{$2d$}{2d}} \label{subsec 2.1: compact boson in 2d}

We will quickly reproduce the $\Z_p$ gauging and the duality defect construction \cite{Choi:2021kmx} in the compact scalar, following the discussion of \cite{Arias-Tamargo:2025xdd}. The starting point is the compact boson at radius $R$, defined with action
\begin{equation}
    S = \frac{R^2}{2} \int d\mathsf X \wedge * d\mathsf X,
\end{equation}
where $\mathsf X$ is a $2\pi$ periodic (compact) scalar. The action has two $0-$form $U(1)$ symmetries, referred to as the momentum and the winding symmetries ${U(1)}^{(0)}_m \times {U(1)}^{(0)}_w$, with corresponding charges $\frac{1}{2\pi} \int *d\mathsf X$ and $\frac{1}{2\pi} \int d\mathsf X$. The two symmetries have a mixed anomaly, reflecting the obstruction to turning on background gauge fields for both symmetries (we will discuss a similar version of this anomaly in the next subsection for Maxwell's theory). It is well known that the path integral of this theory has a T-duality under the exchange $R \mapsto 1/2\pi R$. Let us derive this. We couple a background field $\C$ to the momentum symmetry acting as $\mathsf X \mapsto \mathsf X + \mathsf h$ and $\C \mapsto \C + d \mathsf h$. The combination $d\mathsf X - \C$ is invariant under this transformation, so we take the action to be 
\begin{equation}
    \frac{R^2}{2} \int |d\mathsf X - \C |^2,
\end{equation}
where $|d\mathsf X - \C|^2 = (d\mathsf X - \C) \wedge * (d\mathsf X - \C)$. We will mostly use $|\alpha|^2$ notation, but when appropriate we will write $\alpha^2$ without the absolute sign. To get the T-dual, we further deform this theory to 
\begin{equation} \label{compact scalar deformed with a BF term}
    \frac{R^2}{2} \int |d\mathsf X - \C|^2 + \frac{i}{2\pi} \int \C \wedge d \mathsf Y,
\end{equation} 
with $\mathsf Y$ being another compact scalar. Performing the integral over $\mathsf Y$ sets $d\C = 0$ and the period sum over $\int d\mathsf Y$ enforces the holonomy of $\C$ to be trivial. These conditions fix the gauge invariant data for the field $\C$, and since it is a flat field with trivial holonomies, one can fix away $\C$ and get back to the original compact scalar action. Alternatively, one may first perform the integral over $\C$. There are two ways to do this, one way is as outlined in \cite{Arias-Tamargo:2025xdd}, and the other is in the fashion familiar from the path integral derivations of electric-magnetic duality in 4d abelian gauge theory  \cite{Witten:1995gf, Gukov:2006jk, Deligne:1999qp}. We will frequently perform integrals of this form throughout the paper, and both approaches give the same result. For this section, we follow the first path \cite{Arias-Tamargo:2025xdd} and complete the square for $\C$ in the above action so we have 
\begin{equation}
    \frac{R^2}{2} \int \bigg( \Big|\C - d\mathsf X + \frac{1}{2\pi R^2} * d\Y \Big|^2 - \Big| d\mathsf X - \frac{1}{2\pi R^2} * d\Y \Big|^2 + d\mathsf X^2 \bigg).    
\end{equation} 
Then, we shift $ \C \mapsto \C'$ such that the path integral over $\C'$ is trivial and only affects the normalization \cite{Arias-Tamargo:2025xdd}. The remaining action, upon recollecting the terms, has the form
\begin{equation}
    \frac{1}{8\pi^2 R^2} \int d\mathsf Y^2 + \frac{i}{2\pi} \int d\mathsf X \wedge d \mathsf Y.
\end{equation}
The integral over $\mathsf X$ does not change the path integral since both $d\mathsf Y$ and $d\mathsf X$ have period quantization, so the second term contributes $e^{2\pi i\Z} = 1$. Thus, we are left with a compact scalar at radius $R' = \frac{1}{2\pi R}$. The two descriptions can be derived from the same path integral over \ref{compact scalar deformed with a BF term} (up to some normalization constants), so they are dual. 

With a twist in our deformation, we can obtain a more interesting duality. The topological term $\C \wedge d \mathsf Y$ is still well defined if we multiply it by an integer $p$. Then, the field equation of $\Y$ imposes $\C$ to be a $\Z_p$ gauge field taking values in $H^1(M_2, \Z_p)$ because the holonomies are now constrained to be $\int \C \in \frac{2\pi}{p} \Z$. Since $\C$ is the background field for the shift symmetry, restricting it to a $\Z_p$ gauge field from a $U(1)$ gauge field signals a subgroup gauging $\Z_p \subset U(1)$. It is still easy to perform the integral over $\C$ to obtain a dual theory, which yields the action
\begin{equation}
    \frac{p^2}{8\pi^2 R^2} \int d\mathsf Y^2.
\end{equation}
We thus see that this subgroup gauging acts on the radius as $R' = p/2\pi R$. At the self-dual radius $R = \sqrt{p/2\pi}$, the gauging (duality transformation) leaves the compact scalar invariant. This is precisely the stage at which non-invertible duality defects come up. First, observe that in the action 
\begin{equation}
    \frac{R^2}{2} \int |d\mathsf X - \C|^2 + \frac{ip}{2\pi} \int \C \wedge d \mathsf Y,
\end{equation} 
the field equation of $\mathsf Y$ allows us to construct the topological operators $\eta_\alpha = \exp\big( i \alpha \oint \C \big)$, where $\alpha \in \Z_p$ because $\C$ is a flat $\Z_p$ gauge field. Consider the topological boundary condition $\C \big| = 0$, where we assume that the 2d (Euclidean) spacetime has a boundary. Typically, the surface is divided into left ($x<0$) and right ($x>0$), and $x=0$ is the boundary. The Dirichlet boundary condition is topological since $d\C = 0$, so deforming the boundary costs nothing. Therefore, this condition defines a topological defect $\mathcal D$, on which the $\Z_p$ defects are transparent $\eta_\alpha\big| = 1$.

This duality defect turns out to obey a non-invertible fusion algebra \cite{Choi:2021kmx}. To understand this, we start with the compact scalar $\mathsf X$ at self-dual radius in the entire spacetime, and couple the TQFT $\frac{ip}{2\pi} \int \C \wedge d \mathsf Y$ on the right part of the sheet. From the T-duality, we know this to be another compact scalar, and since the radius is at the self-dual point, the radii of the two scalars are the same. On the boundary, a Chern-Simons coupling between $\mathsf Y$ and $\mathsf X$ exists at level $p$, which plays the role of inducing the correct boundary conditions on the left and the right fields so that the construction respects $\C \big| = 0$. One can see this as follows. Studying the field equation of $\C$ in the above action, the boundary condition $\C \big|$ reveals the mixing condition $d\mathsf X \big| = i* d\mathsf Y \big|$, which is the duality transformation from one compact boson to the other. Considering also the boundary variations from the kinetic terms on both sides, one concludes that the duality defect $\mathcal D$ is realized by the action 
\begin{equation} \label{Action for the duality defect with radius p}
    S = \frac{p}{4\pi} \int_{x<0} d\mathsf X^2 + \frac{p}{4\pi} \int_{x>0} d \mathsf Y^2 + \frac{ip}{2\pi} \int_{x=0} \mathsf X \wedge d \mathsf Y.
\end{equation}
What is the sense in which this gives a non-invertible symmetry? Let us study the fusion algebra of defects. We have $\eta = \exp \big( i\oint \C \big)$ and $\mathcal D$. The $\eta$ generators furnish a representation of $\Z_p$, the quantum dual arising from the $\Z_p$ gauging. Also, from the topological boundary condition, we have $\eta \times \mathcal D = \mathcal D = \mathcal D \times \eta$ since $\eta \big| = 1$. The last fusion to consider is $\mathcal D \times \mathcal D$. As argued in \cite{Choi:2021kmx}, the only consistent form for this fusion that is compatible with the rule $\eta \times \mathcal D = \mathcal D$ reads
\begin{equation}
    \mathcal D \times \mathcal D = \sum_{q \in \Z_p} \eta^q,
\end{equation}
which is non-invertible. In summary, we have the fusion algebra
\begin{equation}
    \eta^p = 1, \quad \eta \times \mathcal D = \mathcal D = \mathcal D \times \eta, \quad \mathcal D \times \mathcal D = \sum_{i \in \Z_p} \eta^i.
\end{equation}
This defines a Tambara-Yamagami category associated to the group $\Z_p$, denoted $\mathsf{TY}(\Z_p)$. Therefore, the topological line operators in the compact scalar theory above obey the fusion algebra of the $\mathsf{TY}(\Z_p)$ category, hence the non-invertibility.

\subsection{\texorpdfstring{$U(1)$}{U(1)} Maxwell in \texorpdfstring{$4d$}{4d}} \label{subsec 2.2: U(1) Maxwell in 4d}

There are many similarities between the construction of T-duality defects in the $2d$ compact boson and that of $S-$duality defects of $4d$ $U(1)$ gauge theory at specific values of the electric coupling constant $e$. We will consider the theory with the theta angle set to zero. The action reads
\begin{equation}
    \frac{1}{2e^2} \int_{M_4} \mathsf F^2,
\end{equation}
where $\mathsf F = d\mathsf A$ with $\mathsf A$ being the $U(1)$ gauge field. There are two $U(1)$ 1-form symmetries, called the electric and magnetic symmetry, with charges $\int * d \mathsf A $ and $\int d\mathsf A$ respectively. There is a mixed 't Hooft anomaly between the two symmetries. To see this, we can try to turn on background gauge fields $\mathsf B_e$ and $\mathsf B_m$ which are 2-form $U(1)$ fields. The global electric symmetry acts by shifting the gauge field $\mathsf A \mapsto \mathsf A + \lambda$, with $\lambda$ being a flat 1-form gauge field with quantized periods $\int \lambda \in 2\pi\Z$. It is analogous to the shift symmetry of the compact boson, and one couples the background field $\mathsf B_e$ by the replacement 
\begin{equation}
    \mathsf F \mapsto \mathsf F - \mathsf B_e,
\end{equation}
so that the gauge transformations $\mathsf A \mapsto \mathsf A + \lambda$, $\mathsf B_e \mapsto \mathsf B_e + d\lambda$ leave the action invariant even when $\lambda$ is not closed, so the 1-form shifts are indeed gauged by virtue of the replacement above. 

For the magnetic symmetry, one couples the background field via the deformation $\propto \int \mathsf F \wedge \mathsf B_m$. Since the magnetic symmetry is a topological symmetry, in the sense that its conservation law $d * j^{(2)}_m = d ( d\mathsf A ) = 0$ holds even for fields not obeying Maxwell's equations, the coupling with the background field is also topological, and indeed is of the form $\int \mathsf B \wedge * j^{(2)}$. 

To couple both the electric and magnetic symmetry backgrounds, we replace $\mathsf F $ with $\mathsf F - \mathsf B_e$ everywhere, including the coupling $\int \mathsf F \wedge \mathsf B_m$, so the action in the presence of both background fields reads 
\begin{equation}
    \frac{1}{2e^2} \int |\mathsf F - \mathsf B_e|^2 + \frac{i}{2\pi} \int (\mathsf F - \mathsf B_e) \wedge \mathsf B_m.
\end{equation}
This action cannot be made invariant under background gauge transformations of both electric and magnetic symmetry by a local counterterm, which is the hallmark of the mixed 't Hooft anomaly. 

Just as in the compact boson, one may gauge a $\Z_p$ subgroup of the electric 1-form symmetry (momentarily, we turn off the $\mathsf B_m$ background). This is done so by inserting $\mathsf B_e$ into a $\Z_p$ TQFT, which enforces it to take values in $H^2(M_4, \Z_p)$. Utilizing the $S-$duality of the path integral \cite{Witten:1995gf, Gukov:2006jk, Deligne:1999qp} one can integrate over $\mathsf B_e$ and get a dual theory. We start with the $\Z_p$ TQFT coupling
\begin{equation} \label{Z_p electric subgroup gauging in U(1) Maxwell}
    \frac{1}{2e^2} \int |\mathsf F - \mathsf B_e|^2 + \frac{ip}{2\pi} \int \mathsf B_e \wedge d \mathsf V,
\end{equation}
where $\mathsf V$ is another $1-$form gauge field. We can take the $\mathsf B_e$ integral as we did for the compact boson \cite{Arias-Tamargo:2025xdd}, or we can proceed as in \cite{Witten:1995gf}. Both are equivalent, but the latter is more practical. In the latter, one uses the gauged shift symmetry $(\mathsf A ,\mathsf B_e) \mapsto (\mathsf A + \lambda,  \mathsf B_e + d\lambda)$ to set $\mathsf A = 0$ by choosing $\lambda = -\mathsf A$, so the action has the simple form $\frac{1}{2e^2} \mathsf B_e^2 + \frac{ip}{2\pi} \mathsf B_e \wedge d\mathsf V$ \footnote{A careful treatment of such manipulations in pure $U(1)$ gauge theory, which also take into account various determinant factors, has been done in \cite{Witten:1995gf}. We will omit the details on determinant factors for the sake of brevity, as they will not be crucial in our discussion.}, and the integration over $\mathsf B_e$ yields 
\begin{equation}
    \frac{e^2p^2}{8\pi^2} \int d\mathsf V^2.
\end{equation}
One can then construct duality defects $\mathcal D_p$ at self-dual electric coupling $e^2 = 2\pi/p$. The construction follows similar steps as in the compact boson, and the action 
\begin{equation}
    \frac{p}{4\pi} \int_{x<0} d\mathsf A^2 + \frac{p}{4\pi} \int_{x>0} d\mathsf V^2 + \frac{ip}{2\pi} \int_{x=0} \mathsf A \wedge d\mathsf V,
\end{equation}
realizes the duality defect. As before, we divide the spacetime into two and the defect action at $x=0$, the Chern-Simons coupling of $\mathsf A$ and $\mathsf V$ at level $p$, follows from the boundary condition $\mathsf B_e \big| = 0$ prior to integrating $\mathsf B_e$ out in equation \ref{Z_p electric subgroup gauging in U(1) Maxwell}. 

Before closing this subsection we would like to note that in both the 2d compact boson and the 4d Maxwell, the above constructions can be generalized for rational values of the coupling constants. For example, when $e^2 = \frac{2\pi q}{p}$ with gcd$(q,p)=1$, one can show that the theory is self-dual under the mixed subgroup gauging $\Z_p^{(1)} \times \Z_q^{(1)} \subset U(1)_e^{(1)} \times U(1)_m^{(1)}$, which is not obstructed by the 't Hooft anomaly due to the gcd$(q,p)=1$ condition. The details of this construction will not be important for the non-compact versions, so we will not review the details and refer to \cite{Thorngren:2021yso}.

\subsection{Condensation Defects} \label{subsec 2.3: Condensations in compact theory}

There is a difference about the 2d construction and the 4d one, due to the fact that the orientation reversal of the duality defect, $\overline{\mathcal D}$, is equal to itself in 2d \cite{Choi:2021kmx}; whereas it is not in 4d. An interesting topological defect can be constructed by placing $\mathcal D_p$ and $\overline{\mathcal D}_p$ on parallel codimension-1 manifolds inside the 4d spacetime, and placing them on top of each other. The resulting object is called a condensation defect \cite{Gaiotto:2019xmp, Roumpedakis:2022aik, Choi:2021kmx, Choi:2022zal, Kaidi:2021xfk}, and is related to gauging the $\Z_p$ symmetry on the codimension-1 manifold. In general, a $q-$form symmetry may be gauged on a codimension-$k$ manifold if there is not an obstruction. The obstruction is sometimes referred to as the $q-$form symmetry being $k-$anomalous \cite{Roumpedakis:2022aik}. The higher anomalies of higher symmetries are classified by cohomology groups (of appropriate rank) over higher-classifying spaces, just as ordinary anomalies of 0-form symmetries are classified by group cohomology over ordinary classifying spaces.

In this subsection, we will review the construction of a condensation defect obtained by composing the $S-$duality defect in 4d pure $U(1)$ gauge theory and its orientation reversal. Later on, we will make parallels with this construction when discussing the 4d pure $\R$ gauge theory. Our discussion will follow \cite{Choi:2021kmx, Choi:2022zal}.\footnote{One can find a clear exposition on condensation defects in 4d gauge theory in the fourth lecture on non-invertible symmetries by Clay C\'ordova as part of the 2025 TASI school \cite{Cordova:2025TASI}.}

Let us take the 4d (Euclidean) spacetime to be of the form $M_3 \times \R$ with $M_3$ being a closed 3-manifold, and let $x \in \R$ parameterize the real line. On $x = 0$ and $x=\varepsilon$, we insert the duality defects $\mathcal D_p$ and $\overline{\mathcal D}_p$ respectively where the electric coupling is at the self-dual point $e^2 = \frac{2\pi}{p}$. Therefore, on the $M_3 \times [0,\varepsilon] \equiv M_3 \times I_\varepsilon$ slice, we are gauging the $\Z_p^{(1)} \subset U(1)^{(1)}_e$ subgroup, which means that we insert the electric 2-form background field $\mathsf B_e$ into a $\Z_p$ TQFT on the slice. Recall that the defects $\mathcal D_p$, $\overline{\mathcal D}_p$ are defined via the Dirichlet boundary conditions on the background field, and for the setup depicted in the figure below, this corresponds to  $\mathsf B_e \big|_{\partial (M_3 \times I_\varepsilon)} = 0$, which means $\mathsf B_e \big|_{M_3 \times \{0\}} = 0 = \mathsf B_e \big|_{M_3 \times \{\varepsilon\}}$. Therefore, on the slab, the background takes values in 
\begin{equation}
    H^2 \big(M_3 \times I_\varepsilon, \partial(M_3 \times I_\varepsilon), \Z_p \big)    
\end{equation}
where the $\partial(M_3 \times I_\varepsilon)$ part ensures that $\B_e$ satisfies the Dirichlet boundary conditions. Therefore, the sum over $\B_e$ in the $M_3 \times I_\varepsilon$ slab is characterized by what is called the relative cohomology group above \cite{Choi:2021kmx}. Using Poincaré-Lefschetz duality, this is equivalent to a sum over the homology 2-cycles $H_2(M_3, \Z_p)$ due to the chain of isomorphisms 
\begin{equation}
    H^2 \big( M_3 \times I_\varepsilon, \partial( M_3 \times I_\varepsilon), \Z_p \big) \simeq H_2 \big( M _3 \times I_\varepsilon, \Z_p \big) \simeq H_2 \big( M_3, \Z_p \big), 
\end{equation}
where in the last one we contracted the slab. This contraction in turn produces a condensation defect $\mathcal C_p(M_3)$, obtained as the schematic limit

\begin{equation}
    \mathcal C_p(M_3) \equiv \lim_{\varepsilon \to 0} \ \mathcal D_p \big( M_3 \times \{0\} \big) \times \overline{\mathcal D}_p \big( M_3 \times \{\varepsilon\} \big).
\end{equation}

\begin{center}
    \resizebox{1\textwidth}{!}{
    \begin{circuitikz}
    \tikzstyle{every node}=[font=\LARGE]
    
    \draw  (0,15.25) rectangle (15,6.5);
    \draw [fill={rgb,255:red,180; green,180; blue,180}] (7.5,15.25) rectangle (10,6.5);
    \draw [color={rgb,255:red,107; green,107; blue,107}, line width=2pt] (7.5,11) -- (7.25,10.75);
    \draw [color={rgb,255:red,107; green,107; blue,107}, line width=2pt] (7.5,11) -- (7.75,10.75);
    \draw [color={rgb,255:red,107; green,107; blue,107}, line width=2pt] (10,10.75) -- (10.25,11);
    \draw [color={rgb,255:red,107; green,107; blue,107}, line width=2pt] (10,10.75) -- (9.75,11);
    \draw [color={rgb,255:red,107; green,107; blue,107}, line width=4pt] (7.5,6.5) -- (7.5,15.25);
    \draw [color={rgb,255:red,107; green,107; blue,107}, line width=4pt] (10,6.5) -- (10,15.25);

    \node [font=\huge, color={rgb,255:red,107; green,107; blue,107}] at (8.7,5.75) {$M_3 \times I_\varepsilon$};
    \node [font=\huge, color={rgb,255:red,107; green,107; blue,107}] at (6.5, 11) {$\mathcal D_p$};
    \node [font=\huge, color={rgb,255:red,107; green,107; blue,107}] at (11,11) {$\overline{\mathcal D}_p$};    

    \draw [line width=2pt, ->, >=Stealth] (16.5,11) -- (18.5,11);    
    \node [font=\Large, color={rgb,255:red,0; green,0; blue,0}] at (17.5,11.75) {slab contraction};

    \draw  (20,15.25) rectangle (35,6.5);
    \draw [fill={rgb,255:red,180; green,180; blue,180}] (27.5,15.25) rectangle (27.75,6.5);
    \draw [color={rgb,255:red,107; green,107; blue,107}, line width=8pt] (27.5,6.5) -- (27.5,15.25);
    \draw [color={rgb,255:red,107; green,107; blue,107}, line width=8pt] (27.75,6.5) -- (27.75,15.25);

    \node [font=\huge, color={rgb,255:red,107; green,107; blue,107}] at (29,11) {$\mathcal{C}_p$};
    \node [font=\Huge, color={rgb,255:red,107; green,107; blue,107}] at (27.5,5.75) {$M_3$};
    \end{circuitikz}
    }
\end{center}

The action localized on $M_3$ that describes this defect is given by \cite{Choi:2022zal, Cordova:2025TASI}
\begin{equation} \label{Z_p condensation defect action}
\begin{aligned}
    S_{\mathcal C_p(M_3)} = \underbrace{\frac{ip}{2\pi} \int_{M_3} \mathsf A \wedge d \mathsf a}_{\mathcal D_p} - \underbrace{\frac{ip}{2\pi} \int_{M_3} \mathsf a \wedge d \mathsf V}_{\overline{\mathcal D}_p} = \frac{ip}{2\pi} \int_{M_3} \mathsf a \wedge ( d \mathsf A - d \mathsf V),
\end{aligned}
\end{equation}
where $\mathsf a$ is a $U(1)$ 1-form gauge field living on the condensation defect, and $\mathsf A$ and $\mathsf V$ are the gauge fields on the left and right side of the defect. The equations of motion for $\mathsf a$ imposes $\frac{p}{2\pi} (d\mathsf A - d\mathsf V) = 0$, which means that the difference $\mathsf A - \mathsf V$ is a flat gauge field having $\Z_p$ valued holonomies, namely $\mathsf A - \mathsf V \equiv \frac{2\pi}{p} \C \in H^1(M_3, \Z_p)$. Hence, upon integrating the $U(1)$ field $\mathsf a$, the remaining degree of freedom is the $\Z_p$ gauge field. The emergence of this field is natural from the point of view of the condensation as gauging the $\Z_p^{(1)} \subset U(1)_e^{(1)}$ on the codimension-1 manifold $M_3$; so that $\C$, localized on $M_3$, is the object that detects this higher gauging. 

To better understand the relation of $\mathcal C_p(M_3)$ and the sum over homology 2-cycles $H_2(M_3, \Z_p)$, observe that on $M_3$, a fixed holonomy for $\C$ on some 1-cycle $\gamma \subset M_3$ uniquely determines a $\Z_p$ symmetry defect $\eta$ on the Poincaré dual surface $S$ (namely, $S$ links $\gamma$ on $M_3$). Note also that since a flat gauge field is characterized by its holonomy, each distinct class in $H^1(M_3, \Z_p)$ determines a holonomy $\oint_\gamma \C$, which in turn fixes an $\eta(S)$ for $S\in H_2(M_3,\Z_p)$. Thus, the sum over the $\Z_p$ gauge field can be recast as a sum over $\eta(S)$ defects (through the Poincaré duality $H^1(M_3,\Z_p) \simeq H_2(M_3,\Z_p)$) for each class in $H^1(M_3,\Z_p)$. Therefore, the condensation defect can be written as \cite{Choi:2021kmx, Cordova:2025TASI, Roumpedakis:2022aik}
\begin{equation} \label{Z_p condensation defect as a sum over 2-homology cycles in Maxwell}
    \mathcal C_p(M_3) \propto \sum_{S \in H_2(M_3, \Z_p)} \eta(S).
\end{equation}
The standard normalization factor in this construction is $1/p$. When we discuss the non-compact case, it is trickier to fix a normalization, but as we will see, one can construct a condensation defect that has a similar expression as the above defect.

At the end of the previous subsection, we mentioned constructions of duality defects when the magnetic as well as the electric symmetry is gauged, and the electric coupling has the rational value $e^2 = \frac{2\pi q}{p}$. Here we additionally note that in that case, one can define duality defects $\mathcal D_{p,q}$ from the self-duality under gauging $\Z_p^{(1)} \times \Z_q^{(1)} \subset U(1)^{(1)}_e \times U(1)^{(1)}_m$. Along similar lines, by fusing $\mathcal D_{p,q}$ with its orientation reversal, one may construct a condensation defect $\mathcal C_{p,q}(M_3)$ that implements the subgroup gauging on the codimension-1 submanifold $M_3$. Since we will not need the details of this construction in the $\R$ gauge theory discussion, we refer to \cite{Choi:2022zal, Roumpedakis:2022aik, Cordova:2025TASI, Choi:2021kmx, Gaiotto:2019xmp} for more details.

\section{Non-Compact Boson in 2d}  \label{sec 3: R scalar in 2d}

We begin with the real-valued scalar field theory in two dimensions. Some of the manipulations we perform have also been carried out in recent works \cite{Argurio:2024ewp, Paznokas:2025epc}. 

We would like to note that due to the infrared divergences in the two-point functions, this is not strictly a well-defined model\footnote{I thank Justin Kulp for pointing this out to me.}. We will overlook this problem and proceed with a formal treatment of the path integral, omitting any discussion on the correlation functions. At the very least, the formal manipulations we perform motivate some interesting questions in the context of bosonic string theory, discussed in subsection \ref{subsec 3.7: small discussion on bosonic string theory}.

\subsection{Gauging the \texorpdfstring{$\R$}{R} Shift Symmetry} \label{subsec 3.1: flat gauging R shift}

We begin with the standard action
\begin{equation}
    \X_\R : \hspace{1cm} \frac{1}{2} \int d X \wedge * d X, \phantom{ \X_\R : \hspace{1cm} }
\end{equation}
where we label this theory as $\X_{\R}$. This model has a 0-form $\R$ shift symmetry acting as $X \mapsto X + h$ for a real valued constant $h$. As opposed to the compact scalar, we have only the shift symmetry, because $\int dX = 0$ for the real-valued scalar so the analogue of winding symmetry does not exist here. 

To enact topological manipulations on this $\R^{(0)}$ We introduce an $\R$-valued 1-form gauge field $C$ and write the gauged action 
\begin{equation}
    \frac{1}{2} \int |dX-C|^2
\end{equation}
which is invariant under the gauged shift transformations $X \mapsto X + h$ and $C \mapsto C + d h$ where $h$ is promoted to a real valued function. Manipulations of continuous symmetries has a very crucial contrast with manipulations of discrete symmetries, which is the fact that a discrete gauge field is always flat, whereas a continuous gauge field can have a nonzero field strength. Thus, whereas a discrete gauging is always a topological manipulation, a continuous gauging may introduce dynamical degrees of freedom. Throughout, we will focus on flat gaugings of continuous symmetries, which are achieved by coupling the relevant background fields into non-compact TQFTs. 

As is familiar from path-integral duality derivations in compact theories reviewed in the previous section, we introduce a topological coupling for $C$ with another $\R$ valued scalar $Y$, of which we denote the path integral as $\phi(\X_{\R})$, $\phi$ being the topological manipulation corresponding the $C \wedge dY$ coupling:
\begin{equation}
    \phi(\X_{\R}) : \hspace{1cm}\frac{1}{2} \int |dX-C|^2 + \frac{i}{2\pi r} \int C \wedge dY, \phantom{ \phi(\X_{\R}) : \hspace{1cm} }
\end{equation}
where $r$ is an arbitrary coefficient. Integration over $Y$ forces $C$ to be a flat connection, and there is no sum over periods $dY$ in contrast with the compact scalar. We can then set $C=0$ and get back to the initial action. Alternatively, we can perform the integral over $C$ first along the lines of \cite{Witten:1995gf, Deligne:1999qp}. Using the gauged shift symmetry, we fix $X=0$, so that the action is of the form $\frac{1}{2} C^2 + \frac{i}{2\pi r} C \wedge dY$. Integrating $C$ amounts to inserting its field equation into the action since the path integral is a Gaussian in $C$. This yields the ``T-dual" action 
\begin{equation}
    \frac{1}{8\pi^2 r^2} \int dY \wedge * dY.
\end{equation}
Since the non-compact boson has no radius parameter, the $1/8\pi^2r^2$ factor can be rescaled to $1/2$ without changing the physics. This is equivalent to choosing $r = 1/2\pi$. Therefore, this operation is not in fact a duality, but it is a reflection of the fact that the non-compact boson remains invariant under the topological manipulation that corresponds to a flat gauging of its $\R$ symmetry. 

Even though this is not a duality transformation, we will discuss that it is possible to define a topological defect along similar lines as in the compact boson duality defect, by employing the half-space gauging. In later sections, we will also argue that while the parameter $r$ seems unphysical as far as the gauging operation is concerned, it will appear physical once we consider the action of the topological defect on local operators $V_s= e^{is X}$ of the noncompact boson. 

That the non-compact boson is invariant under gauging its $\R^{(0)}$ symmetry \emph{topologically} fits nicely with the fact that in 2d, the quantum dual of a 0-form Abelian symmetry $G^{(0)}$ is given by $\mathsf{Rep}(G^{(0)}) = \widehat G^{(0)} = \text{Hom}(G^{(0)}, U(1))$, and $G = \R$ is self-dual in the sense that $\mathsf{Rep}(\R) = \widehat \R = \R$.

\subsection{Constructing Topological Defects in \texorpdfstring{$\R$}{R} Scalar} \label{subsec 3.1: topological defect in non-compact boson}

We have seen that the non-compact boson is invariant under the flat gauging of its shift symmetry. By exploiting this invariance, we will construct a topological defect via the half-space gauging construction. The construction is straightforward and has parallel steps with the constructions outlined in section \ref{sec 2: non-invertible defects in U(1) gauge theory}. 

\begin{figure}[!ht]
    \centering
    \resizebox{0.6\textwidth}{!}{%
    \begin{circuitikz}
    \tikzstyle{every node}=[font=\LARGE]
    \draw  (2.5,0.25) rectangle (22.5,-9.75);
    \draw [ fill={rgb,255:red,0; green,128; blue,128} ] (12.5,0.25) rectangle (22.5,-9.75);
    \draw [ color={rgb,255:red,0; green,64; blue,128}, line width=4pt, short] (12.5,-9.75) -- (12.5,0.25);
    \draw [ color={rgb,255:red,0; green,64; blue,128}, line width=2pt, short] (12.5,-4.75) -- (12.25,-5);
    \draw [ color={rgb,255:red,0; green,64; blue,128}, line width=2pt, short] (12.5,-4.75) -- (12.75,-5);
    \node [font=\LARGE] at (7.5,-10.5) {$M_-$};
    \node [font=\LARGE] at (17.75,-10.5) {$M_+$};
    \node [font=\Huge, color={rgb,255:red,0; green,64; blue,128}] at (12.5,-10.5) {$\mathfrak B$};
    \node [font=\Huge] at (7.25,-4.75) {$\mathcal X_{\mathbb R}$};
    \node [font=\Huge] at (17.75,-4.75) {$\phi (\mathcal X_{\mathbb R} )\simeq \mathcal X_{\mathbb R}$};
    \end{circuitikz}
    }%
    
\caption{We divide the spacetime into two parts, joined along an interface boundary $\mathfrak B$. The left and right hand sides of $\mathfrak B$ are denoted $M_-$ and $M_+$ respectively. We gauge the $\R$ shift symmetry on the $M_+$ part, which produces the same theory. Then, Dirichlet boundary conditions on $C$ near $\mathfrak B$ defines a topological defect since the gauging operation is topological.}

\label{fig: half-space gauging on non-compact boson}
\end{figure}

We start with a 2d Euclidean spacetime\footnote{It is alluring to think of this as a string worldsheet with some collection of scalar fields $X^i$ parameterizing a target space. In subsec \ref{subsec 3.7: small discussion on bosonic string theory}, we touch upon a number of intriguing aspects of our gauging ideas concerning bosonic string theory.} $M_2$, which is divided into left $M_-$ and right $M_+$. We consider a single scalar field $X$ and gauge the $\R^{(0)}$ symmetry in the $M_+$ half, see figure \ref{fig: half-space gauging on non-compact boson}. The left and right parts are described by
\begin{equation} \label{Gauging the R shift symmetry of X}
\begin{aligned}
    \X_\R \big|_{M_-}&: \hspace{3cm} \frac{1}{2} \int_{M_-} dX^2 \\
    \phi(\X_\R) \big|_{M_+}&: \hspace{1cm} \frac{1}{2} \int_{M_+} |dX - C|^2 + \frac{i}{2\pi r} \int_{M_+} C \wedge dY. \phantom{ \phi_r(\X_\R) \big|_{M_+}: \hspace{1cm} }
\end{aligned}
\end{equation}
From the field equation of $Y$, the topological operator $\zeta = \exp \big( i\oint C \big)$ defines an $\R$ symmetry generator. We impose the Dirichlet boundary condition $C \big|_{\mathfrak B} =0$, which is topological due to the flatness of $C$. Upon integrating $C$, we obtain a topological defect $\mathfrak D$ supported on $\mathfrak B$ inside the non-compact boson theory. To understand how $\mathfrak D$ mixes the left and right fields after $C$ is integrated out, we look at the equations of motion for $C$, which then suggest the following mixing conditions on $\mathfrak B$ for the $X$ and $Y$ fields
\begin{equation}
    dX \big|_{\mathfrak B} = \frac{i}{2\pi r} *dY \big|_{\mathfrak B}.
\end{equation}
This is analogous to the compact version. When $C$ is integrated, the resulting defect should respect this boundary condition on the interface. The presence of this defect is going to manifest itself with a defect action localized on $\mathfrak B$, so upon integrating $C$ with Dirichlet boundary conditions we should have the action
\begin{equation}
    S = \frac{1}{2} \int_{M_-} dX^2 + \frac{1}{2} \int_{M_+} dY^2 + \mathcal S_{\mathfrak D}
\end{equation}
(for now, we will set $r=1/2\pi$ since it does not change anything interesting. When we come back to the action on local operators we will restore $r$). The defect action $\mathcal S_{\mathfrak D}$ must be such that i) it is a topological term on $\mathfrak B$, ii) its variation produces the correct topological boundary condition on $X$ and $Y$, and iii) for simplicity, it only contains the fields $X$ and $Y$. On $M_+$, the boundary fluctuations of $Y$ produce $* dY$, and on $M_-$, those of $X$ produce $*dX$. From these, we determine $\mathcal S_{\mathfrak D}[X,Y]$ as to satisfy
\begin{equation}
\begin{aligned}
    \delta Y\big|_{\mathfrak B} &: \quad \frac{\delta \mathcal S_{\mathfrak D}}{\delta Y} + * dY \big|_{\mathfrak B} = 0, \\
    \delta X \big|_{\mathfrak B} &: \quad \frac{\delta \mathcal S_{\mathfrak D}}{\delta X} + * dX \big|_{\mathfrak B} = 0.
\end{aligned}
\end{equation}
Clearly, the 1d TQFT that meets our requirements is given by a BF type coupling between $X$ and $Y$
\begin{equation}
    \mathcal S_{\mathfrak D} = i \int_{\mathfrak B} X \wedge dY.
\end{equation}
Therefore, the action 
\begin{equation}
    \X_{\R,\mathfrak D}: \hspace{2cm} \frac{1}{2} \int_{M_-} dX^2 + \frac{1}{2} \int_{M_+} dY^2 + i \int_{\mathfrak B} X \wedge dY \phantom{ \X_{\R, \mathfrak D}: \hspace{1cm} }
\end{equation}
realizes the topological defect $\mathfrak D$ in the non-compact boson theory. We denote the theory with the insertion of the codimension-1 defect as $\X_{\R,\mathfrak D}$ as in figure \ref{fig: half space gauging at self-dual coupling with r-parameter in X_R} below.
\begin{figure}[!ht]
    \centering
    \resizebox{0.6\textwidth}{!}{%
    \begin{circuitikz}
    \tikzstyle{every node}=[font=\LARGE]
    \draw  (2.5,0.25) rectangle (22.5,-9.75);
    \draw [ fill={rgb,255:red,255; green,255; blue,255} ] (12.5,0.25) rectangle (22.5,-9.75);
    \draw [ color={rgb,255:red,0; green,64; blue,128}, line width=4pt, short] (12.5,-9.75) -- (12.5,0.25);
    \draw [ color={rgb,255:red,0; green,64; blue,128}, line width=2pt, short] (12.5,-4.75) -- (12.25,-5);
    \draw [ color={rgb,255:red,0; green,64; blue,128}, line width=2pt, short] (12.5,-4.75) -- (12.75,-5);
    \node [font=\LARGE] at (7.5,-10.5) {$M_-$};
    \node [font=\LARGE] at (17.75,-10.5) {$M_+$};
    \node [font=\Huge, color={rgb,255:red,0; green,64; blue,128}] at (12.5,-10.5) {$\mathfrak B$};
    \node [font=\Huge, color={rgb,255:red,0; green,64; blue,128}] at (13.5,-4.75) {$\mathfrak D$};
    \node [font=\huge] at (7.25,-4.75) {$\mathcal X_{\mathbb R}$};
    \node [font=\huge] at (17.75,-4.75) {$\mathcal X_{\mathbb R}$};

    \node [font=\huge] at (-0.75,-4.75) {$\mathcal X_{\mathbb R, \mathfrak D} =$};

    \end{circuitikz}
    }%
    
\caption{Due to invariance under gauging $\phi(\X_\R) \simeq \X_\R$, it is possible to define a topological defect $\mathfrak D$ realized by a Chern-Simons like coupling between the fields of $M_-$ and $M_+$.}

\label{fig: half space gauging at self-dual coupling with r-parameter in X_R}
\end{figure}

What is the fusion algebra of defects in the presence of $\mathfrak D$? Since the analogous object in the compact boson gives a non-invertible duality defect, one may anticipate the above defect to obey a non-invertible algebra. The topological defects in the theory are the $\R$ generators $\zeta$ along with the defect $\mathfrak D$. From the Dirichlet boundary condition, $\zeta$ is trivial on the defect so $\zeta \times \mathfrak D = \mathfrak D = \mathfrak D \times \zeta$. 

The important fusion rule to determine is $\mathfrak D \times \mathfrak D$. As reviewed in section \ref{sec 2: non-invertible defects in U(1) gauge theory}, the compact boson version of this fusion involving $\mathcal D_p$ was determined in \cite{Choi:2021kmx} to be that of a $\Z_p$ Tambara Yamagami category $\mathsf{TY}(\Z_p)$, which follows from the compatibility with the rule $\eta \times \mathcal D_p = \mathcal D_p = \mathcal D_p \times \eta$ . With a standard normalization, the quantum dimension for the defect $\mathcal D_p$ reads \cite{Choi:2021kmx}
\begin{equation}
    \langle \mathcal D_p \rangle = \sqrt p.
\end{equation}
Therefore, the algebra
\begin{equation}
    \mathcal D_p \times \mathcal D_p = \sum_{i\in \Z_p} \eta^i
\end{equation}
preserves the quantum dimensions of both sides, since the right hand side contains the direct sum of $p$ many invertible defects, each of which has unit dimension (as invertible defects always do).

Based on this, what can we extract about the fusion algebra of the non-compact defect $\mathfrak D$? First of all, $\mathfrak D$ is defined by employing invariance under gauging the $\R$ shift symmetry; and second of all, the generators of the quantum dual $\R$ symmetry $\zeta$ are trivial on the support of $\mathfrak D$ due to the Dirichlet boundary condition, so $\zeta \times \mathfrak D = \mathfrak D = \mathfrak D \times \zeta$ holds. By observing that $\mathcal D_p$ is defined by employing self-duality under $\Z_p$ gauging and the quantum dual $\Z_p$ generators $\eta$ are transparent on $\mathcal D_p$ due to Dirichlet boundary conditions, one may suggest that $\mathfrak D$ obeys the $\mathsf{TY}(\R)$ algebra by parallel with the fact that $\mathcal D_p$ obeys $\mathsf{TY}(\Z_p)$ algebra.

There are many objections to this expectation. An immediate one is that Tambara-Yamagami fusion categories $\mathsf{TY}(G)$ are defined for finite groups $G$. A recent work \cite{Marin-Salvador:2025stc} on the quantum algebras literature explored the construction of Tambara-Yamagami \emph{tensor} categories for more generic groups, such as $\R$\footnote{When the group no longer has finitely many simple objects, i.e. it is non-compact, the corresponding tensor category is no longer fusion \cite{Marin-Salvador:2025stc}.}. In \cite{Marin-Salvador:2025stc}, it was shown that the non-invertible element $\tau$ of the $\mathsf{TY}(\R)$ category obeys the fusion rule 
\begin{equation}
    \tau \otimes \tau \simeq L^2(\R),
\end{equation}
with $L^2(\R)$ being the space of square integrable real functions. This expression roughly translates as the composition being a direct integral indexed over the simple objects of $\R$, which is reminiscent of the simpler $\mathsf{TY}(\Z_p)$ algebra. To better motivate this connection, one may naively try to regularize the non-compact boson by a compact boson of radius $R$, where the limit $R \to \infty$ washes away all the winding modes, which produces the non-compact scalar. If we require $R^2 = \frac{p}{2\pi}$ with $p$ an integer, we can construct $\mathcal D_p$ defects obeying the $\mathsf{TY}(\Z_p)$ algebra. The $R \to \infty$ limit then corresponds to the discrete limit $p \to \infty$. However, it would be erroneous to identify this limit directly with the non-compact boson with the defect $\mathfrak D$ because in the limit, the quantum dimension
\begin{equation}
    \lim_{p\to \infty} \langle \mathcal D_p \rangle \sim \sqrt p\to \infty
\end{equation}
reflects a countable infinity, since the limit is a discrete one over the integers; whereas the direct integral in the $\mathsf{TY}(\R)$ category implies that the quantum dimension of the non-invertible element $\tau$, assuming such a concept can be defined for a continuous tensor category using some measure theoretic notions capturing the ``count of simple objects", should yield an uncountable infinity, as the right hand side of $\tau \otimes \tau$ contains elements parameterized by $\R$.  

Therefore, even though the $R\to \infty$ limit can be regarded as some regularization of the non-compact boson, requiring the existence of the duality defect puts a constraint on the limit one can take, and that limit is not ``non-compact" enough to produce the topological defect $\mathfrak D$. To better understand the fusion algebra, and in particular the (non-)invertibility of the defect $\mathfrak D_r$, we puruse a different direction, which is to determine its action on the spectrum of local operators, and we will see that $\mathfrak D_r$ is in fact an invertible defect.

\subsection{Action on Local Operators} \label{subsec 3.3: action on local operators non-compact boson defect}

We would like to understand the action of $\mathfrak D$ defect on the local operators of the form $V_s = e^{isX}$ ($s\in\R$) in the non-compact boson. Studying this is important to understand the role of the defect in the theory. For convenience, we write the theory with the insertion of the defect again, this time restoring $r$ factors 
\begin{equation} \label{The non-compact boson defect Lagrangian}
    \X_{\R,\mathfrak D}: \hspace{2cm} \frac{1}{4\pi r} \int_{M_-} dX^2 + \frac{1}{4\pi r} \int_{M_+} dY^2 + \frac{i}{2\pi r} \int_{\mathfrak B} X \wedge dY. \phantom{ \X_{\R, \mathfrak D}: \hspace{1cm} }
\end{equation}
Previously, we set $r=1/2\pi$ as the value of $r$ does not change the physics (observe that we have also rescaled the coefficient for the $X$ kinetic term here). This way of writing is both analogous to the compact defect action \ref{Action for the duality defect with radius p}, and it is easier to see the action of the defect with this presentation. We will argue that different choices of $r$ changes the action of the defect, which reads:
\begin{equation}
    \mathfrak D : \quad V_s \mapsto V_{rs}.
\end{equation}
Therefore, even though different choices of $r$ does not change the non-compact boson theory, it nevertheless plays a role in the action on local operators. We thus add a label to the defect $\mathfrak D_{r}$. 

To derive this action, we study the effect of gauging the $\R$ shift symmetry topologically on a non-compact boson with the insertion of the vertex operator. First, observe inserting the operator at point $y\in M_2$ in the worldsheet of the single non-compact boson theory modifies the path integral as  
\begin{equation}
    \big\langle V_s(y) \big\rangle = \int \mathcal DX \exp \Big( - \frac{1}{4\pi r} \int_{M_2} dX^2 + isX(y) \Big).
\end{equation}
When gauging the $\R$ symmetry, we need to make the replacement $dX \mapsto dX -C$ everywhere, but the term $isX(y) = is \int 
\delta_y \wedge X$\footnote{Here, $\delta_y$ is the Poincaré dual of the point $y$, namely $\int \delta_y \wedge X = \int  \delta^2(x-y) X(x) \ d^2x = X(y)$} creates an issue. We will move around this problem with a trick, and verify that our procedure makes sense by showing that the action of the duality defect $\mathcal D_p$ can correctly be derived in the same way. 

The trick is as follows. Consider the non-compact boson theory on $M_2 = \mathfrak B \times \R$ with a topological defect $\mathfrak D$ insertion, which extends along the $t$ direction, parameterizing the $\R$. Towards the left of $\mathfrak D$, we insert $V_s(y)$, which modifies the action with the term $is X(y)$. We extend a line $L$ towards the point $y$ from the defect $\mathfrak D$, and insert the ``Verlinde"\footnote{It would have been essentially a Verlinde line, had it been a non-trivial operator when inserted on a closed line. } operator $\exp \big( is \int_L dX )$. The line $L$ ends at the point $y$ and the locus of $\mathfrak D$ as depicted in the figure below. If $L$ was an infinite or closed line without boundary, it would have defined a trivial defect since $dX$ is globally defined in the non-compact boson, and the Stokes' theorem makes the operator vanishing. However, since $L$ ends at the point $y$, we have $\exp \big( is \int_L dX \big) = \exp \big( is X(y) \big) = V_s(y)$. Thus, we can write the action on the left of $\mathfrak D_r$ as 

\begin{figure}[!ht]
    \centering
    \resizebox{0.5\textwidth}{!}{%
    \begin{circuitikz}
    \tikzstyle{every node}=[font=\Large]
    \draw [ color={rgb,255:red,64; green,128; blue,128} , fill={rgb,255:red,64; green,128; blue,128}, line width=0.2pt ] (0,11.5) circle (0.1cm);
    \draw [ color={rgb,255:red,64; green,128; blue,128}, line width=2pt, dashed] (0,11.5) -- (6,11.5);
    \draw [ color={rgb,255:red,0; green,64; blue,128}, line width=4pt, short] (6,16) -- (6,7);
    \node [font=\Huge, color={rgb,255:red,0; green,64; blue,128}] at (5.25,14.5) {$\mathfrak{D}_r$};
    \node [font=\Large, color={rgb,255:red,64; green,128; blue,128}] at (0,12.1) {$V_s(y)$};
    \node [font=\Large, color={rgb,255:red,64; green,128; blue,128}] at (2.75,10.75) {$L$};
    \draw [ color={rgb,255:red,64; green,128; blue,128}, line width=2pt, short] (3,11.5) -- (3.25,11.75);
    \draw [ color={rgb,255:red,64; green,128; blue,128}, line width=2pt, short] (3,11.5) -- (3.25,11.25);

    \draw [ line width=1pt ] (9.5,16) rectangle (-4.5,7);
    
    \end{circuitikz}
    }%

\caption{From the defect $\mathfrak D_r$ we stretch out a straight line $L$ ending on the support of the $V_s$ operator. We can then attach the invisible line $\exp\big( is \int_{L} dX \big)$ to $V_s$. This does not make $V_s$ a non-genuine local operator because the line is trivial. }
    
\label{fig: stretching a line L from vertex operator V(y) to the topological defect D_r}
\end{figure}

\begin{equation}
     -S_{V_s} = -\frac{1}{4\pi r} \int dX^2 + is \int_L dX.
\end{equation}
Now, we have $dX$ everywhere in the action. Denoting the Poincaré dual to the open line $L$ as $\delta_L$, we write 
\begin{equation}
     -\frac{1}{4\pi r} \int dX^2 + is \int \delta_L \wedge dX.
\end{equation}
The downside of this trick is that the notion of Poincaré duality for submanifolds with a boundary is mathematically non-trivial to deal with. We will be schematic with our usage of Poincaré dual for open manifolds. The important point is that from a physical point of view, the trick produces the correct results (this we will verify in the compact boson since the action of the duality defect on operators is known in the literature). 

Overlooking the mathematical subtleties, we study the effect of moving the defect $\mathfrak D_r$ past the point $y$. By construction, the effect of this deformation is a gauging of the $\R$ symmetry. So, to understand what happens when $\mathfrak D_r$ is swept past $y$, we should understand what happens when the $\R$ symmetry is gauged in the action with the operator insertion. That is, we would like to understand
\begin{equation}
     -\frac{1}{4\pi r} \int |dX-C|^2 + is \int \delta_L \wedge (dX-C) - \frac{i}{2\pi r} \int C \wedge dY,
\end{equation}
after $C$ is integrated out. If $s=0$, we know very well that integration over $C$ produces the non-compact boson theory $\frac{1}{4\pi r} |dY|^2$. The distributional nature of the additional term should manifest itself as a singularity on the $Y$ description when $s\neq 0$. This should be easy to see for the readers who have some familiarity with Gukov-Witten surface operators in 4d gauge theory and the action of $S-$duality (or more generically $SL(2,\Z)$ duality) on them. These are discussed in detail in \cite{Gukov:2006jk}. We will closely draw analogies from \cite{Gukov:2006jk} in the following.

\begin{figure}[!ht]
    \centering
    \resizebox{1\textwidth}{!}{%
    \begin{circuitikz}
    \tikzstyle{every node}=[font=\LARGE]
    \draw [ color={rgb,255:red,64; green,128; blue,128} , fill={rgb,255:red,64; green,128; blue,128}, line width=0.2pt ] (5,11.5) circle (0.1cm);
    \draw [ color={rgb,255:red,64; green,128; blue,128}, line width=2pt, dashed] (5,11.5) -- (8.75,11.5);
    \draw [ color={rgb,255:red,0; green,64; blue,128}, line width=4pt, short] (8.75,16.5) -- (8.75,6.5);
    \node [font=\Huge, color={rgb,255:red,0; green,64; blue,128}] at (8,14) {$\mathfrak{D}_r$};
    \node [font=\Large, color={rgb,255:red,64; green,128; blue,128}] at (5,12.25) {$V_s(y)$};
    \node [font=\Large, color={rgb,255:red,64; green,128; blue,128}] at (7,10.75) {$L$};
    \draw [ color={rgb,255:red,64; green,128; blue,128}, line width=1pt, short] (6.75,11.5) -- (7,11.75);
    \draw [ color={rgb,255:red,64; green,128; blue,128}, line width=1pt, short] (6.75,11.5) -- (7,11.25);

    \draw [ color={rgb,255:red,0; green,64; blue,128}, line width=4pt, short] (21.25,16.5) -- (21.25,6.5);
    \node [font=\Huge, color={rgb,255:red,0; green,64; blue,128}] at (22,14) {$\mathfrak{D}_r$};
    \draw [ color={rgb,255:red,64; green,128; blue,128}, line width=2pt, dashed] (21.25,11.5) -- (25,11.5);
    \node [font=\Large, color={rgb,255:red,64; green,128; blue,128}] at (23,10.75) {$L'$};
    \node [font=\Large, color={rgb,255:red,64; green,128; blue,128}] at (25.25,12.5) {$V'_s(y) = V_{rs}(y)$};
    \node [font=\Huge, color={rgb,255:red,64; green,128; blue,128}] at (5.75,10.75) {};
    \node [font=\Huge, color={rgb,255:red,64; green,128; blue,128}] at (5.5,10.75) {};
    \node [font=\Huge, color={rgb,255:red,64; green,128; blue,128}] at (5.75,11) {};
    \node [font=\Huge, color={rgb,255:red,64; green,128; blue,128}] at (5.75,11) {};
    \draw [ color={rgb,255:red,64; green,128; blue,128}, line width=1pt, short] (23.5,11.5) -- (23.25,11.75);
    \draw [ color={rgb,255:red,64; green,128; blue,128}, line width=1pt, short] (23.5,11.5) -- (23.25,11.25);
    \draw [ color={rgb,255:red,64; green,128; blue,128} , fill={rgb,255:red,64; green,128; blue,128}, line width=0.2pt ] (25,11.5) circle (0.1cm);
    
    
    \draw [line width=1pt, ->, >=Stealth] (13.75,11.5) -- (16.25,11.5);
    \node [font=\Large] at (15,12.25) {Deform $\mathfrak{D}_r$};
    \draw [ line width=1pt ] (12.5,16.5) rectangle (-2.5,6.5);
    \draw [ line width=1pt ] (32.5,16.5) rectangle (17.5,6.5);
    \end{circuitikz}
    }%
    
\caption{As the topological defect $\mathfrak D_r$ is swept across the $V_s(y)$ insertion, the line $L$ is deformed to $L'$, and the $\mathfrak D_r$ passing over $V_s$ produces a new operator $V'_s(y)$ which we argued to be $V_{rs}(y)$. To this operator, a trivial line $\exp\big( i rs \int_{L'} dY \big)$ is attached. As the operator inserted on $L'$ is transparent, $V'_s(y)$ is also a genuine local operator.}
    
\label{fig: sweeping D_r across V_s(y) gives V'_s(y) = V_rs(y)}
\end{figure}

To explicitly see the appearance of the singularity in the $Y$ description, we proceed to gauge $X = 0$ using the shift symmetry manifest in the above action, and then integrate over $C$ to obtain 
\begin{equation}
    -\frac{1}{4\pi r} \int |\widehat{dY}|^2 = -\frac{1}{4\pi r} \int |dY - 2\pi r s \delta_L|^2.
\end{equation}
The $U(1)$ gauge theory version of this action in 4d with surface operators was discussed in Section 2.4 of \cite{Gukov:2006jk}. In the path integral over $Y$, the contributions come from field configurations obeying the singular boundary conditions near $L$:
\begin{equation}
    dY = 2\pi r s \delta_L + \cdots,
\end{equation}
where the ellipses denote terms that are regular near the line. 

Had $L$ been a closed line and the fields were compact scalars, this boundary condition would have defined a Verlinde-like line operator. In the non-compact boson for closed $L_{\text{closed}}$, this operator is going to be trivial since in the $X$ description this amounts to the insertion  $\int_{L_{\text{closed}}} dX = 0$. However, having an $L$ such that $\partial L = y$ gives an insertion of $V_s$ in the $X$ description. In the $Y$ description, it is easy to see that this gives again the point operator due to the boundary condition, but the parameter being $rs$ this time. Therefore, gauging the $\R$ symmetry acts as $V_s \mapsto V_{rs}$. Equivalently, this is the action of $\mathfrak D_r$ on the vertex operator $V_s(y)$ 

This appears as a $\R^\times = \R / \{ 0 \}$ group action. The existence of the defect labelled by $r \in \R^\times$ can be thought of as an enhancement of the symmetry structure of the non-compact boson, in similar vein to how the $\mathcal D_p$ defect extends the symmetry of compact boson theory to $\mathsf{TY}(\Z_p)$ fusion category. The big contrast in our case is we appear to have a non-compact group extending the ordinary shift symmetry. We also note that since the shifts and $\mathfrak D_r$ both act on vertex operators, they should have a non-trivial product between them. The details of this twist would be encoded in the commutator of the two generators, which reads
\begin{equation}
    \begin{aligned}
    \Big( D^{\text{shift}}_t \times \mathfrak D_{r} \Big) \cdot V_s &= \quad \ D^{\text{shift}}_t \cdot V_{rs}  \quad = \exp \big( i rs t  \big) \ V_{rs}, \\
    \Big( \mathfrak D_r \times D^{\text{shift}}_t \Big) \cdot V_s &= \exp \big( ist \big) \mathfrak D_r \cdot V_s = \exp \big( ist \big) V_{rs}.
\end{aligned}
\end{equation}
The two generators do not commute, and the failure of their commutations is given by a phase of the form 
\begin{equation} \label{Mixing between shift generator and the non-compact topological defect}
    D_t^{\text{shift}} \times \mathfrak D_r = e^{it(r-1)} \mathfrak D_r \times D^{\text{shift}}_t.
\end{equation}
This is very reminiscent of a 't Hooft anomaly, and potentially reflects a deeper topological structure in the non-compact boson theory. One should then expect a non-trivial group product $\R ``\times " \R^\times$ enhancing the ordinary shift symmetry as a result of the half-space gauging defects. 

To gain further insight into the question of what kind of a twisted product captures the above relation, we proceed as follows\footnote{I would like to thank Lorenzo di Pietro for an illuminating discussion on this.}. We first rewrite $r = e^{\mathfrak t}$ so that $\R^\times$ action turns into an $\R$ one since $r \times r'$ becomes $\mathfrak t + \mathfrak t'$ where $\mathfrak t$ can be an arbitrary real number. Then, we write the symmetry generators as exponentials 
\begin{equation}
    D^{\text{shift}}_t \equiv e^{i t Q} \quad ; \quad \mathfrak D_{\mathfrak t} \equiv e^{i \mathfrak t \mathfrak d},
\end{equation} 
where $Q$ and $\mathfrak d$ are the Lie algebra generators. Going back to the relation \ref{Mixing between shift generator and the non-compact topological defect} and expanding to linear order in $t,\mathfrak t$, we get the commutation (observe that $r-1 \approx \mathfrak t$ near the identity of the group, so the phase factor takes the form $e^{it\mathfrak t}$, which makes everything much more transparent)
\begin{equation}
    [\mathfrak d , Q ] = i.
\end{equation}
This is the familiar canonical commutation relation we know from quantum mechanics $[q,p] = i$! Mathematically, this is called a Heisenberg algebra, and the corresponding Heisenberg group fits into a short exact sequence 
\begin{equation}
    1 \to U(1) \to \textsf{Heis}(\R \times \R) \to \R \times \R \to 1,
\end{equation}
where the $U(1)$ reflects the non-commutativity of the two generators in \ref{Mixing between shift generator and the non-compact topological defect}. One of the key properties of this group is that it has, up to isomorphisms, a unique irreducible representation, called the Stone-von Neumann-Mackey representation\footnote{In the TASI 2023 school, the fourth lecture of the Differential Cohomology series by Gregory W. Moore has a nice exposition on Heisenberg groups \cite{Moore:2023TASI}. Recently, \cite{Moore:2025tmt} appeared on arXiv, which is the typed version of these lectures. }. This is essentially the basic quantum mechanics problem of a single particle on the real line, and by choosing a particular basis, one can think of $Q$ and $\mathfrak d$ as position and momentum on the internal space with symmetry $\R\times \R$, which act on states of the Hilbert space. Thinking in terms of Lorentzian signature, since we know the action of these objects on the vertex operators, which span the operator algebra, and borrowing the operator-state correspondence ideas by leveraging the fact that the non-compact boson is a conformally invariant theory, we know the action of these generators on the states of the Hilbert space. In an eigenbasis of one of the symmetries, say the shift symmetry, $Q$ acts to give the eigenvalue, and $\mathfrak d$ acts to shift the $Q$ eigenvalue. Therefore, the Hilbert space is organized into the irreducible representation of the global symmetry group. 

We note that the appearance of the Heisenberg group is actually quite natural, if we regard the single boson as a Wess-Zumino-Witten (WZW) model of target group $\R$. In a generic WZW, the global symmetries consists of left- and right- actions of the group $G_L \times G_R$, with a mixed anomaly. And indeed, the Hilbert space is organized by the representations of the group (more precisely, they are organized by the representations of the affine algebra, which includes the WZW level and has finitely many integrable representations) in a WZW theory. Regarding the non-compact boson as a WZW model, we then capture the correct $\R \times \R$ structure, and the fact that Hilbert space is a representation of $\textsf{Heis}(\R \times \R)$ and not $\R\times \R$ is a consequence of the mixed anomaly. From this point of view, the defect does not change much about what we know in this theory.

We also note that a somewhat similar phenomenon happens in the quantization of topologically non-trivial $(p-1)-$form $U(1)$ gauge theory, where the Hilbert space is related to the representation of $\textsf{Heis} \big( \check H^p(N_{n-1}) \times \check H^{n-p}(N_{n-1}) \big)$ \cite{Moore:2023TASI}, with $\check H$ denoting the $\check{\text{C}}$ech cohomology over the space $N_{n-1}$ (where space-time is of the form $N_n = N_{n-1} \times \R$). Taking $n=2$ and $p=1$, we get the topologically non-trivial 0-form gauge theory, namely a compact boson theory which is also a WZW model, so the similarity is understood.

\subsubsection{Revisiting the Compact Boson}

We will make some comments for the compact boson and see that the same trick of stretching a line $L$ from the support of the vertex operator to the topological defect produces the correct action. The discussion is essentially the same, where we replace $r \mapsto 1/p$ with $p$ being an integer, vertex operators become $\V_n$ with $n\in\Z$, and the duality defect is $\mathcal D_p$. To understand how $\Z_p$ gauging acts in the presence of $\V_n(y)$ operator, which is equivalent to the action of $\mathcal D_p$ on $\V_n$, we stretch a line $L$ from $y$ to the support of $\mathcal D_p$, as in figure \ref{fig: stretching a line L from vertex operator V(y) to the topological defect D_r} with the appropriate replacements. At the level of action, we have 
\begin{equation}
    -S_{\V_n} =  -\frac{p}{4\pi} \int |d\mathsf X -\C|^2 + in \int \delta_L \wedge (d\mathsf X - \C) - \frac{ip}{2\pi} \int \C \wedge d\Y,
\end{equation}
where $\Y$ being the dual boson living at the right of $\mathcal D_p$. Integrating $\C$ produces 
\begin{equation}
    -\frac{p}{4\pi} \int |\widehat{d\Y}|^2 = -\frac{p}{4\pi} \int \Big| d\Y - \frac{2\pi n}{p} \delta_L \Big|^2.
\end{equation}
The field configurations that contribute to the path integral should then obey the following boundary condition near $L$
\begin{equation}
    d\Y = \frac{2\pi n}{p} \delta_L + \cdots.
\end{equation}
This, in turn, is the effect of sweeping $\mathcal D_p$ across $\V_n$. Thinking as in figure \ref{fig: sweeping D_r across V_s(y) gives V'_s(y) = V_rs(y)}, the singular boundary condition is tantamount to inserting  
\begin{equation}
    \V_{n/p}( \partial L') \times \mathsf U_{2\pi n/p} (L')  = \exp \Big( \frac{i n}{p} \int_{L'} d\Y\Big),
\end{equation}
where $\mathsf U_\alpha(L) = \exp \big( \frac{i\alpha}{2\pi} \int d\Y \big)$ is the topological (Verlinde) line operator that generates the winding symmetry. In the non-compact case, the corresponding topological line $U_s = \exp \big( i s \int dY \big)$ is trivial, so even after $\mathfrak D_r$ is swept across $V_s$, the new operator $V'_s = V_{rs}$ is not attached to any lines, hence it remains a genuine local operator. By contrast, the topological line $\mathsf U_\alpha$ is non-trivial in the compact boson case, and in particular, the vertex $\V_{n/p}$ needs a $\mathsf U_{2\pi n/p}$ attachment to be well defined, therefore it is a non-genuine operator. When $n = 0 \text{ mod } p$, the operator $\V_{n/p}$ is well-defined without attaching any lines, so in that case it is a genuine operator. 

This is similar to the mechanism of how sometimes Wilson or 't Hooft lines need a surface attached to them in order to be well defined, in which case they are non-genuine line operators \cite{Aharony:2013hda, Gukov:2006jk, Kapustin:2014gua, tHooft:1977nqb, Alford:1992yx, Deligne:1999qp, Gaiotto:2010be}. We will discuss this in more detail when studying the 4d abelian gauge theory with non-compact gauge group. Before closing this section, we note that in the compact boson the defect $\mathcal D_p$ acts on $\V_n$ as 
\begin{equation}
    \mathcal D_p : \quad \V_n(y) \mapsto \V_{n/p}(\partial L') \times \mathsf U_{2\pi n/p} (L') \quad \quad ( \text{where } \partial L' = y),
\end{equation}
which is a non-genuine operator for $n \neq 0 \text{ mod } p$ and a genuine operator for $n = 0 \text{ mod } p$. On the other hand, the non-compact boson defect acts as 
\begin{equation}
    \mathfrak D_r : \hspace{4mm} \quad V_s(y) \mapsto V_{rs}(y) \times \underbrace{ U_{rs}(L') }_{ \text{trivial line} } \quad \quad \quad \hspace{2mm} (\text{where } \partial L' = y) \hspace{3mm}
\end{equation}
and since the corresponding line operator on the right-hand side is always trivial, the action of the defect $\mathfrak D_r$ always produces a genuine local operator. The fact that $\mathfrak D_r$ does not attach a line when acting is another indicator that it has an invertible action the spectrum, as opposed to our expectation in the previous version of the paper.

Before closing this section, note that in the limit $p \to \infty$, $\mathcal D_p$ annihilates vertex operators $\V_n$ with finite $n$ to the identity operator. For large operators with $n\to \infty$, we can consider the limit to be such that $n/p$ is a fixed rational number. When $m \equiv n/p \in \Z$, so $n = 0 \text{ mod } p$, the action reads $\V_n(y) \mapsto \V_m(y)$ without a line attachment since $\mathsf U_{2\pi m}(L) = 1$ with $m\in \Z$. For $m \in \mathbb Q/\Z$, so $n \neq 0 \text{ mod } p$, the action again attaches a topological line to the vertex operator. This analysis provides an additional perspective on why the $p\to \infty$ limit does not capture the non-compact boson and its defect because the discrete limit is not ``non-compact enough" (recall the discussion with quantum dimensions), and the non-compact boson theory cannot attach any lines whereas the $p\to \infty$ limit of the duality defect can still attach lines to some operators with large quantum numbers. One may suspect that the $p\to \infty$ limit of the duality defect of \cite{Choi:2021kmx} is related to a $\mathbb Q/\Z$ symmetry (or its dual $\mathsf{Rep}(\mathbb Q/\Z) = \hat \Z$ with $\hat \Z$ being the profinite integers \cite{Putrov:2022pua,Perez-Lona:2025add}). This expectation stems from the recent work \cite{Perez-Lona:2025add} which discusses the $p\to \infty$ limit of $\Z_p^{(-1)} \subset U(1)^{(-1)}$ subgroup gauging in the 4d Yang-Mills theory \cite{Unsal:2020yeh} and other setups with instantons, and the suggestion put forward by the authors of \cite{Perez-Lona:2025add} is that this corresponds to the gauging of a $\mathbb Q/\Z$ $(-1)-$form symmetry. It would be interesting if the corresponding statement for the $S-$duality in the Maxwell gauge theory, in the limit $1/e^2 \sim p \to \infty$, is related to this structure.

\subsection{Condensation Defects in the Non-Compact Boson} \label{subsec 3.4: condensations in non-compact boson}

We have seen the action of $\mathfrak D_r$ on the non-compact boson theory. Since it does not attach any lines, the defect is invertible, as opposed to the T-duality defect in the compact boson. To gain another perspective on this, we show that the fusion of $\mathfrak D_r$ with its dual $\overline{\mathfrak D}_r$ gives a trivially acting defect, namely the identity. This is a manifestation of the invertibility of the defect $\mathfrak D_r$. To proceed, we start with the setup in the figure below. 
\begin{figure}[!ht]
    \centering
    \resizebox{0.6\textwidth}{!}{%
    \begin{circuitikz}
    \tikzstyle{every node}=[font=\LARGE]
    \draw  (2.5,0.25) rectangle (22.5,-9.75);
    \draw [ fill={rgb,255:red,255; green,255; blue,255} ] (12.5,0.25) rectangle (22.5,-9.75);

    \draw [ fill={rgb,255:red,0; green,128; blue,128} ] (12.5,-9.75) rectangle (14.5,0.25);

    \draw [ color={rgb,255:red,0; green,64; blue,128}, line width=4pt, short] (12.5,-9.75) -- (12.5,0.25);
    \draw [ color={rgb,255:red,0; green,64; blue,128}, line width=2pt, short] (12.5,-4.75) -- (12.25,-5);
    \draw [ color={rgb,255:red,0; green,64; blue,128}, line width=2pt, short] (12.5,-4.75) -- (12.75,-5);

    \draw [ color={rgb,255:red,0; green,64; blue,128}, line width=4pt, short] (14.5,-9.75) -- (14.5,0.25);
    \draw [ color={rgb,255:red,0; green,64; blue,128}, line width=2pt, short] (14.25,-4.75) -- (14.5,-5);
    \draw [ color={rgb,255:red,0; green,64; blue,128}, line width=2pt, short] (14.5,-5) -- (14.75,-4.755);

    
    \node [font=\Huge, color={rgb,255:red,0; green,64; blue,128}] at (13.5,-10.5) {$\mathfrak B \times I_\varepsilon$};
    \node [font=\Huge, color={rgb,255:red,0; green,64; blue,128}] at (11.5,-4.75) {$\mathfrak D_r$};
    \node [font=\Huge, color={rgb,255:red,0; green,64; blue,128}] at (15.5,-4.65) {$\overline{\mathfrak D}_r$ };

    \end{circuitikz}
    }%
\caption{The defect $\mathfrak D_r$ and its dual inserted along an infinitesimal interval $I_\varepsilon$.}
\label{fig: Condensation defect construction in R scalar}
\end{figure}
In the $\varepsilon \to 0$ limit, we get a condensation defect $\mathfrak C_r$, which describes the gauging of $\mathbb R$ symmetry on the codimension-1 space $\mathfrak B$ with parameter $r$ specifying the action on the spectrum. The resulting defect can be described by an effective interface action as in the compact case \ref{Z_p condensation defect action}, which reads
\begin{equation} \label{non-compact boson trivial condensation defect gauging R on codim-1}
    S_{\mathfrak C_r} = \frac{i}{2\pi r} \int_{\mathfrak B} b_0 (dX - dY),
\end{equation}
where $b_0$ is a non-compact scalar field living on the defect, and $X/Y$ are the non-compact bosons to the left/right of $\mathfrak B$, respectively. Since there is no period quantization condition on $b_0$, the left and right fields are indistinguishable so the defect acts trivially. 

In addition to this trivial condensation, one can discuss another defect defined by an effective interface action of the form \ref{non-compact boson trivial condensation defect gauging R on codim-1}, where we replace $b_0$ with a compact field $\varphi \sim \varphi +2\pi$
\begin{equation}
    \frac{i}{2\pi r} \int_{\mathfrak B} \varphi (d X - dY).
\end{equation}
In this interface action, the field equation of $\varphi$ and the flux sum over $d\varphi$ imposes $\frac{1}{r} (X-Y) \equiv \mathsf x \in H^0(\mathfrak B, \Z)$. The object $\mathsf x$ can be regarded, in very rough terms, as the field strength of a $(-1)-$form gauge field\footnote{Frankly, this naming does not make complete sense since there is no $(-1)-$form object for which the exterior derivative gives $\mathsf x$, but given the mathematical fact that a $p-$form gauge field having quantized periods defines a class in $H^{p+1}(M, \Z)$, we can continue this statement to $p=-1$. This is in similar vein to how one talks of $(-1)-$form symmetries, which may equivalently appear fairly non-sensical at first encounter but has some non-trivial aspects if approached carefully.}. In any case, the condensation is governed by classses $[\mathsf x/ 2\pi] \in H^0(\mathfrak B, \Z)$ which measure the connectedness of the 1d interface $\mathfrak B$. Through the Poincaré duality $H^0(\mathfrak B, \Z) \simeq H_1( \mathfrak B, \Z)$, one may express this as a homology sum 
\begin{equation}
    \mathfrak C_\Z( \mathfrak B ) \propto \sum_{ \gamma \in H_1( \mathfrak B, \Z) } \zeta(\gamma).
\end{equation}
We note in particular that $\mathfrak B$ must have several disconnected parts in order for the defect to be non-trivial, meaning that for ordinary topologies $\mathfrak B = \R, S^1, [0,1]$, we have a trivial object. The mathematical details of how to make sense of the infinite sum are unclear, and we do not discuss it any further. 

\subsection{Subgroup Gaugings of \texorpdfstring{$\R^{(0)}$}{R^0} Symmetry} \label{subsec 3.5: subgroup gaugings in 2d R scalar}

We now study the coupling of $C$ to a compact scalar with period quantization, which is a $U(1)/\R$ type TQFT coupling. As we will see, the coupling $C \wedge d \Y$, with $\Y$ a compact scalar, turns out to gauge a subgroup $\Z \subset \R$ in the context of the real-valued scalar, which has also been discussed in \cite{Argurio:2024ewp}. 

Let us begin with the $\R/U(1)$ TQFT
\begin{equation}
    \frac{i}{2\pi r} \int_{M_2} C \wedge d \Y
\end{equation}
with $C$ and $\Y$ being 1-form $\R$ and 0-form $U(1)$ fields, respectively. The field equations of $\Y$ make $C$ a flat field with period quantization $\oint C \in 2\pi r \Z$. Hence, the operator $\exp \big( i \frac{\alpha}{2\pi} \oint C \big)$ with $\alpha \in [0, 2\pi r)$ generates a $U(1)$ symmetry. 

Consider a QFT having a $0$-form $\R$ symmetry, for which the background field $C$ is turned on. Deforming this theory by coupling $C$ to the $U(1) / \R$ TQFT, we are enacting a topological manipulation on $\R^{(0)}$, which we anticipate to be a topological subgroup gauging since the field equation of the conjugate $U(1)$ field restricts $C$ to be flat field with quantized periods. 

To verify this, we first analyze the quantum dual symmetry that emerges from the deformation. Since the codimension-1 topological defect $\exp \big( i \frac{\alpha}{2\pi} \oint C \big)$ is the generator of the quantum dual symmetry, we infer that the quantum dual symmetry is a $U(1)$ 0-form symmetry. For a 0-form $U(1)$ symmetry to emerge in 2d, the gauging must be of the form $\Z^{(0)} \subset \R^{(0)}$. This is due to the Pontryagin duality relation
\begin{equation}
    \mathsf{Rep}(\Z) = \widehat \Z = U(1).
\end{equation} 
Therefore, the $U(1) / \R$ TQFT coupling has the effect of the subgroup gauging $\Z \subset \R$, out of which a dual $U(1)$ appears.

Getting back to the non-compact boson in 2d, the topological manipulation above is realized by
\begin{equation}
    \hspace{1cm} \frac{1}{2} \int |dX - C|^2 + \frac{i}{2\pi r} \int C \wedge d\Y.
\end{equation}
Along the lines of \cite{Witten:1995gf}, we fix $X=0$ using the shift symmetry and integrate $C$, yielding 
\begin{equation}
    \frac{1}{8\pi^2r^2} \int d\Y \wedge * d\Y
\end{equation}
This is nothing other than a compact scalar action at radius $R = \frac{1}{2\pi r}$. So, gauging the $\Z$ subgroup of shift symmetry produces the compact boson. Observe that if we set $r = 1/p$ with $p\in\Z$, the period quantization reads $\int C \in \frac{2\pi \Z}{p}$, and this give rise to $\Z_p$ defects $\exp\big( i \alpha \int C\big)$ which are transparent at $\alpha = 0 \text{ mod } p$. This clearly describes a $\Z_p$ subgroup gauging, which can also be seen since the resulting action
\begin{equation}
    \frac{p^2}{8\pi^2} \int d\Y^2
\end{equation}
can be obtained from a $\Z_p$ gauging in the compact boson at radius $R= 1$. More generally, the TQFT coupling that imposes the period quantization $\int C \in \frac{2\pi r\Z}{p}$ describes the same gauging for a boson at radius $R= r$.

\subsection{Explanation via the SymTFT} \label{subsection 3.6: SymTFT explanation in R scalar}

The manipulations we discussed in the previous section can be understood from the SymTFT for $\R$ symmetries studied in \cite{Antinucci:2024zjp}. We will briefly go over the choice of topological boundary conditions in the SymTFT proposed and studied in \cite{Antinucci:2024zjp}. We will have a more extended discussion on this in section \ref{subsec 5.2: Lagrangian Algebras of the R SymTFT and a Generalization of the R(d-1) Gauging}, when discussing manipulations on $(-1)-$ and $(d-1)-$form $\R$ symmetries. 

For the $\R^{(0)}$ shift symmetry of the real scalar on $M_2$, the SymTFT reads
\begin{equation}
    \mathcal Z = \frac{i}{2\pi} \int_{M_2 \times I} C \wedge d \lambda,
\end{equation}
with $\lambda$ an $\R$ 1-form field, and $C$ couples to the shift symmetry on the left end of the slab $\mathfrak B_{\text{phsy}} \equiv M_2 \times \{ 0\}$. Field equations in the bulk kill all local degrees of freedom, and the interesting observables are given by 
\begin{equation}
    \zeta_{\alpha}(M_1) = e^{i \alpha \oint_{M_1} C} \quad ; \quad U_{\beta}(M_1) = e^{i\beta \oint_{M_1} \lambda},
\end{equation}
where $\alpha, \beta \in \R$. The braiding relation between these defects reads \cite{Antinucci:2024zjp}
\begin{equation}
    \Big\langle \zeta_\alpha(M_1) \ U_\beta(N_1) \Big\rangle = \exp \Big( 2\pi i \alpha \beta \times \text{Link}(M_1,N_1) \Big)
\end{equation}
with Link$(M_1,N_1)$ being the topological linking number of $M_1$ and $N_1$ on the slab $M_2 \times I$.

The topological boundary conditions for this SymTFT consist of the standard Dirichlet and Neumann boundary conditions, and an additional topological boundary condition that we will discuss. The Dirichlet-Neumann boundary conditions realize the gauging of $\R$ shift symmetry we performed previously. 

Fixing the Dirichlet boundary condition on the right end of the slab $\mathfrak B_{\text{top}} \equiv M_2 \times \{ 1 \}$,
\begin{equation}
    C \big|_{\mathfrak B_{\text{top}}} = 0,
\end{equation}
we have that $\zeta_\alpha$ ends on the right boundary, and consequently appears as point operators upon slab contraction. These are the charged objects under the $\R^{(0)}$ symmetry on $M_2$, which in turn is generated by the $U_\beta$ operators that are topological on the right boundary, so upon slab contraction they act as the symmetry generators on $M_2$.

On the other hand, fixing the Neumann boundary condition 
\begin{equation}
    \lambda \big|_{\mathfrak B_{\text{top}}} = 0,
\end{equation}
we have $U_\beta$ operators ending on and $\zeta_\alpha$ operators being topological on the right end of the slab, so upon contraction, $U_\beta$ become local operators charged under the symmetry generated by $\zeta_\alpha$. This corresponds to gauging the $\R$ symmetry of the real-valued scalar, upon which the quantum dual $\R$ symmetry appears. We therefore see the invariance of the non-compact scalar under gauging its shift symmetry as a manifestation of the fact that in the SymTFT, both the Dirichlet and Neumann boundary conditions at $\mathfrak B_{\text{top}}$ produce the same symmetry.

Additionally, there are a family of topological boundary conditions in the $\R$ SymTFT discussed in \cite{Antinucci:2024zjp} (see also \cite{Argurio:2024ewp} for details on such boundary conditions). This is given by a choice of Lagrangian algebra, which trivializes $\zeta_n$ and $U_m$ with $n,m\in \Z$ on the topological boundary. The remaining coset classes of operators $\zeta_\alpha, U_\beta$ which are topological on the boundary are labelled by $\R/\Z \simeq U(1)$, so the contraction of the slab produces $U(1) \times U(1)$ symmetry, both of which are realized as 0-form symmetries on $M_2$ since the topological operators on $\mathfrak B_{\text{top}}$ are supported on lines. Furthermore, due to the unsplittability of the exact sequence \cite{Antinucci:2024zjp, Tachikawa:2017gyf}
\begin{equation}
    0 \to 2\pi \Z \to \R \to U(1) \to 0,
\end{equation}
the two $U(1)$s have a mixed anomaly. This choice of boundary condition then clearly gives the compact boson in $M_2$ when we choose the mixed Lagrangian algebra, which is compatible with the subgroup gauging $\Z \subset \R$ that we discussed in the previous subsection. 

This mixed topological boundary condition can be generalized to a family of Lagrangian algebras as follows. We may trivialize $\zeta_{n \mathrm R}$ and $U_{m \mathrm R^{-1}}$ on the boundary, which still constitute a Lagrangian algebra and hence give topological boundary conditions. This parameter corresponds, for example, to a scaling of the radius in the resulting compact scalar theory. The special case of $\mathrm R = q \in \Z$ gauges a $\Z_q$ subgroup of the $U(1)$ symmetry since $U_{\frac{m}{q}}$ ends on the topological boundary when $m = 0 \text{ mod } q$. We have realized this boundary condition by a local TQFT coupling at the end of the previous subsection. 

The choices $\mathrm R \in \mathbb Q$ realize various discrete subgroup gaugings in the compact boson. We will discuss these cases in more detail in section \ref{subsec 5.2: Lagrangian Algebras of the R SymTFT and a Generalization of the R(d-1) Gauging}, and will propose a local TQFT coupling that realizes these boundary conditions.

Lastly, we would like to point out a contrast between the discrete gauging $\Z_p \subset U(1)$ and the gauging $2\pi \Z \subset \R$. In the former, the relevant exact sequence reads 
\begin{eqnarray}
    0 \to \Z_p \to U(1) \to U(1) \to 0,
\end{eqnarray}
which means that as a topological space, $U(1) / \Z_p$ is still a circle\footnote{A nice physics explanation for this fact along the lines of Yang-Mills theory is the following. Gauging a $\Z_p \subset U(1)$ subgroup of the $(-1)-$form instanton symmetry modifies the periodicity of the theta angle from $\theta \sim \theta + 2\pi$ to $\theta \sim \theta + 2\pi/p$, as explained in \cite{Seiberg:2010qd, Tanizaki:2019rbk}. The subgroup gauging modifies the periodicity of the theta angle, but it is still an angular parameter, so topologically we still have a circle.}. On the other hand, the quotient $\R/ 2\pi \Z \simeq U(1)$ has a different topology than $\R$. This distinction between the two gaugings is interesting, and may provide deep topological insights when making contrasts in the study of compact/non-compact gauge theories.

\subsection{Bosonic String Theory} \label{subsec 3.7: small discussion on bosonic string theory}

In this subsection, we comment on various applications for the ideas we developed up to this point in the paper. We do not go into technical details, and merely attempt to inspire some curious questions in bosonic string theory. 

\subsubsection{Compactifications in the Target Space}

We can gain further intuition on the compactification of the gauge group from the target space point of view as follows. Starting from the real line $\R$, compactifying it to a circle $S^1$ is a manipulation on the 1-manifold $\R$ that changes its topology. As such, for a sigma model with target space $\R$, there must be a topological manipulation in the internal space mapping it to a sigma model with target space $S^1$. Based on our construction above, we infer that for a QFT with target space $\R$, the topological manipulation that converts the target space to an $S^1$ is a subgroup gauging of the $\R$ shift symmetry, controlled by the $\R/U(1)$ TQFT above.

Consider a 2d sigma model on $\Sigma_2$ consisting of some scalar fields $X^i$ parameterizing a target space $M$
\begin{equation}
    X : \quad \Sigma_2 \to M.
\end{equation}
We may regard $\Sigma_2$ as a string worldsheet. Taking $M = \R^d$, we can perform a topological manipulation on this sigma model via the gauging above, and change the target space into the $d$-torus $\T^d = \R^d / (2\pi \Z)^d \simeq (S^1)^d$. Furthermore, we have the freedom to choose different gauging parameters in each direction, so we can in general obtain $\mathsf T^d \equiv \prod_{i} \R / (2\pi r_i \Z) \simeq \prod_i S^1_{r_i}$, where $S^1_{r_i}$ denotes the circle of radius $r_i$. Another possibility is gauging (compactifying) only for $q$ directions, giving a ``cylinder" $\R^{d-q} \times \T^{q}$. The most generic scenario is then $\R^{d-q} \times \mathsf T^{q}$ where $\mathsf T^{q}$ is obtained by quotiening each of the $q-$directions with an arbitrary parameter, hence there are $\R^q$ many different $\mathsf T^{q}$ spaces with labels $\vec{r} = (r_1, \cdots, r_q) \in \R^q$\footnote{It would be efficient to define an equivalence relation $\sim$ such that if any $r_i$ and $r_j$ are exchanged, $\mathsf T^{q}$ remains the same since we cannot genuinely distinguish different directions of $\R^q$. Then $\sim$ is related to $S_q$, the $q-$permutation group, and there are $\R^q / S_q$ many different $\mathsf T^{q}$.}. Thus, our topological manipulation above can be used to modify the topology of the target space of a sigma model as outlined.

Much more interesting is considering a half-space gauging in the string worldsheet. For simplicity, consider the worldsheet of a closed string to be the cylinder $\R \times S^1$. Dividing the circle into two by $\theta < \pi$ and $\theta \geq \pi$ for $\theta \in [0,2\pi)$, we can gauge some of the $\R$ symmetries in the $\theta \geq \pi$ half of the worldsheet, thus producing a target space of the form $\R^{d-q} \times \mathsf T^{q}$ there. While the procedure is straightforward, this is conceptually remarkable because starting from a string propagating in $\R^d$, we can couple a TQFT on half of the sigma model, which makes it such that the same string at different points on its body perceives the target space with different topologies! For example, whereas the left part of the string $\theta < \pi$ sees the world as a flat $d-$dimensional infinite space, the right part of that \emph{same} string may see the world in a different, in particular compact, topology. Even more so, we can divide the string worldsheet into many slices and perform various topological manipulations and insert various topological defects on the interfaces. For example, in two consecutive sheets $W_1$ and $W_2$ defined by $\theta_0 < \theta \leq \theta_1$ and $\theta_1 < \theta \leq \theta_2$ respectively, we can insert non-compact/compact theories, with corresponding topological defects at the $\theta = \theta_1$ point. All of these setups can be regarded as the non-compact boson coupled to various TQFTs on different subspaces, and appropriate boundary conditions on the interfaces. See \cite{Bharadwaj:2024gpj} for discussions on the implications of topological defects for bosonic string theory in the toroidal compactification. 

While we do not study the implications of the said constructions for string theory in detail, we remark that understanding the subgroup gaugings and insertions of topological defects in the target space physics may be illuminating along the lines of string compactifications, stringy singularities, and dualities.

\subsubsection{Quantum Field Theory and the Topology of Spaces}
\small{*This section briefly deviates from the main line of development of $\R$ symmetries in the paper. }

Motivated by the discussion in this subsection above, one may also consider the ideas of global manipulations in a QFT to gain insight into the topology of manifolds. To do so, we consider a more generic sigma model 
\begin{equation}
    X : \quad \Sigma_d \to M,
\end{equation}
with $\Sigma_d$ being a $d-$dimensional world-history of the corresponding physical object. For example, $d=1$ describes point particles propagating in $M$ governed by some sigma model action, $d=2$ describes strings propagating in $M$, and in general we have a brane of $d-$dimensional support propagating inside $M$, with the sigma model action providing a world-history formalism to describe these objects. The action is given by 
\begin{eqnarray}
    S \propto \int_{\Sigma_d} g_{\mu\nu}(X) d X^\mu \wedge *_{\Sigma_d} dX^\nu ,
\end{eqnarray}
where $g_{\mu\nu}$ is a metric on the manifold $M$, which has local coordinates $X^\mu$, and $*_{\Sigma_d}$ is the Hodge star on the world-history of the propagating brane-like object $\Sigma_d$. 

Depending on $M$, there will be certain symmetry transformations on the target space, leaving the action invariant. For example, in $\R^D$, the relevant group of transformations is given by $O(D)$, so the action has this group of transformations as a symmetry on the target space. Suppose we turn on an $O(D)$ adjoint valued gauge field $\mathcal A$ coupled to the corresponding Noether current $\mathcal J$
\begin{equation}
    S \supset \int_{M_d} \text{Tr} \ \mathcal A \wedge * \mathcal J,
\end{equation} 
with $\text{Tr}$ denoting the Cartan-Killing form on the Lie algebra. To enact a topological manipulation on this sigma model, we may insert $\mathcal A$ into a TQFT. $\mathcal A$ being valued in a non-abelian group, we no longer have the simple $b \wedge da$ type of theory, but one with a twisted covariant derivative $d_{\mathcal A} \mathcal A \equiv d \mathcal A + \mathcal A \wedge \mathcal A$, and where the conjugate field carries group indices. This type of non-abelian BF theory, studied in detail in \cite{Horowitz:1989ng}, was recently proposed \cite{Antinucci:2024zjp, Brennan:2024fgj, Bonetti:2024cjk} as the SymTFT for $0-$form non-abelian symmetries. To be precise, the proposal of \cite{Bonetti:2024cjk} is to start from a Yang-Mills theory, dualize as in \cite{Witten:1992xu}, and take the limit where the Yang-Mills coupling goes to zero, which reproduces the non-Abelian BF theory. At the end, one ends up with the TQFT
\begin{eqnarray}
    \int_{M_d \times I} \text{Tr} \ \mathcal G_{d-1} \wedge d_{\mathcal A} \mathcal A.
\end{eqnarray}
From the SymTFT principle for abelian symmetries, we have the set of topological variants of a global symmetry to be captured by the codimension-$(-1)$ TQFT. If one believes the same is true for a $0-$form non-abelian symmetry, one would expect that the set of distinct topological manipulations on the symmetry with current $\mathcal J$, of which the background field is $\mathcal A$, to be governed by the above type of TQFT on the slab.

Manipulating the symmetries and obtaining the global variants in this setup has a geometric meaning, as we are considering a sigma model where the symmetry transformations can be viewed as deformations on the target space. Thus, different global structures on the sigma model, expected to be captured by a SymTFT of the form above, may correspond to distinct target space topologies. One can therefore obtain topological information about manifolds from the distinct global variants of the same symmetry, all known to the SymTFT. For the abelian case, this does happen. Starting from an $\R$ target space, we can produce an $S^1$ target space by gauging the corresponding shift symmetry. However, this being true for the abelian case is by no means enough to assert that the same should hold for the non-abelian case. In any event, I believe it is an interesting question to investigate due to its relevance in String Theory and sigma models, where such questions give rise to richer geometric/topological structures. The theory indepdendence of the SymTFT may play a role as an organizing principle to understand these richer geometric/topological structures in a broader and more abstract way, just as group theory and in particular the abstract study of unitary irreducible representations organize, in a kinematical way, the spectrum of a QFT with the specified symmetry structure.

\subsubsection{Topological Defects of the Bosonic String}

In the sigma model with target space $\R^D$, which is relevant to the bosonic string theory in $D=26$, one may define a generalization of the defects $\mathfrak D_r$ acting as $\mathbb R^\times$: $V_s \mapsto V_{rs}$. We achieve this generalized defect by considering a setup of the form 
\begin{equation}
    \frac{R_{ij}}{4\pi}  \int_{\Sigma_+} dX^i \wedge *  dX^j + \frac{R_{ij}}{4\pi} \int_{\Sigma_-} dY^i \wedge * dY^j + \frac{i R_{ij} }{2\pi} \int_{\mathfrak B} X^i \wedge dY^j, 
\end{equation}
which is a natural generalization of \ref{The non-compact boson defect Lagrangian}. We note that in gauging the $\R^D$ shift symmetry on the right part of the worldsheet $\Sigma_+$, we turn on $C^i$ background fields all of which obey the Dirichlet boundary condition on the interface $\mathfrak B$, therefore the defect obtained by integrating $C^i$ over flat configurations is topological.

By redefining the fields in the target space appropriately, the symmetric rank-2 form $R : \mathbb R^D \otimes \mathbb R^D \to \R$ can be generically put into a diagonal form (assuming invertibility) with all diagonal terms being 1. That all of the diagonals can be made 1 is the same statement as $r$ being an unphysical parameter (when there are no operator insertions) in the case $D=1$. We will keep the entries to be generic real numbers $r_i \in \R^\times$ 
\begin{equation}
    R_{ij} = 
    \begin{pmatrix}
        r_1 & & & \\ 
         & r_2 & & \\ 
         & & \ddots & & \\
         & & & & r_D 
    \end{pmatrix}
\end{equation}
in anticipation that they will determine the action of the operator.

To derive the action of the generalized defect, we insert a vertex operator of momentum $k_i$ into a non-compact boson theory and gauge the shift symmetry, namely we study the path integral 
\begin{equation}
    \Big\langle V_{k_i} (y) \Big\rangle = \int \mathcal DX \exp \Big( - \frac{R_{ij}}{4\pi} \int  dX^i \wedge *  dX^j + i k_i X^i(y) \Big).
\end{equation}
As before, we employ the trick of writing the operator insertion as a trivial operator inserted on an open line $\exp \big( i k_i \int_L dX^i  \big)$ ($\partial L = y$), and then turn on backgrounds $C^i$ which are in turn entered into a BF term 
\begin{equation}
    - \frac{R_{ij}}{4\pi} \int (dX^i - C^i) \wedge *  (dX^j - C^j) + i k_i \int (dX^i - C^i) \wedge \delta_L +  \frac{i R_{ij}}{2\pi} \int C^i \wedge dY^j.
\end{equation}
Proceeding with the integral we get the Lagrangian in terms of $Y^i$:
\begin{equation}
    -\frac{R_{ij}}{4\pi} \int \widehat{dY}^i \wedge * \widehat{dY}^j \equiv - \frac{R_{ij}}{4\pi}  \int \big(dY^i - 2\pi (k')^i \ \delta_L \big) \wedge * \big( dY^j - 2\pi (k')^j \delta_L \big),
\end{equation}
where $(k')^i = \big( r_1^{-1} k_1, r_2^{-1} k_2, \cdots, r_D^{-1} k_D \big)$\footnote{Note that $r_i^{-1}$ is due to the fact that the positioning of $r_i$ parameters is opposite to that in \ref{The non-compact boson defect Lagrangian}}. The regularity of the configurations require the boundary condition 
\begin{equation}
    dY^i = 2\pi (k')^i \ \delta_L + \cdots,
\end{equation}
which in turn gives an operator insertion 
\begin{equation}
    \exp \Big( i (k')^i \int_L dY^i \Big) = \exp \Big( i (k')^i \ Y^i(\partial L) \Big).
\end{equation}
This determines the action of the topological defect $\mathfrak D$, which is now labelled by $(\R^\times)^D$:
\begin{equation}
    \mathfrak D_{R_{ij}} : \quad V_{k_i} \mapsto V_{ (k')^i}.
\end{equation}
As in the single boson case, we can look at the interplay of this defect with the shift generators (while mapping $r = e^\mathfrak{t}$). Denoting the generators of shift symmetry as $Q^i$ and those of the half-space gauging defect as $\mathfrak d_j$ with $i,j = 1,2,\cdots, D$, we have the algebra $[Q^i , \mathfrak d_j ] = i \delta^i_j$, which in turn gives a $\textsf{Heis}(\R^D \times \R^D) $ group. Given that what we actually have is a WZW model of group $\R^D$, this fact is quite natural and not surprising. It would be fascinating if these defects had some non-trivial use to understand the target space dynamics of sigma models and the bosonic string theory.

\section{\texorpdfstring{$\R$}{R} Maxwell Theory in 4d} \label{sec 4: R maxwell}

\subsection{Gauging the 1-form Symmetry} \label{subsec 4.1: gauging 1-form R symmetry}

The story here with $4d$ Maxwell theory with non-compact gauge group is similar to that of the real-valued scalar in $2d$. For the $\R$ Maxwell theory in $4d$, there is no global magnetic symmetry since $\int F = \int dA = 0$, so the only global symmetry is a 1-form $\R$ symmetry (in addition to the $\Z_2$ conjugation acting as $A \mapsto -A$ which we omit). The action reads\footnote{As in the non-compact boson, we will label theories and their topological manipulations to better connect them with figures. The $\R$ Maxwell is denoted as $\mathcal Y_\R$, and topological manipulations again as maps $\phi: \mathcal Y_\R \mapsto \mathcal Y'$.}
\begin{equation}
    \mathcal Y_\R : \hspace{2cm} \frac{1}{4\pi r} \int F^2. \phantom{ \mathcal Y_\R : \hspace{2cm} }
\end{equation}
We introduced the parameter $r$ as in the non-compact boson. For the Maxwell action, the ordinary choice is $r = 1/2\pi$, but we will keep $r$ unfixed, in anticipation that a topological defect may be constructed by gauging the 1-form $\R$ symmetry which acts depending on the value of $r$. 

To gauge the 1-form symmetry, we extend the gauge invariance $A \mapsto A + d\epsilon$ to an exotic gauge invariance \cite{Witten:1995gf}, or equivalently to a 1-form symmetry transformation in modern terms, by introducing a 2-form field $G$ with the transformation rules
\begin{equation} \label{1-form gauge transformation}
\begin{aligned}
        A &\mapsto A + h, \\
        G &\mapsto G + dh,
\end{aligned}
\end{equation}
with $h$ being a real-valued 1-form. For $h$ exact, we recover the ordinary gauge transformation. Defining $\mathcal F = F -G$, which is invariant under the transformations above, we write the action as
\begin{equation}
    \frac{1}{2} \int \mathcal F^2.
\end{equation} 
Following typical derivations in electric-magnetic duality of gauge theory \cite{Deligne:1999qp, Witten:1995gf, Gukov:2006jk}, we deform the theory with a BF coupling for $G$ along with a new 1-form $\R$ gauge field $V$ having field strength $W = dV$ (we mostly use the conventions of \cite{Witten:1995gf})
\begin{equation} \label{Action for the gauged R 1-form symmetry}
    \phi(\mathcal Y_\R) : \hspace{2cm} \frac{1}{4\pi r} \int \mathcal F^2 + \frac{i}{2\pi r} \int G \wedge W, \phantom{ \phi(\mathcal Y_\R) : \hspace{2cm} }
\end{equation}
where we fixed the coefficient of the BF coupling for convenience. Integrating $V$ out sets $dG = 0$, and this makes $G$ a trivial gauge field that can be set to $0$. This gives back the original $\R$ gauge theory. On the other hand, we can first integrate over $G$. Using the 1-form gauge transformation \ref{1-form gauge transformation}, we fix $A=0$ and solve the field equation of $G$ as $*G = -iW$ paralleling \cite{Witten:1995gf, Deligne:1999qp}. Performing the Gaussian integration then yields
\begin{equation}
    \frac{1}{4\pi r} \int W^2.
\end{equation}
Hence, the $\R$ gauge theory is invariant under gauging the 1-form $\R$ symmetry, which is a non-compact version of $S-$duality.

From the perspective of TQFT coupling, we observe that the quantum dual symmetry, following from the $V$ field equation, is generated by $\zeta_\alpha = \exp \big( i \alpha \oint G \big)$ with $\alpha \in \R$. This demonstrates that one can talk about the notion of a non-compact version of ``$S-$duality'', in the sense that the deformations for $S-$duality in $U(1)$ gauge theory plays the same role as gauging the $\R^{(1)}$ shift symmetry \emph{topologically}. Upon the gauging, a quantum dual $\R^{(1)}$ symmetry appears since $\widehat \R = \R$ and in 4d, flat gauging of a 1-form symmetry produces another 1-form symmetry

\subsection{Construction of Topological Defects in \texorpdfstring{$\R$}{R} Maxwell} \label{subsec 4.2: topological defect in R Maxwell}

Using the invariance under topological manipulation above, we can construct a topological duality defect in $\R$ Maxwell theory. We proceed as in subsection \ref{subsec 3.1: topological defect in non-compact boson} with the idea of half-space gauging. We divide the spacetime into two $M_4 = M_- \cup M_+$, joined along a boundary $\mathscr B = \partial M_+$, and couple the BF term $\frac{i}{2\pi r} \int G \wedge W$ only on $M_+$. Performing the $G$ integration on $M_+$ with Dirichlet boundary conditions on the interface $G \big|_{\mathscr B} = 0$, we end up with a topological defect since Dirichlet boundary condition is invariant under deformations of the locus of the interface.

We reproduce the steps quickly. More detailed explanations are available in section \ref{sec 3: R scalar in 2d}. The setup is described by
\begin{figure}[!ht]
    \centering
    \resizebox{0.5\textwidth}{!}{%
    \begin{circuitikz}
    \tikzstyle{every node}=[font=\LARGE]
    \draw (2.5,0.25) rectangle (22.5,-9.75);
    \draw [ fill={rgb,255:red,128; green,64; blue,64} ] (12.5,0.25) rectangle (22.5,-9.75);
    \draw [ color={rgb,255:red,64; green,0; blue,0}, line width=4pt, short] (12.5,-9.75) -- (12.5,0.25);
    \draw [ color={rgb,255:red,64; green,0; blue,0}, line width=2pt, short] (12.5,-4.75) -- (12.25,-5);
    \draw [ color={rgb,255:red,64; green,0; blue,0}, line width=2pt, short] (12.5,-4.75) -- (12.75,-5);
    \node [font=\LARGE] at (7.5,-10.5) {$M_-$};
    \node [font=\LARGE] at (17.75,-10.5) {$M_+$};
    \node [font=\Huge, color={rgb,255:red,64; green,0; blue,0}] at (12.5,-10.5) {$\mathscr B$};
    \node [font=\Huge] at (7.5,-4.75) {$\mathcal Y_{\mathbb R}$};
    \node [font=\Huge] at (17.75,-4.75) {$\phi (\mathcal Y_{\mathbb R} )\simeq \mathcal Y_{\mathbb R}$};
    \end{circuitikz}
    }%
\caption{The half-space gauging procedure for the non-compact Maxwell theory.}
\label{fig: half space gauging with r-parameter in Y_R}
\end{figure}
\begin{equation}
\begin{aligned}
    \mathcal Y_\R \big|_{M_-} &: \hspace{2.5cm} \frac{1}{4\pi r} \int_{M_-} F^2 ,  \phantom{ \mathcal Y_\R \big|_{M_-} : \hspace{1cm}  }\\ 
    \phi(\mathcal Y_\R) \big|_{M_+}&: \hspace{1cm} \frac{1}{4\pi r} \int_{M_+} \mathcal F^2 + \frac{i}{2\pi r} \int_{M_+} G \wedge W.  \phantom{ \phi_r(\mathcal Y_\R) \big|_{M_+}: \hspace{1cm} }
\end{aligned}
\end{equation}
The Dirichlet boundary condition $G \big|_{\mathscr B} = 0$ gives rise to a topological defect $\mathscr D$ upon performing the $G$ integral. This will manifest itself as a localized topological term as before, so upon integration we end up with the defect action
\begin{equation}
    \mathcal Y_{\R, \mathscr D}: \hspace{1cm} \frac{1}{4\pi r} \int_{M_-} F^2 + \frac{1}{4\pi r} \int_{M_+} W^2 + \mathcal S_{\mathscr D}. \phantom{ \mathcal Y_{\R, \mathscr D}: \hspace{1cm} }
\end{equation}
What is the form of $\mathcal S_{\mathscr D}$? To understand this, we look at the field equation of $G$ before integration:
\begin{equation}
    G = F - i * W.
\end{equation}
Then, to obey the Dirichlet boundary conditions, the left and right fields should mix on $\mathscr B$ so that
\begin{equation} \label{Dirichlet boundary condition imposing mixing condition on fields}
    F \big|_{\mathscr B} = i* W \big|_{\mathscr B}.
\end{equation}
This boundary condition characterizes the defect $\mathscr D$.
\begin{figure}[!ht]
    \centering
    \resizebox{0.6\textwidth}{!}{%
    \begin{circuitikz}
    \tikzstyle{every node}=[font=\LARGE]
    \draw  (2.5,0.25) rectangle (22.5,-9.75);
    \draw [ fill={rgb,255:red,255; green,255; blue,255} ] (12.5,0.25) rectangle (22.5,-9.75);
    \draw [ color={rgb,255:red,64; green,0; blue,0}, line width=4pt, short] (12.5,-9.75) -- (12.5,0.25);
    \draw [ color={rgb,255:red,64; green,0; blue,0}, line width=2pt, short] (12.5,-4.75) -- (12.25,-5);
    \draw [ color={rgb,255:red,64; green,0; blue,0}, line width=2pt, short] (12.5,-4.75) -- (12.75,-5);
    \node [font=\LARGE] at (7.5,-10.5) {$M_-$};
    \node [font=\LARGE] at (17.75,-10.5) {$M_+$};
    \node [font=\Huge, color={rgb,255:red,64; green,0; blue,0}] at (12.5,-10.5) {$\mathscr B$};
    \node [font=\Huge, color={rgb,255:red,64; green,0; blue,0}] at (13.5,-4.75) {$\mathscr D$};
    \node [font=\huge] at (-1,-4.75) {$\mathcal Y_{\mathbb R, \mathscr D}=$};
    \node [font=\huge] at (7.25,-4.75) {$\mathcal Y_{\mathbb R}$};
    \node [font=\huge] at (17.75,-4.75) {$\mathcal Y_{\mathbb R}$};
    \end{circuitikz}
    }%
\caption{The invariance under gauging the 1-form shift symmetry leads to a topological defect $\mathscr D$ supported on the interface $\mathscr B$.}
\label{fig: half space gauging at self-dual coupling with r-parameter in Y_R}
\end{figure}
It is then easy to see that the defect action that gives the correct mixing has the form
\begin{equation}
    \mathcal S_{\mathscr D} = \frac{i}{2\pi r} \int_{\mathscr B} A \wedge d V.
\end{equation}
Then, looking at the boundary variations of the total action
\begin{equation}
\begin{aligned}
    \delta A \big|_{\mathscr B} &: \phantom{W} \frac{1}{2\pi r} *F \big|_{\mathscr B} + \frac{\delta \mathcal S_{\mathscr D}}{\delta A} = 0 ,\\
    \delta V \big|_{\mathscr B} &: \phantom{F} \frac{1}{2 \pi r} *W \big|_{\mathscr B} + \frac{\delta \mathcal S_{\mathscr D} }{\delta V} = 0,
\end{aligned}
\end{equation}
we easily conclude that $\mathcal S_{\mathscr D}$ is the consistent choice that respects the Dirichlet boundary condition $G\big|_\mathscr B = 0$ \ref{Dirichlet boundary condition imposing mixing condition on fields}.

\subsection{Action on Line Operators} \label{subsec 4.3: action on line operators non-compact maxwell defect}

We now undertake the study of how the defect $\mathscr D$ acts on the Wilson lines of the theory, which will parallel section \ref{subsec 3.3: action on local operators non-compact boson defect}. For non-compact gauge group, the operators are defined as 
\begin{equation}
    W_s(\gamma) = e^{is \oint_\gamma A}, \quad \quad (s \in \R),
\end{equation}
and we take $\gamma$ to be a closed cycle embedded inside spacetime $\gamma \subset M_4$. To understand the action of the defect on this operator, we will gauge the 1-form symmetry in the presence of the line defect $W_s(\gamma)$. This insertion modifies the action as 
\begin{equation}
    \langle W_s(\gamma) \rangle = \int \mathcal D A \exp \Big( -\frac{1}{4\pi r} \int dA^2 + is \oint_\gamma A \Big).
\end{equation}
To gauge $\R^{(1)}$, we need to replace $dA \mapsto dA - G$. Analogous to the compact boson, we must recast the Wilson line insertion in terms of a surface integral over $dA$. In doing so, we stretch a surface $S$ from the locus of the defect (the boundary $\mathscr B$) toward $\gamma$, namely $\partial S = \gamma$, and insert the trivial operator 
\begin{equation}
    U_s = \exp \Big( is\int F \Big).
\end{equation}
On a closed surface $S_{\text{closed}}$, we have $ U_s(S_{\text{closed}}) = 0$ since $F = dA$ globally and $\int_{S_{\text{closed}}} dA = 0$ follows from Stokes' theorem. For an open surface $S$, this operator is interpreted as a trivial surface operator attached to a Wilson line 
\begin{equation}
    U_s(S) = \exp\Big( is \int_{\partial S} A\Big) = W_s(\partial S = \gamma).
\end{equation}
In compact gauge theory, this is expressed as follows. A Wilson line $\W_s(\gamma)$ in compact gauge theory always carries such topological surfaces attached to it, but depending on the parameter $s$, the attached surface may be trivial. Surface operators ending on lines may be viewed as the world-sheets of Dirac strings \cite{Gukov:2006jk}, and if the operator obeys the Dirac quantization condition the worldsheet has no physical effect and is invisible (just as the Dirac string of a monopole is an unphysical construct since it does not induce any Aharonov-Bohm phases). Coming back to non-compact gauge theory, there is no Dirac quantization condition (as there are no monopoles, which is a statement about the topology of $\R$ bundles over manifolds: They are always trivial bundles), so the surfaces are always transparent. This distinction shall be important when making comparisons with the action of duality defects $\mathcal D_p$ in $U(1)$ Maxwell and the topological defect $\mathscr D_r$ in $\R$ Maxwell theory.

Coming back to gauging $\R^{(1)}$ symmetry in the presence of $W_s(\gamma)$, we write the action as 
\begin{equation}
    -S_{W_s} = - \frac{1}{4\pi r} \int F^2 + i s \int \delta_S \wedge F,
\end{equation}
where $\delta_S$ is a distributional object (the Poincaré dual) that localizes the integral on $S$ (where $\partial S = \gamma$). We will overlook the mathematical subtleties of defining a Poincaré dual for open submanifolds and proceed with physical intuition. At the end of this subsection we will verify that this procedure captures the correct action for the $\mathcal D_p$ defect on $U(1)$ Maxwell.

Gauging corresponds to the replacement $F \mapsto \mathcal F = F-G$ and inserting $G$ into a non-compact BF TQFT:
\begin{equation}
    -\frac{1}{4\pi r} \int \mathcal F^2 + is \int \delta_S \wedge \mathcal F - \frac{i}{2\pi r} \int G \wedge dV.
\end{equation}
Integration over $G$ then produces 
\begin{equation}
    -\frac{1}{4\pi r} \int |\widehat{dV}|^2 = -\frac{1}{4\pi r} \int |dV - 2\pi r s \delta_S\big|^2.
\end{equation}
The delta term in the action prescribes a singularity on $W =dV$ near the surface $S$, which means that $W$ has the form
\begin{equation}
    \frac{1}{2\pi} dV = rs \delta_S + \cdots,
\end{equation}
where ellipsis denote regular terms. This is a surface-like singularity and indeed corresponds to the insertion of 
\begin{equation}
    U_{rs}(S) = \exp \Big( irs \int_S  dV \Big) = \exp \Big( i rs \oint_{\gamma} V \Big) \quad \quad (\gamma = \partial S).
\end{equation}
Thus, we see that gauging the 1-form symmetry of $\R$ Maxwell acts on the Wilson line as $W_s \mapsto W_{rs}$. This in turn is equivalent to the effect of sweeping $\mathscr D_r$ across the line defect $W_s$ since $\mathscr D_r$ is constructed by gauging the 1-form symmetry on half of the space. 

Thus we have the (invertible) action
\begin{equation}
    \mathscr D_r : \quad  W_s \mapsto W_{rs},
\end{equation}
where we infer the invertibility from the fact that the right hand side is a genuine line operator, as in the non-compact boson. As in the non-compact boson, we expect a mixing between this defect and the 1-form shift generators. In the non-compact boson, both defects are supported on lines, so both are $0-$form symmetries and they extend to a Heisenberg group; whereas in here, the defect $\mathscr D_r$ lives on a volume, so there is a mixing of the form $ \R^{(1)} `` \times " \R^{*(0)} $. Understanding the twist in the product structure that replaces the Heisenberg group for the non-compact boson (which likely has a higher-group structure) may be very interesting to study, at least from a mathematical point of view. 

\subsubsection{\texorpdfstring{$U(1)$}{U(1)} Gauge Theory Revisited}

Let us verify that our trick of wrapping an open surface between $\gamma$ and the locus of the defect $\mathscr B$ correctly reproduces the action of duality defects in compact gauge theory. After inserting the Wilson line $\W_n(\gamma) = \exp \big( i n\oint_\gamma \mathsf A \big)$ with $n\in \Z$ into the action at self-dual electric coupling $e^2 = 2\pi/p$, we extend a trivial surface $\mathscr U_{2\pi n} = \exp \big( i n \int d \mathsf A \big)$\footnote{For general $\alpha$, $\mathscr U_\alpha = \exp \big( \frac{i\alpha}{2\pi} \int d\mathsf A\big)$ is not trivial in compact gauge theory. However, due to the Dirac quantization condition $\frac{1}{2\pi} \int d\mathsf A \in \Z$, the operator is trivial for $\alpha \in 2\pi \Z$, which is a manifestation of the fact that $\mathscr U_\alpha$ generates the $U(1)^{(1)}$ magnetic symmetry. } from the support of $\mathcal D_p$ defect to $\gamma$. Sweeping the duality defect $\mathcal D_p$ across $\W_n(\gamma)$ has the effect of gauging $\Z_p^{(1)} \subset U(1)^{(1)}$ for the electric 1-form symmetry, so understanding the $\Z_p$ subgroup gauging in the presence of $\W_n$ gives the action of $\mathcal D_p$. We perform this gauging as follows
\begin{equation}
    -S_{\W_n} = -\frac{p}{4\pi} \int \mathscr F^2 + in \int \delta_S \wedge \mathscr F + \frac{ip}{2\pi} \int \B_e \wedge d \V,  \quad \quad (\mathscr F = d\mathsf A - \B_e),
\end{equation}
where $\V$ is the dual 1-form gauge field (sometimes also referred to as the dual photon). Integration over $\B_e$ gives 
\begin{equation}
    -\frac{p}{4\pi} \int |\widehat{d\V}|^2 = -\frac{p}{4\pi} \int \Big| d \V - \frac{2\pi n}{p} \delta_S \Big|^2.
\end{equation}
Such a singularity defines a surface operator \cite{Gukov:2006jk}, since the field configurations that contribute to the path integral must obey, for finiteness, the boundary condition
\begin{equation}
    \frac{1}{2\pi} d\V = \frac{n}{p} \delta_S + \cdots,
\end{equation}
near the surface. This singularity near $S$ corresponds to an insertion of 
\begin{equation}
    \exp \Big( \frac{in}{p} \int_S d\V \Big) \equiv \mathscr U_{2\pi n/p} (S) \times \W_{n/p} (\partial S),
\end{equation}
which is a non-genuine Wilson line operator having an improperly quantized parameter. The consequence of this is that the attached surface operator does not obey the Dirac quantization condition, and therefore the worldsheet of the Diriac string, the surface operator, is visible and is required to make the configuration well-defined. The surface is invisible if and only if $n = 0 \text{ mod } p$ which constitute a subset of the spectrum of extended observables, all the rest of the spectrum $n \neq 0 \text{ mod } p$ has an attached Dirac surface. The most crucical difference between the action of $\mathscr D_r$ on $W_s$ and that of $\mathcal D_p$ on $\W_n$ is this: Namely $\mathcal D_p$ action typically attaches a physical surface, whereas $\mathscr D_r$ never attaches a surface (or, what is the same, it attaches a surface but the surface is invisible due to topology of $\R$ bundles). 

We remark that the action of the defect on lines:
\begin{equation}
    \mathcal D_p : \quad \W_n(\partial S) \mapsto \mathscr U_{2\pi n/p} (S) \times \W_{n/p} (\partial S),
\end{equation}
is indeed the equivalent to what has been discussed in the original work \cite{Choi:2022zal} in which $\mathcal D_p$ was constructed. One could say that the left operator also has an attached surface $\mathscr U_{2\pi n} (S)$, which is a trivial defect so one has a genuine line, whereas the right hand side has a physical surface for generic values of $n$.

\subsection{Condensation Defects in \texorpdfstring{$\R$}{R} Gauge Theory}\label{subsec 4.4: condensation defects in R Maxwell}

We reviewed the construction of condensation defects in compact gauge theory obtained by gauging the $\Z_p^{(1)}$ subgroup of the electric symmetry on a codimension-1 space in section \ref{sec 2: non-invertible defects in U(1) gauge theory} (see also the references \cite{Gaiotto:2019xmp, Roumpedakis:2022aik, Choi:2021kmx, Choi:2022zal, Kaidi:2021xfk} for more details). We now discuss condensation defects in the non-compact gauge theory. The first such defect to consider is
\begin{equation}
    \mathscr C_r \equiv \mathscr{D}_r \times \overline{\mathscr D}_r,
\end{equation}
where this expression means that we insert $\mathscr D$ and its dual parallel to each other by a distance $\varepsilon$, and take the limit $\varepsilon \to 0$ to get the condensation defect $\mathscr C_r$. This produces a trivial object\footnote{In the previous version of the paper, there was a naive discussion on condensation defects with some errors. I would like to thank the anonymus JHEP refree for challenging that discussion in the previous version, which pushed me to make these points more precise.}. To see this, we just observe the defect action \ref{Z_p condensation defect action} and write the corresponding version for gauing the $\R$ symmetry on the interface $\mathscr B$:
\begin{equation}
    S_{\mathscr C_r(\mathscr B)} = \underbrace{\frac{i}{2\pi r} \int_{\mathscr B} A \wedge d a}_{\mathscr D_r} - \underbrace{\frac{i}{2\pi r} \int_{\mathscr B}  a \wedge d V}_{\overline{\mathscr D}_r} = \frac{i}{2\pi r} \int_{\mathscr B}  a \wedge ( d A - d V),
\end{equation}
where $a$ is a 1-form $\R$ gauge field localized on $\mathscr B$, and $A,V$ are the gauge fields living to the left and right of $\mathscr B$, respectively. The field equation of $a$ sets $d(A-V) = 0$, and there are no period quantizations since $\int da = 0$ due to the fact that $a$ is an $\R$ gauge field. We then have $A = V$ up to some gauge transformation, meaning the left and right fields are gauge equivalent. This in particular implies that $W_s(L)$ operator is unchanged as $\mathscr C_r$ is swept across $L$, and therefore, the condensation $\mathscr C_r$ gives a trivial defect. This in turn implies the invertibility of $\mathscr D_r$.

\subsubsection{Another Condensation?}

The triviality of $\mathscr C_r$ above is due to the fact that $A-V$ does not obey any period quantization condition due to the field equations of the defect field. This in turn is a consequence of choosing the defect degree of freedom to be an $\R$-gauge field $a$. If we make the replacement $a \mapsto \mathsf a$ where the sans-serif field is $U(1)$ and obeys the Dirac quantization condition $\int d\mathsf a \in 2\pi \Z$, then it gives something potentially interesting. In this case, the defect action reads 
\begin{equation} \label{Defect action for C_Z on non-compact theory}
    S_{\mathscr C_{\Z}} = \frac{i}{2\pi r} \int_{\mathscr B} \mathsf a \wedge ( dA - dV ).
\end{equation} 
This time, the difference $A-V$ is a non-trivial gauge field. The flux sum over $d\mathsf a$ enforces $\int (A-V) \in 2\pi r \Z$, which means that $\frac{1}{r} (A-V) = \mathsf c \in H^1(\mathscr B, \Z)$. Recall that a $p-$form $U(1)$ gauge field with appropriate Dirac quantization condition has field tensor taking values in the integer cohomology $H^{p+1}(M, \Z)$. With this perspective, we regard $\mathsf c$ as the strength of a 0-form $U(1)$ gauge field, namely a compact scalar: $\mathsf c = d \varphi$ (locally) and $\varphi \sim \varphi + 2\pi$. Therefore, this new defect $\mathscr C_\Z(\mathscr B)$ acts non-trivially on Wilson lines. Due to the local field equations of $\mathsf a$, $dA = dV$, so the codimension-1 defect does not act on any of the local operators in the spectrum, since the gauge invariant operators are all related to the field strength, and that is invariant under the action of the defect. 

The fact that the codimension-1 defect $\mathscr C_\Z(\mathscr B)$ acts on lines but is porous to local operators (using the wording of \cite{Roumpedakis:2022aik}) is a characteristic of condensations. However, the nature of this topological defect is not immediately clear, and requires further developing. We do not pursue this task in this paper, but just remark that the condensation, and therefore the corresponding gauging on $\mathscr B$, should be governed by a sum over classes $[\mathsf c/2\pi ] \in H^1(\mathscr B, \Z)$, which using Poincaré duality $H^1( \mathscr B, \Z ) \simeq H_2(\mathscr B, \Z)$, may schematically be expressed as a homology sum 

\begin{equation}
    \mathscr C_\Z ( \mathscr B) \propto \sum_{S \in H_2(\mathscr B, \Z)} \zeta (S).
\end{equation}
For instance, if $\mathscr B = \mathbb T^3$, a three torus, then the correpsonding homology group is $\Z^{\oplus 3}$ and therefore there is an infinite sum to regularize, and along with that, one has to deal with the issue of the normalization factor. Both of these are highly non-trivial problems to overcome.

\subsection{Subgroup Gaugings of \texorpdfstring{$\R^{(1)}$}{R^1} Symmetry} \label{subsec 4.5: subgroup gauging in 4d R maxwell}

For completeness, we now discuss the coupling of the TQFT 
\begin{equation}
    \frac{i}{2\pi r} \int G \wedge d\V,
\end{equation}
with $\V$ a $U(1)$ 1-form gauge field, to the $\R$ Maxwell theory. As in the $\R$ scalar, we expect a topological manipulation enacted by this $U(1)/\R$ TQFT to compactify the gauge group from $\R$ to $U(1)$ by gauging a $2\pi \Z \subset \R$ subgroup. See \cite{Argurio:2024ewp,Paznokas:2025epc} for related manipulations.

We proceed in the standard order. $\V$ being a $U(1)$ field, its field strength $\W = d \V$ obeys the period quantization 
\begin{equation}
    \frac{1}{2\pi} \int \W \in \Z.
\end{equation}
Then, the integral over $\V$ consists of a continuous part and a discrete sum over the line bundles $\mathcal L$ with connection $\V_{\mathcal L}$ \cite{Witten:1995gf, Deligne:1999qp, Gukov:2006jk}
\begin{equation}
    \int \mathcal D\V = \sum_{\mathcal L} \int \mathcal D \V_{\mathcal L},
\end{equation}
Now consider the coupling of this term in $\R$ Maxwell theory, with action
\begin{equation} \label{Gauging the Z subgroup of R 1-form symmetry via W coupling}
    \hspace{1cm} \frac{1}{2} \int \mathcal F^2 + \frac{i}{2\pi r} \int G \wedge \W. 
\end{equation}
The continuous part of the integral over $\V$ enforces $G$ to be closed $dG = 0$, and the sum over the periods of $\W$ (i.e., the sum over the bundles) quantizes the periods $\oint G \in 2\pi r\Z$. We fix $A=0$ and write $G = r d\B$ locally, with $\B$ a $U(1)$ gauge field, which satisfies the conditions coming from the integral over $\V$. Thus, the $\V$ integral converts an $\R$ gauge theory to a $U(1)$ gauge theory 
\begin{equation}
    \frac{r^2}{2} \int d \B^2.
\end{equation}
The choice $r^2 = 1/e^2$ recovers the usual normalization. 

Alternatively, we can first integrate over $G$, giving the condition $*G = - \frac{i}{2\pi r} \W$ as in the case where $\V$ is $\R$ valued, but this time with the crucial difference that period quantization of $\V$ enforces  $\int *G$ to be quantized\footnote{The factor of $i$ may cause some confusion. In order to avoid that, it would be healthier to perform the duality transformation in the Lorentzian signature where there is no $i$, and then convert back to Euclidean.}. Inserting the solution of the field equation into the path integral upon performing the $G$ integral, we again land on a $U(1)$ gauge theory 
\begin{equation}
    \frac{1}{8\pi^2 r^2} \int d\V^2.
\end{equation}
The choice $r^2 = 1/e^2$ makes it manifest that the $d\B^2$ description and the $d\V^2$ description are $S-$dual to each other. The ordinary derivation of $S-$duality starts from a $U(1)$ theory path integral \cite{Witten:1995gf}, but the manipulation we described above also demonstrates that the two path integrals describe the same physics, as they descend from the same non-compact gauge theory where $\Z^{(1)} \subset \R^{(1)}$ is gauged. 

We note that the choice $r = 1/p$ where $p \in \Z$ has a non-trivial effect on period quantization $\int C \in \frac{2\pi \Z}{p}$. Namely, the topological defect $\exp \big( i \alpha \int C \big)$ realizes a $\Z_p$ subgroup gauging of the electric 1-form symmetry, since $\alpha = 0 \text{ mod } p$ is a trivial surface. Later on, we will develop a local TQFT coupling that also realizes subgroup gaugings of magnetic symmetries, and more generally mixed gaugings of electric and magnetic symmeteis that are not obstructed by the mixed 't Hooft anomaly. 

\textbf{Turning On A $\theta$ Angle From Gauging}

In the $\R$ Maxwell theory, the $\theta$ term is trivial even in the quantum theory. However, the subgroup gauging $\Z \subset \R$ gives a $U(1)$ Maxwell in which the theta angle can have non-trivial effects. What kind of a modification in our gauging procedure gives rise to a $\theta$ angle for the resulting $U(1)$ theory?

The answer is straightforward. Without changing the field equations of $\V$, we can add a theta term and consider the twisted gauging with a theta parameter
\begin{equation} \label{Gauging with theta angle for the W field}
    \hspace{1cm} \frac{1}{2e^2} \int \mathcal F^2 + \frac{i}{2\pi} \int G \wedge \W + \frac{i\theta}{8\pi^2} \int \W \wedge \W.
\end{equation}
The field equation of $\V$ still reads $dG=0$, but the integration over $\V$ is now different. The $G$ integral, on the other hand, is still easy and gives a description in terms of $\V$
\begin{equation}
    \hspace{1cm} \frac{e^2}{8\pi^2} \int \W \wedge * \W + \frac{i\theta}{8\pi^2} \int \W \wedge \W,
\end{equation}
which indeed is a $U(1)$ gauge theory with theta angle. Defining the complexified coupling 
\begin{equation}
    \tau = \frac{e^2}{2\pi} i + \frac{\theta}{2\pi},
\end{equation}
we can write this action as \cite{Witten:1995gf,Deligne:1999qp,Gukov:2006jk}
\begin{equation} \label{W gauge field with theta angle written in terms of W+ and W-}
    S = \frac{i\overline \tau}{4\pi} \int \W_+^2 - \frac{i\tau}{4\pi} \int \W_-^2,
\end{equation}
where we defined the (anti-)self dual projections $\W_{\pm} = \frac{1}{2} (\W \pm *\W)$. This action enjoys the $S-$duality acting as $\tau \mapsto -1/\tau$, along with the periodicity of the theta angle $\tau \mapsto \tau +1$ (provided $M_4$ has spin structure, otherwise only $\tau \mapsto \tau +2$ leaves the path inetgral invariant) which together extends to an $SL(2,\Z)$ action. Therefore, by turning on a $\theta$ angle for the conjugate field in the BF coupling, we can obtain a $U(1)$ Maxwell at coupling $\tau$ starting from an $\R$ Maxwell \ref{Gauging with theta angle for the W field}. This was also noted in \cite{Paznokas:2025epc}. 

\subsection{SymTFT Perspective}

These manipulations are again encoded in the SymTFT for $\R$ symmetries \cite{Antinucci:2024zjp}. The action on the slab reads
\begin{equation}
    \mathcal Z = \frac{i}{2\pi} \int_{M_4 \times I} G \wedge d \lambda,
\end{equation}
with $G$ and $\lambda$ being 2-form $\R$ gauge fields, and $G$ restricted to the left boundary ${\mathfrak B_{\text{phys}}} \equiv M_4 \times \{ 0\}$ is coupled to the 1-form symmetry of Maxwell. In the bulk, the interesting observables are 
\begin{equation}
    \zeta_\alpha(M_2) = e^{i\alpha \oint_{M_2} G} \quad ; \quad U_\beta(M_2) = e^{i \beta \oint_{M_2} \lambda},
\end{equation}
with the braiding given by the phase factor $e^{2\pi i \alpha\beta}$ when the support of the two topological defects has nontrivial linking. We again have three classes of topological boundary conditions \cite{Antinucci:2024zjp}.

Fixing the Dirichlet boundary condition on the topological boundary $\mathfrak B_{\text{top}} \equiv M_4 \times \{ 1\}$ 
\begin{equation}
    G \big|_{\mathfrak B_{\text{top}}} = 0,
\end{equation}
we have $\zeta_\alpha$ as the line operators charged under the $\R$ symmetry generated by $U_\beta$ upon slab contraction. The Neumann boundary condition
\begin{equation}
    \lambda \big|_{\mathfrak B_{\text{top}}} = 0,
\end{equation}
changes the picture and gauges the $\R$ symmetry, which reflects the invariance of the $\R$ Maxwell theory under gauging $\R^{(1)}$ symmetry discussed in section \ref{sec 4: R maxwell}. 

Finally, we can let the operators $\zeta_n$ and $U_m$ end on the boundary, after which the remaining topological operators acting on the integer-charged operators are given by $U_\alpha$ and $\zeta_\beta$ respectively, with $\alpha, \beta \in \R / 2\pi\Z$. This generates two $U(1)$ 1-form symmetries on the resulting theory in $M_4$ upon slab contraction, which we interpret as the $U(1)$ Maxwell theory \cite{Antinucci:2024zjp}. Thus, the subgroup gauging outlined in the previous subsection is an explicit realization of this boundary condition from a codimension-0 coupling. 

Extending the mixed boundary conditions by terminating $\zeta_{\mathrm R n}$ and $U_{\mathrm R^{-1} m}$ on the topological boundary, we are scaling the radii in the compact theories. The special choices $\mathrm R \in \mathbb Q$ describe various mixed subgroup gaugings of the $U(1)^{(1)} \times U(1)^{(1)}$ symmetry group that is realized on the physical boundary upon compactification with mixed boundary conditions. In particular, $\mathrm R = p \in \Z$ describes a $\Z_p$ subgroup gauging of the electric 1-forn symmetry, which we have realized by a local TQFT coupling in the previous subsection. Realizations of other boundary conditions for arbitrary choices of $\mathrm R$ will be realized by a local TQFT deformation in section \ref{sec 6: p-form non-compact gauge theory in arbitrary dimension}, which can be done for any $p-$form gauge theory in arbitrary dimensions.

\section{Subgroup Gaugings of \texorpdfstring{$\R^{(-1)}$}{R(-1)} and \texorpdfstring{$\R^{(d-1)}$}{R(d-1)} Symmetries} \label{sec 5: subgroup gauging of (-1) form R symmetries}

In this section, we explore the idea of gauging $\R-$valued $(-1)-$form and its quantum dual $(d-1)-$form symmetries. We start from $(-1)-$form symmetries. 

A $(-1)-$form symmetry \cite{Seiberg:2010qd,Tanizaki:2019rbk,Cordova:2019jnf,Cordova:2019uob,McNamara:2020uza,Aloni:2024jpb} is defined by a 0-form ``Noether current'' $j_0$ of which the conservation law is a trivial statement $d *j_0 = 0$ since a top form is always closed. The $(-1)-$form symmetry charges are typically integrated over the entire space-time, so as opposed to ordinary symmetries they do not have topological defects or charged objects\footnote{Since $(-1)-$form symmetries are not composed, there is no need for a group structure, and the corresponding structure can be an arbitrary set \cite{McNamara:2020uza}. }. However, they are not completely redundant. $(-1)-$form symmetries can be coupled to background fields \cite{Seiberg:2010qd,Tanizaki:2019rbk,McNamara:2020uza,Aloni:2024jpb}, gauged, and be spontaneously broken \cite{Aloni:2024jpb}. Further discussions on $\R^{(-1)}$ symmetries can be found in \cite{Perez-Lona:2025add}.

Given any local operator $\mathcal O$, $\int *\mathcal O$ defines a $(-1)-$form symmetry charge. Gauging this ``symmetry'' corresponds to considering a space-time dependent coupling coefficient. For example, in 4d Yang-Mills theory, instantons are associated to $U(1)$ $(-1)-$form symmetries, the charge being $\int \nu \equiv \frac{1}{8\pi^2} \int \text{Tr} F \wedge F$. This couples to a theta angle as 
\begin{equation}
    i \theta \int \nu,
\end{equation}
and since the symmetry is $U(1)$, the object we couple to $\nu$, the $\theta$ parameter, is $2\pi$ periodic. Gauging this corresponds to 
\begin{equation}
    i \int \theta(x) \wedge \nu.
\end{equation}
By a $\Z_p$ TQFT coupling, one can modify the instanton sectors of Yang-Mills theory so that only the instantons having a charge $p \Z$ contribute in the path integral. This modifies the periodicity of the theta angle to $\theta \sim \theta + 2\pi /p$. 

In earlier sections, motivated by the $\Z_p \subset U(1)$ gauging, we devised a $2\pi \Z \subset \R$ gauging procedure via a $U(1)/ \R$ TQFT coupling 
\begin{equation}
    \frac{i}{2\pi} \int b \wedge d \mathsf L,
\end{equation}
with $b$ an $\R$ scalar field and $\mathsf L$ a $U(1)$ gauge field. What happens if we couple this TQFT to a $(-1)-$form symmetry background field? For $(-1)-$form $\R$ symmetries, the background field $b$ is a $0-$form scalar field, and $\mathsf L$ is a $(d-1)$-form $U(1)$ gauge field.

Considering a single term in the Lagrangian, the corresponding symmetry is an $\R$ $(-1)-$form symmetry since its integral is real valued. To be concrete, take the free $\R$ scalar with Lagrangian top form $\mathcal L = \frac{1}{2} * (dX \wedge * dX)$. We write the action as
\begin{equation}
    S = \frac{g}{2\pi} \int \mathcal L,
\end{equation}
where we introduced some factors for convenience in what follows. Since $d \mathcal L = 0$, we have a $(-1)-$form symmetry, and the action is real-valued, so it is an $\R$ symmetry. We can turn on a background field to this by coupling an $\R$-valued scalar field $b$ 
\begin{equation}
    S = \frac{g}{2\pi} \int \mathcal L + \frac{1}{2\pi} \int b \wedge \mathcal L.
\end{equation}
Introducing a $(d-1)$ form $\R$ field $L$ which interacts with $b$ inside a BF term, we study the action\footnote{In this section, we take the spacetime manifold to have a Lorentzian signature, so the factors of $i$ drop from the action}
\begin{equation}
    S = \frac{g}{2\pi} \int \mathcal L + \frac{1}{2\pi} \int b \wedge \mathcal L - \frac{1}{2\pi} \int b \wedge d L.
\end{equation}
Field equation of $L$, $db = 0$, fixes $b$ to a constant $b_0$, and since there is no period quantization, $b_0 \in \R$. So we have the quantum dual $\R^{(d-1)}$ form symmetry. Thus, integrating over $L$ just shifts $g\mapsto g + b_0$ and the functional integral of $b$ becomes an ordinary integral over the $b_0 \in \R$ values. On the other hand, the field equation of $b$ imposes 
\begin{equation}
    \mathcal L = dL,
\end{equation}
which means that $\int \mathcal L = 0$ since $\int dL = 0$ for a real gauge field. Integrating over $b$ gives the path integral 
\begin{equation}
    Z = \int \mathcal D L\mathcal D X \ e^{ \frac{ig}{4\pi} \int (dX)^2 } \ \delta(S) ,
\end{equation}
where we can drop the trivial sum over $L$. Since this is a free theory, the classical solutions satisfy $S(X_{cl}) = 0$. Moreover, as the path integral is Gaussian, the saddle points, namely the classical solutions, suffice to solve the integral. Thus, we end up with 
\begin{equation}
    Z = \int \mathcal D X_{cl},
\end{equation}
which is a trivial integral over the space of classical solutions. 

Along similar lines, we can couple a $U(1)$ $\mathsf L$ field, in which case the period sum over $d \mathsf L$ fixes $\mathsf b \in 2\pi \Z$, which corresponds to a shift $g \mapsto g + 2\pi n$. The corresponding sum over $b$, $\sum_{n \in \Z} e^{i n S} \sim \delta(S)$, selects out the configurations on which the action is $2\pi\Z$, so we end up with a trivial theory. On the other hand, if we integrate out $b$ first, we have the quantization condition on the action 
\begin{equation}
    \int \mathcal L = \int d \mathsf L \in 2\pi \Z,
\end{equation}
which gives the same condition we obtained by first integrating $b$. Thus, the remaining path integral weight reads $e^{i gn}$, and the sum over fields reduces to a discrete sum over $n$ since the TQFT coupling selects out only those field configurations for which $\int \mathcal L = 2\pi \Z$. Hence, the resulting path integral is again trivial.

Having had these cavalier discussions, we will now employ the SymTFT tools in attempting to understand the manipulations of $(-1)-$form symmetries. 

\subsection{SymTFT for \texorpdfstring{$(-1)-$}{(-1)-}form \texorpdfstring{$\R$}{R} Symmetries}

The exotic notion of $(-1)-$form symmetry allows a SymTFT description \cite{Antinucci:2024zjp, Brennan:2024fgj,  Aloni:2024jpb, Lin:2025oml, Yu:2024jtk}. Using the topological boundary conditions, we can gain insight into $(-1)-$form symmetries, of which the current itself is the Lagrangian. 

Consider an $\R^{(-1)}$ symmetry in $M_d$ with current $\mathcal L_{b}$, which we think of as a possible term in the Lagrangian. Turning on an $\R$ valued background scalar field $b$, we consider the codimension-$(-1)$ coupling of the $\R/\R$ TQFT as the SymTFT of this $\R^{(-1)}$ symmetry:
\begin{equation} \label{SymTFT for the (-1)-form R symmetry in d dimensions}
    \mathcal Z = \frac{i}{2\pi} \int_{M_d \times I} b \wedge d \Lambda,
\end{equation}
with $\Lambda$ a $d-$form $\R$ gauge field. From the topological boundary conditions of this SymTFT on the right part of the slab, the QFT on $M_d$ can have the following symmetries \cite{Antinucci:2024zjp, Aloni:2024jpb}
\begin{equation} \label{Lagrangian algebras for the R SymTFT}
\begin{aligned}
    \text{Dirichlet: } \quad \quad \quad \quad b \big|_{\mathfrak B_{\text{top}}} = 0 \quad \quad \quad \quad &\Longrightarrow \quad \quad \quad \quad \R^{(-1)} \\
    \text{Neumann:} \hspace{3.1mm} \quad \quad \quad \Lambda \big|_{\mathfrak B_{\text{top}}} = 0 \quad \quad \quad \quad &\Longrightarrow \quad \quad \quad \quad \R^{(d-1)} \\
    \text{Mixed: } \hspace{4mm} \quad \quad \zeta_n \big|_{\mathfrak B_{\text{top}}} = 1 = U_m \big|_{\mathfrak B_{\text{top}}} \quad & \Longrightarrow \quad U(1)^{(-1)} \times U(1)^{(d-1)}
\end{aligned}
\end{equation}
with $\zeta_n = e^{i n b}$, $U_m = e^{im \oint \Lambda}$ and $n,m \in \Z$. We will discuss the mixed boundary condition of \cite{Antinucci:2024zjp} in more detail in section \ref{subsec 5.2: Lagrangian Algebras of the R SymTFT and a Generalization of the R(d-1) Gauging}. 

From the wisdom of SymTFT, we learn about two intriguing possibilities given an $\R^{(-1)}$ symmetry. First, starting from any real-valued term in the Lagrangian defining an $\R^{(-1)}$, it should be possible to obtain an $\R^{(d-1)}$ symmetry. This symmetry divides the theory into universes (in the language of \cite{Komargodski:2020mxz, Tanizaki:2019rbk, Hellerman:2006zs}) labelled by a real number, which is the value of $b$. The would-be universes are then separated by $\R$ domain walls. Realizing this manipulation on the $\R^{(-1)}$ symmetry is easily achieved by coupling the 0-form background field into a $\R/\R$ TQFT, which gives a flat gauging prescription of $\R^{(-1)}$, yielding a $\R^{(d-1)}$ symmetry 

Second, and more interestingly, the mixed topological boundary conditions of the SymTFT inform us that we can have topological instanton sectors from $U(1)^{(-1)}$ and decomposed universes separated by $U(1)^{(d-1)}$ domain walls. An example of a theory realizing $U(1)^{(-1)} \times U(1)^{(d-1)}$ symmetries is the $d=2$ compact Maxwell theory \cite{Aloni:2024jpb}. 

Since the $\R/\R$ TQFT for the $(-1)-$form $\R$ symmetry can equivalently be viewed as the SymTFT for a $\R^{(d-1)}$ symmetry, we can also start from such a symmetry and according to the mixed boundary conditions of SymTFT, we must be able to obtain $U(1)^{(-1)} \times U(1)^{(d-1)}$ with a topological manipulation of the $(d-1)-$form symmetry as well.

Directed by the mixed topological boundary condition, we then ask the question: what is a 2d QFT with $\R^{(-1)}$ or $\R^{(1)}$ symmetry such that gauging a $2\pi \Z \subset \R$ subgroup produces the 2d $U(1)$ Maxwell theory? We may attempt to answer this question using the codimension-0 realization of $\Z \subset \R$ gauging via a $U(1)/\R$ TQFT coupling. Based on all the discussions up to this point, we naturally guess that this is achieved by starting from the 2d $\R$ Maxwell theory 
\begin{equation} \label{2d R Maxwell action}
    \frac{1}{2e^2} \int_{M_2} |da|^2,
\end{equation}
having the 1-form $\R$ symmetry generated by $*da$, acting on the Wilson lines $\exp \big( i r \oint a \big)$ with $r \in \R$. We gauge this 1-form shift symmetry 
\begin{equation}
    \frac{1}{2e^2} \int_{M_2} |da -c|^2.
\end{equation}
The 2-form field $c$ is always closed in $2d$, and we cannot insert it into a BF coupling on $M_2$. We need to get creative to obtain a $U(1)$ Maxwell from a manipulation of this theory. Observe that in a flat gauging, the essential ingredient is a sum over the \emph{flat} background fields $c$, so our deformation should achieve this in order for it to describe a topological (or what is the same, flat) gauging. The simplest way to do this is to impose this as a constraint via a Lagrange multiplier, which amounts to adding a term of the form\footnote{In the previous version of the paper, we considered this deformation where instead of the real valued $b$ one has a compact scalar obeying a periodicity condition $\mathsf b \sim \mathsf b +2\pi$. As correctly pointed out by the anonymous referee, the terms in such a deformation would not be compatible with the periodicity, or in other words, the 0-form analogue of the large gauge invariance for $\mathsf b$ (see the next footnote). Changing $\mathsf b$ in the previous version to the real valued $b$ in this version washes away that problem, since the local field equation remains the same, and the fact that there is no longer a discrete sum over the periods of the field $\mathsf b$ (or equivalently, there are no large gauge transformations) is not an issue since at the end of the manipulation one still ends up with a $U(1)$ gauge field $\mathsf a$. The couplings $b \wedge c$ and $b \wedge d \mathsf a$ are well defined now, and the latter coupling is of the $\R/U(1)$ TQFT form that we studied in many parts of the paper.} \footnote{ \label{footnote on large gauge transformations for compact scalar fields} To clarify the relation between $\mathsf b \sim \mathsf b +2\pi$ periodicity and the 0-form analogue of large gauge invariance, recall that for a 1-form $U(1)$ gauge field, a large gauge transformation is given as a shift by a flat $U(1)$ field $\lambda$ with the period quantization $\int_{\Sigma_1} \lambda \in 2\pi \Z$. If one schematically regards a compact scalar as a 0-form $U(1)$ gauge field, the analogues of large gauge transformations are then shifts by flat 0-form fields $\varphi$ with a ``period quantization" $\int_{\Sigma_0} \varphi \in 2\pi\Z$ (here, $\Sigma_0$ should appropriately be thought of as a boundary of a finite interval, see footnote \ref{footnote: Period quantization over a 0-manifold}). Flatness of a 0-form field (namely $d \varphi$ being 0 locally) is equivalent to saying that it is locally constant, and the path integral being invariant under the ``large gauge transformation" $\mathsf b \mapsto \mathsf b + \varphi$ amounts to the identification $\mathsf b \sim \mathsf b + 2\pi$ since the value of $\varphi$ at any two arbitrary points can only differ by $2\pi \Z$.}
\begin{equation}
    \frac{i}{2\pi} \int_{M_2} b ( c - d \mathsf a), 
\end{equation}
where $b$ is a real-valued scalar field playing the role of a Lagrange multiplier, and $\mathsf a$ is a $U(1)$ gauge field. One may perhaps interpret this coupling in another way: In essence, we define $b$ as a conjugate of the background field $c$ through the term $b c$, and then insert $b$ into an $\R/U(1)$ TQFT since $c$ cannot be entered into any 2d Lagrangian with derivatives whereas the ``conjugate" $b$ can very well be inserted into a topological action. 

Clearly, the first point of view where $b$ is a Lagrange multiplier whose field equation enforces $c$ to be an exact form with constrained periods $\int_{\Sigma_2} c \in 2\pi \Z$ is the correct way to look at the deformation; the second interpretation in the above paragraph is just a cavalier description with the purpose of making the role of the $\R/U(1)$ TQFT coupling in our manipulation more apparent.

After this deformation, our Euclidean path integral $\int \mathcal Da \mathcal Dc \mathcal Db \mathcal D \mathsf a \ e^{-S[a,c,b,\mathsf a]}$ has the action
\begin{equation} \label{The Lagrangian for the topological gauging of Z subgroup for 1-form shift symmetry in 2d R Maxwell}
    S[a,c,b,\mathsf a] = \frac{1}{2e^2} \int_{M_2} |da-c|^2 + \frac{i}{2\pi} \int_{M_2} b (c-d\mathsf a).
\end{equation}
We expect this path integral to be equivalent to a $U(1)$ Maxwell theory having $U(1)^{(-1)} \times U(1)^{(1)}$ symmetries. Let us first check that these symmetries indeed appear from the simple action above.

First of all, the equations of motion of $b$ make $c$ a flat field with periods $\int c \in 2\pi\Z$, therefore we have the codimension-0 topological defect $\exp \big( \frac{i\alpha}{2\pi} \int c \big)$ generating a $U(1)^{(-1)}$ symmetry, which can be regarded as the quantum dual symmetry upon the flat gauging of the subgroup $2\pi \Z^{(1)} \subset \R^{(1)}$. This space-filling defect coincides with $\exp \big( \frac{i\alpha}{2\pi} \int d\mathsf a \big)$ due to the $b$ equations of motion, and when $c$ is integrated out, it is clearly the $(-1)-$form symmetry generator of the 2d pure $U(1)$ gauge theory. Additionally, as a result of the local equations of motion for $\mathsf a$ ($db = 0$) plus the period sum over $\mathsf a$ fluxes (or equivalently the large gauge invariance under shifts $\mathsf a \mapsto \mathsf a + \lambda$ with $\lambda$ a flat 1-form gauge field having quantized periods as discussed in the footnote \ref{footnote on large gauge transformations for compact scalar fields}), we have the topological local operator $\exp \big( \frac{i\beta}{2\pi} b \big)$, which defines a $U(1)^{(1)}$ symmetry. This decomposes the theory into universes \cite{Sharpe:2022ene} labeled by the expectation value of $b$ in that universe. The period sum of $\mathsf a$ implies that the difference between two expectation values of $b$ at any two universes must be integer-gapped, which is a manifestation of the fact that the symmetry is $U(1)^{(1)}$. As we will see momentarily, upon performing the integrations over $c$ and $b$, the topological local operator $\exp \big( \frac{i\beta}{2\pi} b \big)$ will morph into the 1-form symmetry generator $\exp \big( \frac{i\beta}{2\pi} *da \big)$ of $U(1)$ Maxwell theory.

Having seen the correct $U(1)^{(-1)} \times U(1)^{(1)}$ symmetry structure appearing after our deformation, we now confirm our anticipation that this manipulation results in the $U(1)$ gauge theory upon taking the integrals. As before, we use the gauged shift symmetry to fix $a=0$ and then perform the quadratic integral over $c$, resulting in 
\begin{equation}
    \frac{e^2}{8\pi^2} \int b^2 - \frac{i}{2\pi} \int b \wedge d\mathsf a.
\end{equation}
Since the $b$ field equation is $ b \propto *d\mathsf a$, we can see that the topological local operator obtained by the exponential of $b$ becomes the 1-form symmetry generator of $U(1)$ Maxwell in $d=2$ once $b$ is integrated out. And indeed, the quadratic integral over $b$ will produce the familiar action of the Maxwell theory
\begin{equation}
    \frac{1}{2e^2} \int_{M_2} |d\mathsf a|^2.
\end{equation}
We can also turn a theta angle by adding a counterterm in \ref{The Lagrangian for the topological gauging of Z subgroup for 1-form shift symmetry in 2d R Maxwell} that only contains $\mathsf a$, so that the manipulations are unchanged. Thus, adding the counterterm $\frac{i\theta}{2\pi} \int_{M_2} d\mathsf a$ in our manipulation turns up a theta angle in the resultant theory, and since it is a theta term, it does not change the equations of motion of $\mathsf a$ so the exponential of $b$ will still be topological in the presence of this counterterm.

This concludes our attempt at realizing the mixed boundary condition of the SymTFT for $(-1)-$form and $(d-1)-$form symmetries when $d=2$. 

To generalize the steps above to a generic dimension $d$, consider a $p = d-1$ form $\R$ gauge theory with action 
\begin{equation}
    \frac{1}{2g^2} \int_{M_d} |da_{(d-1)}|^2.
\end{equation}
We have a $(d-1)-$form symmetry generated by $*da_{(d-1)}$, which acts on the codimension-1 walls $\exp\big( ir \oint a_{(d-1)})$ with $r \in \R$. The SymTFT for this symmetry reads as in \ref{SymTFT for the (-1)-form R symmetry in d dimensions}. The mixed boundary conditions in that SymTFT inform us about the existence of a codimension-0 TQFT coupling which converts the $\R$ valued $p=d-1$-form gauge theory to a $U(1)$ valued one, which has the $U(1)^{(-1)} \times U(1)^{(d-1)}$ symmetry with mixed anomaly. The explicit deformation of the $(d-1)-$form gauge theory is governed by the path integral over the action 
\begin{equation}
    \frac{1}{2g^2} \int_{M_d} |da_{(d-1)} - c_d |^2 + \frac{i}{2\pi} \int_{M_d}  b ( c_d - d \mathsf a_{(d-1)} ) + \frac{i\theta}{2\pi} \int_{M_d} d \mathsf a_{(d-1)},
\end{equation}
where $c_d$, $b$, and $\mathsf a_{(d-1)}$ are $\R$ valued $d-$form, $\R$ valued $0-$form (the Lagrange multiplier), and $U(1)$ valued $(d-1)-$form gauge fields, respectively. As before, the $b$ field equation ensures $d c_{d} = 0$ in a non-trivial way so that $c_d$ is locally exact, and due to the period quantization of $d\mathsf a_{(d-1)}$, the topological operator $\exp \big( i \frac{\alpha}{2\pi} \oint c_{d} \big)$ defines a $U(1)^{(-1)}$ symmetry, signalling the $\Z^{(d-1)} \subset \R^{(d-1)}$ subgroup gauging. And as before, due to the $b$ equations of motion, the topological defect coincides with the $(-1)-$form symmetry generator, $\exp \big( \frac{i\alpha}{2\pi} \int_{M_d} d\mathsf a_{d-1} \big)$, of the $(d-1)-$form $U(1)$ gauge theory. Similarly, the $\mathsf a$ field equation $db = 0$ plus the flux sum gives a topological local operator $\exp \big( \frac{i\beta}{2\pi} \ b \big)$ which defines a $U(1)^{(d-1)}$ symmetry since the invariance under large gauge shifts $\mathsf a_{d-1} \mapsto \mathsf a_{d-1} + \lambda_{d-1}$ imposes the vacuum expectation value of $b$ in two arbitrary universes to differ by integer gaps. Thus, the symmetry structure $U(1)^{(-1)} \times U(1)^{(d-1)}$ is apparent, by complete parallel with the 2d Maxwell discussion.

We can also easily perform the integrals to explicitly obtain the $U(1)$ $(d-1)-$form gauge theory action. Using the gauged $(d-1)-$form shift symmetry, we fix $a_{(d-1)} = 0$, and integration over $c_d$ gives an action Gaussian in $b$, integration of which yields the desired Lagrangian:
\begin{equation}
    \frac{1}{2g^2} \int_{M_d} |d\mathsf a_{(d-1)}| ^2 + \frac{i\theta}{2\pi} \int_{M_d} d \mathsf a_{(d-1)}.
\end{equation}
We note that the $d=1$ case is related to the particle on the real line. Here, the real valued gauge field $a_{d-1} \equiv X$ is a scalar field living on a 1d spacetime, which describes a point particle of position $X$. Gauging the $\R^{(0)}$ symmetry as outlined above, we enact a topological manipulation on this theory, which result in the particle on a ring (sometimes also referred to as the Aharonov-Bohm particle) with $U(1)^{(-1)} \times U(1)^{(0)}$ symmetry. 

\subsection{Lagrangian Algebras of the \texorpdfstring{$\R$}{R} SymTFT and a Generalization of the $\R^{(d-1)}$ Gauging} \label{subsec 5.2: Lagrangian Algebras of the R SymTFT and a Generalization of the R(d-1) Gauging}

In this section, we will generalize the gauging operation for $\R^{(d-1)}$ symmetries that was outlined in the previous section. To motivate this, we first revisit the choice of topological boundary conditions in \ref{Lagrangian algebras for the R SymTFT}. In fact, the mixed boundary conditions at the last row of equation \ref{Lagrangian algebras for the R SymTFT} are not the most general ones. In particular, one can make an appropriate rescaling by a parameter $\mathrm R$ such that the boundary conditions still give a Lagrangian algebra \cite{Antinucci:2024zjp}. In terms of the defects $\zeta_\alpha = \exp \big( i \alpha b \big)$ and $U_\beta = \exp \big( i \beta \oint \Lambda \big)$ (where $\alpha$ and $\beta$ can be any real numbers in the bulk), the topological boundary conditions read
\begin{equation}
    \zeta_{n \mathrm R} \big|_{\mathfrak B_{\text{top}}} = 1 = U_{m \mathrm R^{-1}} \big|_{\mathfrak B_{\text{top}}},
\end{equation}
for $n,m \in \Z$, and $\mathrm R$ is an arbitrary real number. As explained in \cite{Antinucci:2024zjp}, the SymTFT does not measure the radius of the $U(1)$s when we contract the slab, but it can distinguish the two radii arising from two different topological boundary conditions. Additionally, when $\mathrm R$ is fixed to an integer $\mathrm R = q \in \Z$, this corresponds to a $\Z_q^{(d-1)} \subset U(1)^{(d-1)}$ subgroup gauging in the physical theory upon slab contraction. Alternatively, fixing $\mathrm R = \frac{1}{k} \in \frac{1}{\Z}$ trivializes $\zeta_{\frac{n}{k}}$ and $U_{mk}$ on the topological boundary, therefore it describes a $\Z_k^{(-1)} \subset U(1)^{(-1)}$ gauging. 

Going even further from the choices where $\mathrm R$ is an integer or 1 divided by an integer, we can consider where $\mathrm R$ is a rational number $\mathrm R = \frac{q}{k} \in \mathbb Q/\Z$ where $\text{gcd}(q,k)=1$. \footnote{The quotient by $\Z$ just means $k \neq 1$. When $k=1$, $\mathrm R \in \Z$ which has already been discussed in the text.} In that case, the defects $\zeta_{\frac{nk}{q}}$ and $U_{\frac{mq}{k}}$ are trivialized on the topological boundary. Since $n,m,k,q$ are all integers, we have $\zeta_{\frac{N}{q}}$ and $U_{\frac{M}{k}}$ trivialized on the boundary for $N,M\in \Z$, and this means that when $\mathrm R = \frac{q}{k} \in \mathbb Q/\Z$, the corresponding boundary conditions describe $\Z_k^{(-1)}  \times \Z_q^{(d-1)}\subset U(1)^{(-1)} \times U(1)^{(d-1)}$ gauging. Such boundary conditions are obviously also allowed for the SymTFT of an $\R^{(p)}$ symmetry, of which the spectrum consists of the extended observables $\zeta_\alpha(M_{p+1}) = \exp \big( i \alpha \oint b_{p+1} \big)$ and $U_\beta(M_{d-p}) = \exp \big( i \beta \oint \Lambda_{d-p} \big)$. In the topological boundary, trivializing $\zeta_{n \mathrm R}$ and $U_{m \mathrm R^{-1}}$ where $\mathrm R = \frac{q}{k} \in \mathbb Q/\Z$ describes, by parallel with the $p=-1$ case, a $\Z_k^{(p)} \times \Z_q^{(d-p-2)} \subset U(1)^{(p)} \times U(1)^{(d-p-2)}$ subgroup gauging.

The existence of these choice of Lagrangian algebra shows, in particular, that the $\R^{(p)}$ SymTFT includes subgroup gaugings of the form $\Z_k^{(p)} \times \Z_q^{(d-p-2)} \subset U(1)^{(p)} \times U(1)^{(d-p-2)}$ as a global realization for a $p-$form $\R$ symmetry. Since the two $U(1)$s have a mixed 't Hooft anomaly, the $\text{gcd}(k,q) =1$ condition is necessary to gauge without obstruction, and indeed the fact that $\R/\R$ TQFT allowing this boundary condition is in par with the fact that the above subgroup gauging bypasses the mixed anomaly. In particular, the cases $d=2, p=0$ and $d=4, p=1$ are naturally related to the duality defects of the $2d$ compact boson and $4d$ $U(1)$ Maxwell theory \cite{Choi:2021kmx, Choi:2022zal}. In \cite{Argurio:2024ewp}, this was used to construct various condensation defects in the $\R$ SymTFT by gauging the $\R \subset \R \times \R$ subgroup, and upon slab contraction these condensations generalize the duality defects $\mathcal D_p$ of \cite{Choi:2021kmx}. Therefore, it is possible to define a T-duality defect at arbitrary radius of the compact boson. The Maxwell version of these condensations has been developed in \cite{Paznokas:2025epc}.

Can we realize this more general choice of Lagrangian algebra, namely the topological boundary conditions, in our deformation \ref{The Lagrangian for the topological gauging of Z subgroup for 1-form shift symmetry in 2d R Maxwell}? Simple enough, we consider modifying the Lagrange multiplier term as 
\begin{equation} \label{Generalized gauging for R(d-1) symmetry}
    \frac{i}{2\pi} \int_{M_d} b \Big( c_d - \frac{1}{\mathrm R} \ d\mathsf a_{d-1} \Big).
\end{equation}
Since the $b d\mathsf a$ coupling is of the $\R/U(1)$ TQFT form, we can arbitrarily rescale its coefficient (we chose to introduce the parameter as $1/\mathrm R$ for convenience in what follows). Consequently, the $b$ field equation enforces $\int c_d \in 2\pi \mathrm R^{-1} \Z$ and the $\mathsf a$ field equation along with the sum over fluxes imposes $\int_{\Sigma_0} b \in 2\pi \mathrm R \Z$ \footnote{ \label{footnote: Period quantization over a 0-manifold} Here, the integral over $\Sigma_0$ may appear troubling. We properly think of $\Sigma_0$ as a boundary of a 1d manifold $M_1$, which means that $M_1$ has the topology of either an interval with one end open or an interval with both ends open (for other 1d topologies, the boundary is either empty or a collection of intervals with one or two ends). In the former case, one has a single point $\Sigma_0 = \{ p_0 \}$, so $\int_{\Sigma_0} b = b(p_0)$, whereas in the latter, $\Sigma_0 = \{p_0\} \cup \{p_1\}$ and thus $\int_{\Sigma_0} b = b(p_1) - b(p_0)$ and the minus sign is due to the orientation in the boundary operation $\partial M_1 = \Sigma_0$. The condition $\int_{\Sigma_0} b \in 2\pi \mathrm R \Z$ thus means that the value of $b$ at any point in spacetime is of the form $2\pi n\mathrm R$. This is compatible with the difference $b(p_1) - b(p_0)$ also being quantized in the same units, because $b(p_1) - b(p_0) = 2\pi \mathrm R (n_1 - n_0) \equiv 2\pi\mathrm R n' \in 2\pi \mathrm R \Z$. }. When we perform the integrals over $c_d$ and $b$, the resulting theory gets a scaled radius (for example in the $d=1$ case, this corresponds to changing the radius of the ring on which the Aharonov-Bohm particle lives).

Fixing $\mathrm R = q \in \Z$ in the Lagrange multiplier term above, we can also realize the $\Z_q \subset U(1)$ gauging that was reviewed above. To see this, note that the periods are now constrained as $\int c \in \frac{2\pi}{q}\Z$ and $\int_{\Sigma_0} b \in 2\pi q \Z$, and therefore one has an emergent $\Z_q$ symmetry generated by the topological space-filling operators $\exp \big( i \eta \int c \big)$. When $\eta = 0 \text{ mod } q$, this is a transparent operator, hence the $\Z_q$. Since we have a $(-1)-$form dual symmetry, the operation with $\mathrm R = q$ should be interpreted as a subgroup gauging of the $(d-1)-$form $U(1)$ symmetry.

We could also fix $\mathrm R = \frac{1}{q} \in \frac{1}{\Z}$. The period quantization conditions are $\int c \in 2\pi q \Z$ and $\int_{\Sigma_0} b \in \frac{2\pi}{q} \Z$. This time, the topological local operator $\exp \big( i \eta \int_{\Sigma_0} b \big)$ is trivial when $\eta = 0 \text{ mod } q$, so $\eta \in \Z_q$ exhausts the distinct generators. Moreover, because the topological local operator generates an emergent $(d-1)-$form symmetry, we should understand the operation with $\mathrm R = q \in \Z$ as gauging a subgroup of the $(-1)-$form $U(1)$ symmetry. 

Finally, we could fix $\mathrm R =\frac{q}{k} \in \mathbb Q /\Z$. In that case the periods are quantized as $\int c \in \frac{2\pi k}{q} \Z$ and $\int_{\Sigma_0} b \in \frac{2\pi q}{k} \Z$. As a result, the topological operators $\exp \big( i \eta \int c \big)$ and $\exp \big( i \theta b \big)$ are transparent when $\eta = 0 \text{ mod } q$ and $\theta = 0 \text{ mod } k$ respectively. Consequently, the operation with $\mathrm R \in \mathbb Q/\Z$ gauges $\Z_k^{(-1)} \times \Z_q^{(d-1)} \subset U(1)^{(-1)} \times U(1)^{(d-1)}$ as in the SymTFT picture. 

Therefore, the deformation of a $(d-1)-$form $\R$ gauge theory by the term \ref{Generalized gauging for R(d-1) symmetry}, which carries a label $\mathrm R$, realizes the mixed boundary conditions for the SymTFT. The parameter $\mathrm R$ for the SymTFT Lagrangian algebra and the deformation parameter in \ref{Generalized gauging for R(d-1) symmetry} are identified, so to each choice of Lagrangian algebra with label $\mathrm R$ there corresponds a deformation \ref{Generalized gauging for R(d-1) symmetry} for the $(d-1)-$form gauge theory. Since the SymTFT result is theory independent, one may be tempted to expect the manipulation \ref{Generalized gauging for R(d-1) symmetry} to realize the choice of Lagrangian algebra with parameter $\mathrm R$ for any theory $\mathcal X$ having a $\R^{(d-1)}$ symmetry, the $d-$form background field of which is inserted into a term \ref{Generalized gauging for R(d-1) symmetry} with the same parameter $\mathrm R$. It is straightforward to see this in the non-compact $(d-1)-$form gauge theory, but it probably would not be as easy to observe for an interacting theory. Another issue is that quantum field theories with $\R^{(d-1)}$ symmetries do not seem to be very abundant. In any case, since the global realizations of these symmetries include topologically rich structures depending on $\mathrm R$, further study of the non-compact symmetries in higher-form gauge theories may yield new insights about symmetries and anomalies of $U(1)$ and $\Z_N$ type symmetries.

\section{\texorpdfstring{$p$-}{p-}form Gauge Theory in Arbitrary Dimensions} \label{sec 6: p-form non-compact gauge theory in arbitrary dimension}

\subsection{Gaugings}

As a closing of our study of $\R$ symmetries in Abelian gauge theory and the topological manipulations in them, we will consider the most general such model, which is the theory of a $p-$form $\R$ valued gauge field $a_p$ in $d-$dimensions, described by the Lagrangian
\begin{equation}
    \frac{1}{2 g^2} \int_{M_d} |da_p|^2.
\end{equation}
We can gauge the $p-$form $\R$ shift symmetry 
\begin{equation}
    \frac{1}{2 g^2} \int |da_p - c_{p+1}|^2 + \frac{i}{2\pi r} \int c_{p+1} \wedge d b_{d-p-2}
\end{equation}
to get the $(d-p-2)-$form non-compact gauge theory
\begin{equation}
    \frac{g^2}{8\pi r^2} \int |db_{d-p-2}|^2.
\end{equation}
Additionally, we could gauge the $\Z \subset \R$ subgroup via the coupling
\begin{equation}
        \frac{1}{2 g^2} \int |da_p - c_{p+1}|^2 + \frac{i}{2\pi r} \int c_{p+1} \wedge d \mathsf b_{d-p-2},
\end{equation}
which produces the $(d-p-2)-$form $U(1)$ gauge theory
\begin{equation}
    \frac{g^2}{8\pi^2 r^2} \int |d \mathsf b_{d-p-2}|^2.
\end{equation}
We also remark that $r=1/p$ describes a $\Z_p^{(d-p-2)} \subset U(1)^{(d-p-2)}$ subgroup gauging.

As we glossed over in section \ref{subsec 5.2: Lagrangian Algebras of the R SymTFT and a Generalization of the R(d-1) Gauging}, there are a family of topological boundary conditions in the $\R$ SymTFT that give rise to $U(1)^{(p)} \times U(1)^{(d-p-2)}$ symmetry structure. This family of Lagrangian algebras are parameterized by a real number $\mathrm R$. The special choices $\mathrm R = \frac{q}{k}$ describe a $\Z_k^{(p)} \times \Z_q^{(d-p-2)} \subset U(1)^{(p)} \times U(1)^{(d-p-2)}$ gauging. We now discuss a realization of these boundary conditions for all parameters $\mathrm R$ in terms of a local TQFT coupling, which is motivated by our considerations of gauging $(d-1)-$form $\R$ symmetries in the section \ref{subsec 5.2: Lagrangian Algebras of the R SymTFT and a Generalization of the R(d-1) Gauging}.

We explain this deformation by first writing down the natural generalization to the local TQFT coupling \ref{Generalized gauging for R(d-1) symmetry} for a $(d-p-2)-$form symmetry in $d-$dimensions\footnote{We are making a slight change in our conventions, but this is no problem since whatever we discuss for the $(d-p-2)-$form gauge theory is related to the $p-$form gauge theory by gauging the corresponding shift symmetry.}, which reads
\begin{equation} \label{Generalized gauging in p-form gauge theory in d dimensions}
    \frac{i}{2\pi} \int_{M_d} b_{p+1} \wedge \Big( c_{d-p-1} - \frac{1}{\mathrm R} d \mathsf a_{d-p-2} \Big),
\end{equation}
where the $b$ and $c$ are $\R$ gauge fields of indicated rank, and $\mathsf a$ is a $U(1)$ gauge field. 

In the case $p=-1$, the field $b$ is a non-compact scalar so it does not have any gauge transformations, but for $p > -1$ we need to check for the invariance under gauge transformations of $b$ as well as for the invariance under $c$ gauge transformations and $\mathsf a$ gauge transformations. The second term is just a $\R/U(1)$ BF coupling between $b$ and $\mathsf a$, so it does not raise any question marks about gauge invariance. The gauge invariance of $b\wedge c$ is not obvious at the onset, so we discuss this point now. 

We assume a closed $M_d$ for simplicity. Under the $b$ gauge transformation $\delta b_{p+1} = d\kappa_{p}$ (with $\kappa$ a real valued $p-$form field), the change in the coupling is given by $\frac{i}{2\pi} \int d\kappa_p \wedge c_{d-p-1}$. With an integration by parts, we see that gauge invariance is attained for a closed $c$: $dc_{d-p-1} = 0$. Since the field equation of $b_{p+1}$ imposes precisely this condition due to $d^2 \mathsf a = 0$, the Poincaré's lemma, for on shell configurations the action is invariant under $\delta b_{p+1} = d \kappa_p$. In the $p-$form abelian gauge theory coupled to \ref{Generalized gauging in p-form gauge theory in d dimensions}, the path integral gets contributions only from fields obeying the equations of motion, so the gauge invariance under $\delta b_{p+1} = d\kappa_p$ is satisfied. 

Along similar lines, under the gauge transformations $\delta c_{d-p-1} = d \kappa_{d-p-2}$, the invariance of the $b\wedge c$ term depends on the flatness of $b$. Since $\mathsf a$ field equation imposes $b$ to be flat, the path integral is invariant under $\delta c = d\kappa$ as well. The invariance under $\delta \mathsf a = d \alpha$ is easy to see due to Poincaré's lemma. The invariance under large gauge shifts $\delta \mathsf a_{d-p-2} = \lambda_{d-p-2}$ imposes \cite{Brennan:2024fgj}\footnote{Easily, one sees that the coupling changes as $-\frac{i}{2\pi \mathrm R} \int b_{p+1} \wedge d\lambda_{d-p-2} = -\frac{i\Z}{\mathrm R} \int b$ and the exponential of this expression is 1 only if the following equation in the text holds. }
\begin{equation}
    \int b \in 2\pi \mathrm R \Z.
\end{equation} 

Having settled the gauge invariance issue, we now explore the consequences of the coupling \ref{Generalized gauging in p-form gauge theory in d dimensions} in $(d-p-2)-$form abelian gauge theory. The Lagrangian description is 
\begin{equation}
    \frac{1}{2g^2} \int | d a_{d-p-2} - c_{d-p-1} |^2 + \frac{i}{2\pi} \int b_{p+1} \wedge \Big( c_{d-p-1} - \frac{1}{\mathrm R} d\mathsf a_{d-p-2} \Big).
\end{equation}
This is a simple theory, all the variables are at most quadratic so the integrations are straightforward. We recall that our purpose is to realize all possible choices of Lagrangian algebras of \cite{Antinucci:2024zjp} by a local coupling, so at the end of the day we would like to identify the parameter $\mathrm R$ in the above action with the parameter in the SymTFT that we have also denoted as $\mathrm R$ (we have introduced the $\mathrm R$ factor in \ref{Generalized gauging in p-form gauge theory in d dimensions} such that it coincides with the parameter in the SymTFT). Therefore, we expect the above action to be a $U(1)$ gauge theory, and some values of $\mathrm R$ should describe subgroup gaugings. 

We concentrate on the second part of the above theory. The $b$ field acts as a Lagrange multiplier setting $dc=0$ as well as $\int c \in \frac{2\pi \Z}{\mathrm R}$. On the other hand, the equations of motion for $\mathsf a$, equipped with large gauge invariance, enforces $db=0$ and $\int b \in 2\pi \mathrm R \Z$. The appearance of $\mathrm R$ in the nominator and the denominator of the periods of $b$ and $c$ is the crucial effect this coupling achieves. 

Due to the flatness conditions and the period quantizations, different values of $\mathrm R$ achieve subgroup gaugings. Suppose $\mathrm R = q \in \Z$. Then the topological operator $\exp \big( i \alpha \int c_{d-p-1} \big)$ generates a $\Z_q^{(p)}$ symmetry since $\alpha = 0 \text{ mod } q$ is a trivial $(d-p-1)-$surface. When $\mathrm R = 1/k \in 1/\Z$, we conversely have $\exp \big( i \alpha \int b_{p+1} \big)$ as a $\Z_k^{(d-p-2)}$ generator since $\alpha = 0 \text { mod } k$ is a trivial $(p+1)-$surface. Lastly, when $\mathrm R = \frac{q}{k} \in \mathbb Q/ \Z$ (we assume gcd$(q,k)$=1), the operators $\exp \big( i \alpha \int c_{d-p-1} \big)$ and $\exp \big( i \beta \int b_{p+1} \big)$ are trivial at $\alpha = 0 \text{ mod } q$ and $\beta = 0 \text{ mod } k$ respectively. 

From the $b$ field equation $c \sim d\mathsf a$ and the $c$ field equation $b \sim *c \sim *d\mathsf a$, we can intuitively regard these two surfaces as the electric and magnetic topological operators of the compact gauge theory. Indeed, upon integrating $b$, we end up precisely at the $(d-p-2)$ compact gauge theory which has the $U(1)^{(p)} \times U(1)^{(d-p-2)}$ symmetry structure. The corresponding subgroup gaugings of the two global symmetries are controlled by the parameter $\mathrm R$ in the non-compact theory. Furthermore, inspection of the SymTFT boundary conditions and the discussion here reveals that the two $\mathrm R$ parameters can be identified! This is so because when $\mathrm R = q/k \in \mathbb Q/\Z$, the corresponding subgroup gauging is $\Z_k^{(p)} \times \Z_q^{(d-p-2)}$ for both the local coupling and the SymTFT Lagrangian algebra choice \cite{Antinucci:2024zjp,Argurio:2024ewp}.

\section{Conclusions \& Outlook}

Motivated by the recent interest towards non-compact TQFTs as SymTFT for continuous symmetries, we have studied the local coupling of the $\R/\R$ and $U(1)/\R$ TQFTs to a QFT, in particular a variety of different abelian non-compact gauge theory models. It seems that there are many curiosities that arises from the study of topological aspects of non-compact abelian gauge theory, and we hope that these ideas may inspire new developments in the future, especially in the context of (bosonic) string theory and sigma models with target spaces having $\R^D$ factors. Some thoughts along these lines are outlined in section \ref{subsec 3.7: small discussion on bosonic string theory}. 

We first note that following the SymTFT results \cite{Antinucci:2024zjp,Brennan:2024fgj}, the works \cite{Argurio:2024ewp, Paznokas:2025epc} have already explored some of the manipulations in this paper (particularly $\Z \subset \R$ gauging, and condensations of $\R$ symmetries on codimension-1). In those works, these manipulations were used to understand topological defects of compact theories. As it turns out, manipulating $\R$ symmetries is a convenient tool to generalize duality defects of compact gauge theories at arbitrary values of coupling constants, not only rational ones \cite{Kaidi:2021xfk,Choi:2021kmx,Thorngren:2021yso}. This is the content of the papers \cite{Argurio:2024ewp, Paznokas:2025epc}. Our focus has been to stay in the non-compact gauge theory zone, and explore tangential ideas while trying to overcome the obstacles of working with non-compact groups. 

Importantly, our gauging manipulations, realized by local TQFT couplings, cover the entire family of topological boundary conditions of the non-compact SymTFT discussed in \cite{Antinucci:2024zjp}. This we achieved in the most generic way for a $p-$form pure gauge theory in section \ref{sec 6: p-form non-compact gauge theory in arbitrary dimension}, which presumably is the most straightforward realization of an $\R^{(p)}$ symmetry in a $d-$dimensional continuum QFT. As we have observed in different versions of Abelian gauge theory, the TQFT couplings prescribe various gaugings of $\R$ groups, such as the flat $\R$ gauging, the $\Z \subset \R$ subgroup gauging, and a further $\Z_p$ subgroup gauging in the resulting compact theory, all of which are encoded in the SymTFT \cite{Antinucci:2024zjp}. Going back to the table in the introduction, we fill in the question marks as
\begin{equation*}
\begin{aligned}
    \text{Codimension-}(-1) \text{ coupling of } \Z_p \text{ TQFT} \quad &\mapsto \quad \text{SymTFT for } \Z_p \\
    \text{Codimension-} \phantom{-} 0 \phantom{-} \text{ coupling of } \Z_p \text{ TQFT} \quad &\mapsto \quad \text{Gauging } \Z_p \subset U(1) \\ 
    \phantom{asdfge} \\
    \text{Codimension-}(-1) \text{ coupling of } \R/\R \text{ TQFT} \quad &\mapsto \quad \text{SymTFT for } \R \\
    \text{Codimension-} \phantom{-} 0 \phantom{-} \text{ coupling of } \R/\R \text{ TQFT} \quad &\mapsto \quad \text{Gauging } \R \\ 
    \phantom{asdfge}\\
    \text{Codimension-}(-1) \text{ coupling of } U(1)/\R \text{ TQFT} \quad &\mapsto \quad \text{SymTFT for } U(1) \\
    \text{Codimension-} \phantom{-} 0 \phantom{-} \text{ coupling of } U(1)/\R \text{ TQFT} \quad &\mapsto \quad \text{Gauging } \Z \subset \R
\end{aligned}
\end{equation*}
Employing the $\R/\R$ TQFT coupling on codimension-0, and exploiting the invariance of particular abelian gauge theories under flat gauging of their non-compact shift symmetries, we constructed topological defects via the half-space gauging. An initial expectation was for these defects to obey a $\mathsf{TY}(\R)$ like fusion algebra, but from the action of the defects it is clear that they have an invertible action, so cannot obey $\mathsf{TY}(\R)$ rules. We learned that the defects act as $\R^\times$ generators, and they have a non-trivial mixing with the shift generators. These point to an intriguing topological structure in the non-compact boson theory, which extends the symmetry to a $\textsf{Heis}(\R \times \R)$ group, which is a natural structure to appear if one regards the non-compact boson as a WZW model with $G = \R$. 

Going back to the fusion of the defects, one could also talk about condensation defects gauging the shift symmetry on codimension-1. As we have seen, this produces trivially acting condensation defects, which is compatible with the invertibility of the defects. Apart from the condensation obtained by higher-gauging $\R$, we additionally discussed condensations related to $\Z\subset \R$ gaugings on a codimension-1 wall. The nature of these object are not clear at this stage, and we did not try to understand them within this paper. It would be nice to develop these defects further and understand their properties. 

It is also interesting to ponder about the possible constraints on dynamical questions with systems possessing $\R^{(p)}$ symmetries, which are also invariant under flat $\R^{(p)}$ gaugings so they allow for the type of defects we constructed in this paper. In any case, $p-$form abelian gauge theory does not seem to be an appropriate context for understanding dynamical constraints, since the $\R$ valued shifts would be preserved only for deformations by local operators of the form $\mathcal O^i$ where $\mathcal O = (f_{p+1})^2$ and with $f_{p+1} = da_p$ being the field strength of the $p-$form field. For deformations with $i>1$, the invariance under flat gauging does not hold (at least not straightforwardly) so these deformations can be thought to produce a nonzero tension for the topological defect, which makes it non-topological. 

Another point of discussion concerns the $(-1)$-form and $(d-1)-$form $\R$ symmetries. The SymTFT for these edge cases point out, from the mixed boundary conditions, a topological manipulation such that by gauging a $\Z \subset \R$ subgroup, one obtains a $U(1)^{(-1)} \times U(1)^{(d-1)}$ symmetry. We provided explicit realizations of this gauging in the context of non-compact $p-$form gauge theory in $d=p+1$ dimensions. We gave a prescription to gauge the $\Z$ subgroup of the $\R^{(d-1)}$ symmetry, yielding the $U(1)$ $(d-1)-$form gauge theory that has the correct symmetries. Motivated by this manipulation, we provided an extension for an arbitrary rank gauge theory in any dimensions. These local realizations are all in one-to-one correspondence with the family of Lagrangian algebras for the SymTFT of a $p-$form $\R$ symmetry \cite{Antinucci:2024zjp}.

Given that the SymTFT is a theory independent construction, one should suspect that there are other explicit cases where one can manipulate a $\R^{(-1)}$ or $\R^{(d-1)}$ (or in general an $\R^{(p)}$ symmetry) to obtain two $U(1)$ symmetries, one instantonic and one leading to decomposition (or the corresponding statement for $p-$form symmetry). It would be interesting to understand these manipulations in other models. Relatedly, another piquant question to ask is what theories other than the pure abelian gauge theory has the $U(1)^{(p)} \times U(1)^{(d-p-2)}$ symmetry structure with the mixed 't Hooft anomaly. 

\bibliographystyle{JHEP}
\bibliography{biblio.bib}

\end{document}